\documentclass[titlepage]{article}
\usepackage[cp1250]{inputenc}
\usepackage[polish]{babel}
\usepackage[MeX]{polski}
\usepackage{indentfirst}
\usepackage{changepage}
\usepackage{enumerate}
\usepackage{graphicx}
\usepackage{geometry}
\usepackage{amsfonts}
\usepackage{amsmath}
\usepackage{caption}
\usepackage{amsthm}
\usepackage{color}
\usepackage{float}
\usepackage{url}
\usepackage{color,soul}
\usepackage{slashed}
\usepackage{bbold}
\usepackage{colortbl}
\usepackage{hyperref}

\newcommand{\pustastrona}
	{
	\newpage
	\null
	\thispagestyle{empty}
	\addtocounter{page}{-1}
	\newpage
	}

\definecolor{col:bckgrnd}{rgb}{0.7, 1.0, 0.3}
\definecolor{col:uwaga}{rgb}{0.0, 0.0, 0.7}
\definecolor{col:dopoprawy}{rgb}{0.3, 0.5, 0.0}
\definecolor{col:gray}{gray}{0.7}
\definecolor{col:lightgray}{gray}{0.8}
\newcommand{\xxx}[1]{{\color{col:uwaga}\bf\sethlcolor{col:bckgrnd}\hl{#1}}}

\newcommand{\kom}[1]{}

\newcommand{\pd}{\partial}
\newcommand{\LHS}{\text{LHS}}

\newcommand{\oddo}[2]{\Big{|}_{#1}^{#2}}
\newcommand{\M}{\mathcal{M}}
\newcommand{\G}{\mathcal{G}}
\newcommand{\Lag}{\mathcal{L}}
\newcommand{\F}{\mathcal{F}}
\newcommand{\D}{\mathcal{D}}
\newcommand{\C}{\mathcal{C}}
\newcommand{\R}{\mathcal{R}}
\newcommand{\Z}{\mathbb{Z}}
\newcommand{\mom}{p}
\newcommand{\pres}{\mathcal{P}}

\newcommand{\Ric}{\mathcal{R}}

\newcommand{\gev}{\text{~GeV}}
\newcommand{\e}[1]{\cdot 10^{#1}}
\newcommand{\rad}{\text{rad}}
\newcommand{\mat}{\text{mat}}

\newcommand{\mpl}{m_\text{Pl}}
\newcommand{\vev}[1]{\langle #1 \rangle}
\newcommand{\p}{\prime}
\newcommand{\mref}{m_\text{ref}}
\newcommand{\lsim}{\raisebox{-0.13cm}{~\shortstack{$<$ \\[-0.07cm]$\sim$}}~} 
\newcommand{\FLRW}{Friedmann-Lema{\^\i}tre-Robertson-Walker}
\newcommand{\micromegas}{\protect\url{micrOMEGAs}}
\newcommand{\Vd}[2]
{\left(\begin{array}{@{}c@{}}
#1\\#2\\
\end{array}\right)}
\newcommand{\Md}[4]
{\left[\begin{array}{@{}cc@{}}
	#1&#2\\#3&#4\\
\end{array}\right]}
\newcommand{\arxiv}[1]{\href{https://arxiv.org/abs/#1}{\UrlFont arXiv:#1}}

\let\stdforall\forall
\renewcommand{\forall}[1]{\mathop{\mathlarger{\mathlarger{\mathlarger{\mathlarger\stdforall}}}}_{#1}\quad}

\begin{document}

\begin{titlepage}
	\begin{center}
		{\LARGE\textbf{University of Warsaw}\\Faculty of Physics\\}
		\vspace{2cm}
		{\Large\textbf{Michał Iglicki}}\\
		{\large Student's book no.: 332475}\\
		\vspace{2cm}
		{\LARGE\textbf{Vector-fermion dark matter}}\\
		\vspace{1cm}
		{\large second cycle degree thesis\\field of study: Physics\\speciality: Theoretical Physics\\within College of the Inter-Faculty Individual Studies in Mathematics\\and Natural Sciences}\\
		\vspace{2cm}
	\end{center}
	\begin{flushright}
		{\large The thesis written under the supervision of\\{\bf prof. dr hab. Bohdan Grządkowski}\\Institute of Theoretical Physics,\\Faculty of Physics, University of Warsaw}
	\end{flushright}
	\begin{center}
		\vspace{2cm}
		{\large Warsaw, September 2017\\\normalsize(last edition: April 2018)}
	\end{center}
\end{titlepage}

\thispagestyle {empty}

		\noindent{\bf Oświadczenie kierującego pracą}\\

		\noindent Potwierdzam, że niniejsza praca została przygotowana pod moim kierunkiem i~kwalifikuje się do przedstawienia jej w~postępowaniu o~nadanie tytułu zawodowego.\\\\

		\noindent Data \hfill Podpis kierującego pracą\\

		\noindent{\bf Statement of the Supervisor on Submission of the Thesis}\\

		\noindent I hereby certify that the thesis submitted has been prepared under my supervision and I declare that it satisfies the requirements of submission in the proceedings for the award of a degree.\\\\

		\noindent Date \hfill Signature of the Supervisor:\\
\\~\\
		\noindent{\bf Oświadczenie autora pracy}\\

		\noindent Świadom odpowiedzialności prawnej oświadczam, że niniejsza praca dyplomowa 
została napisana przez mnie samodzielnie i~nie zawiera treści uzyskanych w~sposób niezgodny z~obowiązującymi przepisami.\\

		\noindent Oświadczam również, że przedstawiona praca nie była wcześniej przedmiotem procedur związanych z~uzyskaniem tytułu zawodowego w~wyższej uczelni.\\

		\noindent Oświadczam ponadto, że niniejsza wersja pracy jest identyczna z~załączoną wersją elektroniczną.\\\\

		\noindent Data \hfill Podpis autora pracy
\\~\\
		\noindent{\bf Statement of the Author on Submission of the Thesis}\\

		\noindent Aware of legal liability I certify that the thesis submitted has been prepared by myself and does not include information gathered contrary to the law.\\

		\noindent I also declare that the thesis submitted has not been the subject of proceedings resulting in the award of a university degree.\\

		\noindent Furthermore I certify that the submitted version of the thesis is identical with its attached electronic version.\\\\

		\noindent Date \hfill Signature of the Author of the thesis

\pagebreak
\thispagestyle{empty}
~\\\\\\\\\\\\

	\begin{center}
		\large{\textbf{Summary}}
	\end{center}
	\noindent W pracy zbadano proste rozszerzenie Modelu Standardowego. Grupa cechowania Modelu Standardowego została rozszerzona o dodatkową grupę $U(1)_X$. Model wprowadza dodatkową cząstkę Higgsa oraz trzy cząstki (bozon cechowania i dwa fermiony Majorany) mogące być kandydatami na ciemną materię. Praca zawiera również dyskusję wyprowadzenia równania Boltzmanna wraz ze szczegółową analizą założeń. Równanie zostało wykorzystane w celu zbadania zachowania się gęstości ciemnej materii w czasie w zależności od wartości parametrów modelu.

	\begin{center}
		\large{\textbf{Key words}}
	\end{center}

		\noindent ciemna materia, równanie Boltzmanna, fizyka poza Modelem Standardowym

	\begin{center}
		\large{\textbf{Area of study (codes according to Erasmus Subject Area Codes List)}}
	\end{center}

		\noindent 13.2 Physics

	\begin{center}
		\large{\textbf{The title of the thesis in Polish}}

	\noindent Wektorowo-fermionowa ciemna materia
	\end{center}

\pustastrona

\renewcommand{\contentsname}{Contents}
\tableofcontents
\pagebreak
\setcounter{section}{-1}
\begin{section}{Introduction}
There exists solid experimental evidence for existence of dark matter (DM) (see section \ref{sec:dm1}). According to majority of current models, DM should consist of massive, non-relativistic, stable particles that do not interact with electromagnetic field. In the Standard Model of fundamental interactions there is no proper candidate for such a particle, therefore investigation of dark matter can lead to new beyond Standard Model physics.\\

The standard approach to the dark matter problem is the hypothesis of so called WIMP's, i.e. weakly interacting massive particles. The WIMP model explains observed large (cosmological) scale effects considered to be a consequence of the existence of dark matter. However, at galactic scales there exist some discrepancies between observations and simulations of the WIMP models (see section \ref{sec:dm2}), which lead to the conclusion that dark matter might consist of more than one component.\\

In this thesis we provide a simple, but QFT-consistent and renormalizable model of dark matter, which is an extension of the Standard Model. We extend the Standard Model gauge group with an additional $U(1)_X$ group. The model provides additional Higgs particle, as well as two or three (depending on the values of the model parameters) candidates for dark matter. We investigate the influence of the model parameters on the shape of solutions of the Boltzmann equations that describe dependence of the number density of a given kind of particles on temperature of the Universe (which can be treated as a time scale).\\

We also provide a detailed derivation of the Boltzmann equation, including analysis of all the assumptions (what is usually at least partially omitted in cosmology textbooks).\\

Within this project, a \protect\url{C++} code that numerically solves a set of Boltzmann equations for multi-component dark matter has been developed. In contrast to \protect\url{micrOMEGAs}, a package popular among dark matter physicists, our code deals with 3 component dark matter as well as with 2 component case.
\begin{subsection}{Notation and conventions}
Throughout the whole thesis, following conventions are used:
	\begin{itemize}
	\item Physical constants: $\hbar$ (Dirac constant), $c$ (velocity of light in vacuum) and $k_B$ (Boltzmann constant) are set to 1. Therefore all the physical quantities are expressed in units of energy taken with appropriate power (the usual unit is gigaelectronovolt, $1\gev\approx~1.78\e{-36}\text{~kg}\cdot c^2\approx 1.602\e{-19}\text{~J}$):
	\begin{table}[H]\begin{center}\begin{tabular}{|c|c|c|}
	\hline
	&corresponding\\
	quantity&power of energy\\
	\hline
	mass, momentum, temperature&1\\length, time&-1\\
	\hline
	$G$ (Newton's constant)&-2\\
	$\Lambda$ (cosmological constant)&2\\
	\hline
	\end{tabular}\end{center}\end{table}
	\item Greek indices run from 0 to 3 (or, equivalently, over time and all spatial coordinates) while latin indices from 1 to 3 (over spatial coordinates only).
	\item Einstein summation convention: when an index is repeated, we implicitly sum over all its possible values, e.g. $A^{\alpha\mu}B_{\mu\beta}\equiv\sum_{\mu=0}^3 A^{\alpha\mu}B_{\mu\beta}$.
	\item Sign conventions (the same as in Kolb \& Turner \cite{bib:kt}):
		\begin{itemize}
		\item for the metric tensor we assume $(+,-,-,-)$ signature,
		\item the Riemann tensor, Ricci tensor and Ricci scalar are defined as
			\begin{gather*}
			R^\alpha_{\;\beta\mu\nu}=\Gamma^\alpha_{\;\beta\nu,\mu}-\Gamma^\alpha_{\;\beta\mu,\nu}+
			\Gamma^\alpha_{\;\mu\lambda}\Gamma^\lambda_{\;\beta\nu}
			-\Gamma^\alpha_{\;\nu\lambda}\Gamma^\lambda_{\;\beta\mu}\;,\qquad
			R^\alpha_{\;\beta}=R^\mu_{\;\alpha\mu\beta}\;,\qquad
			\Ric=R^\mu_{\;\mu}\;,
			\end{gather*}
		\item the Einstein equation takes the following form:
			$$R_{\mu\nu}-\frac{1}{2}g_{\mu\nu}\Ric=+8\pi GT_{\mu\nu}+g_{\mu\nu}\Lambda\;,$$
		\item the Christoffel symbols are defined as follows:
			$$\Gamma^\alpha_{\;\beta\gamma}=
			\frac{1}{2}g^{\alpha\mu}(g_{\mu\beta,\gamma}+g_{\mu\gamma,\beta}-g_{\beta\gamma,\mu})\;.$$
		\end{itemize}
	\item When opposite is not explicitly said, $t_0$ denotes the current time (,,now''). The current values of other time-dependent cosmological quantities, like scale factor or density of the Universe, are also denoted with index $0$ (e.g. $\rho_0=\rho(t_0)$, $\dot R_0\equiv \dot R(t_0)$).
	\item To avoid confusion, momentum is denoted as $\mom$, while $\pres$ denotes pressure.
	\item The scale factor $R$ in the FLRW metric has dimension of length, while the comoving distance $r$ and the curvature $k$ are dimensionless.
	\item In chapter \ref{sec:vfdm} we use some common notation shortcuts:
	\begin{align*}
	\bar\psi&\equiv\psi^\dag\gamma_0\quad\text{($\psi$ -- a fermionic field)}, & \slashed a&\equiv\gamma^\mu a_\mu \quad \text{($a$ -- a vector field)},
	\end{align*}
	where $\gamma^\mu$ denotes the $\mu$-th Dirac matrix.
	\end{itemize}
	\end{subsection}
\end{section}
\begin{section}{Dark matter -- definition and motivation}\label{sec:dm}
Dark matter is a kind of matter which interacts gravitationally, but not electromagnetically (so it is ,,dark''). Existence of dark matter is confirmed by many observations of various kinds, some of them are described below. The total amount of dark matter should be around five times bigger than that of ordinary matter. Currently, the most often considered cosmological model is the cold dark matter model with a cosmological constant, often abbreviated to $\Lambda$CDM. The coldness (i.e. low velocity) of dark matter is needed to fit the structure forming data (see \cite{bib:structure}). In the Standard Model there is no proper candidate for cold dark matter (it should be massive, stable and not interacting with photons)\footnote{One of hypothetical models of dark matter is so called \emph{MACHO}, which comes from \emph{Massive Compact Halo Object}. It assumes that the lacking mass comes from macroscopic objects, such as asteroids or even planets, which do not emit observable radiation. However, theoretical considerations show that the possible contribution of baryons for the total amount of dark matter is far too small to explain observed effects, see e.g. \cite{bib:macho}.}, dark matter is considered to consist of so called \emph{weakly interacting massive particles} or \emph{WIMP's}, not yet discovered kind of heavy, stable and ,,dark'' elementary particles.
	\begin{subsection}{Existence of dark matter}\label{sec:dm1}
		\begin{subsubsection}{Rotation curves}
		The earliest argument for existence of dark matter is analysis of the galaxies' and galaxy clusters' rotation curves. This was first formulated by Fritz Zwicky in 1933 \cite{bib:zwicky}. The idea is that rotation of galaxies or galaxy clusters should lead them to fall apart because of the centrifugal force\footnote{Velocities of such motions are small compared to $c$, so use of Newtonian approximation is fully justified.}. To prevent this, the centrifugal force should be balanced by the gravitational force:
		\begin{align}\label{eq:rotcurve}
		G\frac{M(r)\Delta m}{r^2}=\frac{\Delta mv(r)^2}{r}\Rightarrow v(r)=\sqrt{G\frac{M(r)}{r}}.
		\end{align}
		Here $v(r)$ denotes the velocity of the rotational movement of a given element of galaxy of mass $\Delta m$, situated in distance $r$ from the galaxy center. $M(r)$ stands for the total mass of the galaxy enclosed within the ball of radius $r$; the ordinary matter contribution can be estimated by observations -- since it is electrically charged, it emits radiation that can be detected.

Mass of observed matter does not explain strength of gravitation that would prevent the falling apart, therefore Zwicky suggested existence of some invisible (that is, not interacting electromagnetically) ,,dark matter'' (\emph{dunkle Materie}) which contributes to mass of galaxies. These observations were confirmed many times, nowadays they are presented using so called \emph{rotation curves} -- dependence between distance from the center of the galaxy and rotation velocity (see fig. \ref{fig:rotcurve}).
			\begin{figure}[H]
			\begin{center}
				\includegraphics[width=.5\textwidth]{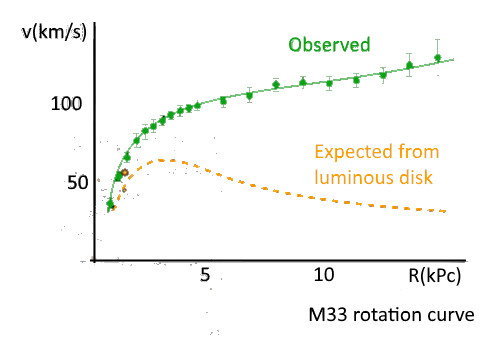}
				\caption{Rotation curve of the M33 galaxy. The green line corresponds to observed distribution of rotational velocity, the orange one -- to distribution derived from eq. (\ref{eq:rotcurve}) using mass of observed luminous matter as $M(r)$. Based on \cite{bib:rotcurve}}
				\label{fig:rotcurve}
			\end{center}
			\end{figure}
		\end{subsubsection}
		\begin{subsubsection}{Gravitational lensing}
		The most direct proof for the existence of dark matter are observations of so called \emph{gravitational lensing effect}. Since gravitational field interacts with light, deflecting its trajectory, analysis of observed images can be used to reproduce the mass distribution in the observed area. It appears often that the distribution of luminous matter is not sufficient to explain the observed lensing effect. Therefore it is assumed that the lacking part of mass that produces reconstructed gravitational field is dark matter. One of the most spectacular examples is the Bullet Cluster (cluster 1E 0657-558), see fig. \ref{fig:bullet}. 
			\begin{figure}[H]
			\begin{center}
				\includegraphics[width=.6\textwidth]{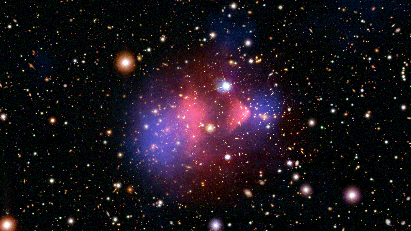}
				\caption{The Bullet Cluster. Pink and blue colouring is artificially added. Pink corresponds to the luminous matter distribution (observed in X-ray range), blue is the distribution reconstructed from the gravitational lensing effects. Source: \protect\url{https://apod.nasa.gov/apod/ap060824.html}}
				\label{fig:bullet}
			\end{center}
			\end{figure}
		This picture is interpreted as an image of a collision of two galaxy clusters (one coming from left and one from right). Electromagnetically interacting luminous matter (pink) slowed down, while dark matter (blue), whose interactions are much weaker, passed through.
		\end{subsubsection}
		\begin{subsubsection}{Non-uniformity of the CMBR}
		The amount of non-baryonic matter in the Universe can be derived from observations of the cosmic microwave background fluctuations, what is described with more details in \cite{bib:dm}. The main idea is that the primordial density anisotropies have attracted both baryonic and dark matter. Since matter was ionized that time, increasing of the density led to increased photon pressure, which repelled ordinary matter interacting with photons, but did not repel dark matter. The ratio between the amounts of baryonic and dark matter influences the multipole expansion of the present spectrum of the CMBR fluctuations, which is measured with a very good accuracy, e.g. by Planck \cite{bib:planck}.
		\end{subsubsection}
	\end{subsection}
	\begin{subsection}{Multi-component dark matter}\label{sec:dm2}
		Observations in big scales, such as non-uniformity of the CMBR, are perfectly explained by the one component WIMP models. However, at galactic scales there appear problems that can be solved by introducing dark matter self-interactions, or even, more than one DM component. These small scale discrepancies, as well as possible solutions, are described in numerous sources, e.g. \cite{bib:multidm1,bib:multidm2}. Here we briefly describe the most often mentioned problems.
		\begin{subsubsection}{The core-cusp problem}
		N-body simulations based of one-component, collisionless cold dark matter models (see e.g. \cite{bib:cusp}) indicate that density profile of the dark matter halo should steeply decrease with the distance from the centers of the galaxies. However, observations lead to constraints that these profiles have rather flat core (see \cite{bib:cusp2}).
		\end{subsubsection}
		\begin{subsubsection}{The missing satellites problem and the too-big-to-fail problem}
		Simulations indicate also that the dark matter haloes of the big galaxies, such as Milky Way, should be surrounded by a large number of smaller sub-haloes, attracting ordinary matter and therefore leading to formation of satellite galaxies. According to \cite{bib:multidm1}, our Galaxy should have several hundreds of subhaloes, while the known number of satellite galaxies is of the order of ten (this is called the \emph{missing satellites} problem). Possible solution is that most of the subhaloes are too small to attract the sufficient amount of baryonic matter. However, size of the largest predicted subhaloes should cause formation of larger satellite galaxies than observed. This discrepancy is known as the \emph{too-big-to-fail} problem.
		\end{subsubsection}
	\end{subsection}
\end{section}
\begin{section}{Basics of cosmology}
\begin{subsection}{Cosmological principle}\label{sec:cosmo}
Cosmology studies the Universe as a whole, therefore it operates with very large scales of distance. The size of our Galaxy is ca. 30 kpc (diameter), the sizes of galaxy clusters are usually of order of 10 Mpc. The scales considered in comology are of order of 100 Mpc and higher. In such large scales the Universe seems to be isotropic and homogeneous (equivalently: isotropic from each point of space) -- these conditions are the usual assumption in cosmological studies. We call this assumption the \emph{cosmological} (or \emph{Copernican}) \emph{principle}.

The main experimental argument for the cosmological principle are observations of the cosmic microwave background radiation (CMBR). This radiation consists of photons that decoupled from matter when the Universe was approximately 300~thousand years old, and has thermal distribution corresponding to temperature $2.725$~K. The observed spatial distribution of this temperature is homogeneous up to about $0.01\%$.
			\begin{figure}[H]
			\begin{center}
				\includegraphics[width=.6\textwidth]{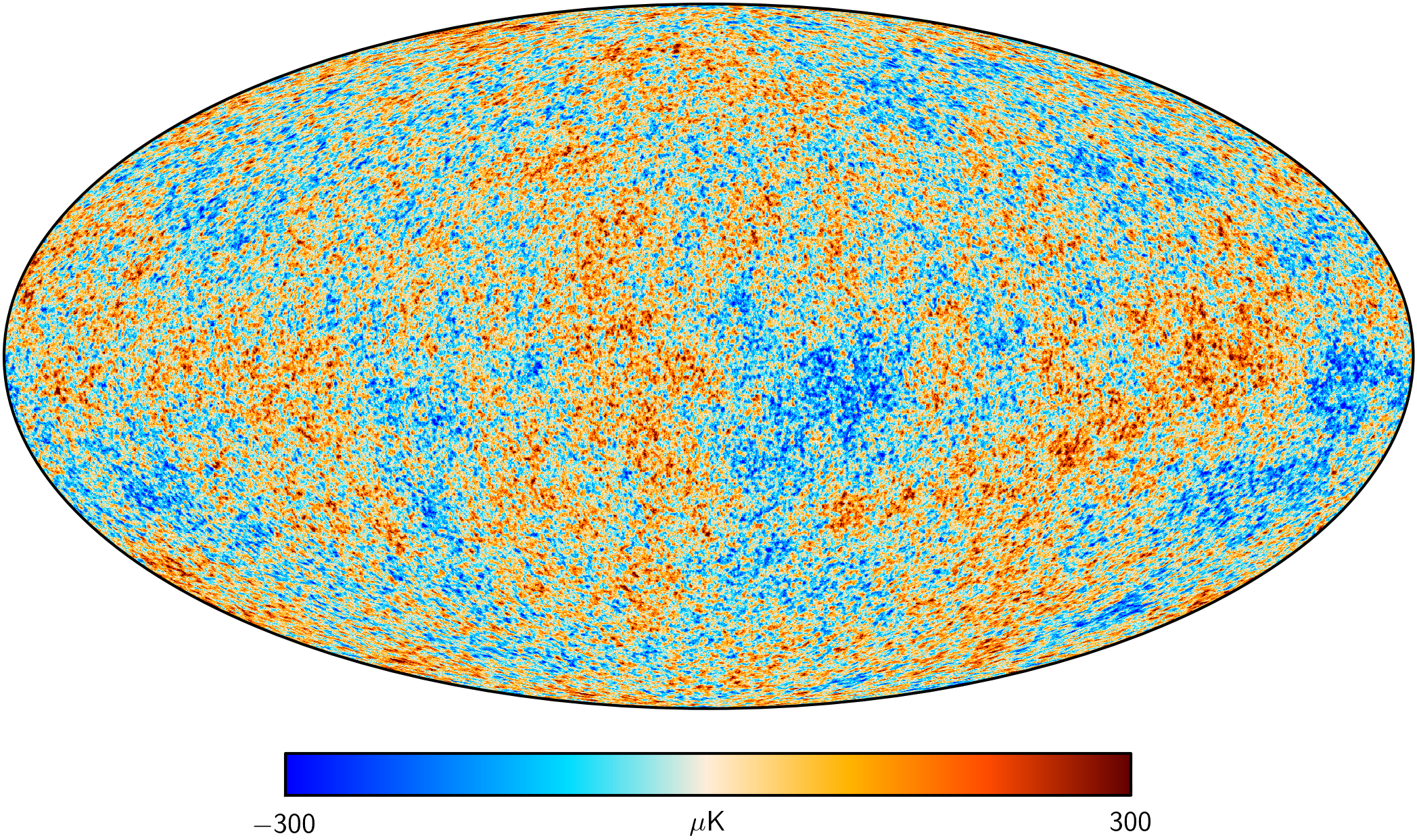}
				\caption{Planck mission data of the Cosmic Microwave Background. Source: \protect\url{http://sci.esa.int/science-e-media/img/61/Planck_CMB_Mollweide_4k.jpg}}
			\end{center}
			\end{figure}
The cosmological principle has three main mathematical consequences, which will be used throughout the thesis:
	\begin{enumerate}
	\item The metric of the Universe should be isotropic and spatially homogeneous.
	\item For any kind of particles, their phase-space distribution function, $f(\mom^\mu,x^\mu)$, should not depend on the spatial position and the direction of momentum. Hence, it depends on time and energy only: $f(\mom^\mu,x^\mu)=f(E,t)$.
	\item The energy-momentum tensor of the Universe takes the same form as for the perfect fluid\footnote{Actually, this is not the only possibility -- sometimes the imperfect fluid with bulk viscosity is considered (only shear viscosity has to vanish because of isotropy), see e.g. \cite{bib:imperfect1,bib:imperfect2,bib:imperfect3}. Perfect fluid is the simplest possibility and is sufficient to describe radiation and dust, which are considered the main components of the Universe.}.
	\end{enumerate}
\end{subsection}
\begin{subsection}{The cosmic time}
The Universe is not isotropic in \emph{all possible} frames. The cosmological principle states only that there \emph{exists} such a frame and the Universe can be treated as a perfect fluid in \emph{this specific frame} only. We will call an observer who is stationary in this frame a \emph{fundamental observer}.

\emph{Cosmic time} is the time measured by the fundamental observer. Observers moving relatively to the fundamental one measure different time and for them the Universe is not isotropic (they will observe larger density of the Universe in direction of their movement and smaller density in opposite direction).

Whenever in this thesis \emph{time} is mentioned, it should be understood as the cosmic time. We will denote it simply by $t$.
\end{subsection}
\begin{subsection}{Friedmann-Lema\^\i tre-Robertson-Walker metric}\label{sec:flrw}
According to the cosmological principle, the metric of the Universe should be isotropic and homogeneous. In 1935 H. P. Robertson and A. G. Walker showed that the most general metric satisfying these assumptions is so called \emph{Friedmann-Lema\^\i tre-Robertson-Walker metric}, which is just a Euclidean metric with spatially-constant \emph{curvature} $k(t)$ and global time-dependent \emph{scale factor} $R(t)$ added:
$$ds^2=dt^2-R^2(t)\left(\frac{dr^2}{1-k(t)r^2}+r^2(d\theta^2+\sin^2\theta d\phi^2)\right).$$
Note that for non-zero curvature one can always rescale it to $\pm 1$:
$$r\rightarrow|k|^\frac{1}{2}r\,,\quad R\rightarrow|k|^{-\frac{1}{2}}R\,,\quad k\rightarrow|k|^{-1}k=\pm 1.$$
Since the sign of the curvature factor remains constant all the time (what is discussed in \cite{bib:curv}), we can use such variables in which $k(t)\equiv 1$ or $k(t)\equiv-1$.

The ratio between the time derivative of the scale factor and the scale factor itself is called the \emph{Hubble parameter} and denoted by $H$:
$$H\equiv\frac{\dot R(t)}{R(t)}.$$
The current value of the Hubble parameter is, according to the Planck data \cite{bib:planck}, $H_0\approx 67\frac{\text{km}/\text{s}}{\text{Mpc}}$.
\end{subsection}
\begin{subsection}{Expansion of the Universe, redshift and the Hubble law}
The Universe expands, what is described by changes of the scale factor $R(t)$. The important relation of observational cosmology connected with the expansion is so called \emph{redshift equation}, that describes change of the wavelength of light emitted when the scale factor was equal to $R_e$ and observed now, with the scale factor equal to $R_0$:
\begin{align}\label{eq:z}
z\equiv\frac{\lambda'-\lambda}{\lambda}=\frac{R(t_0)}{R(t_e)}-1.
\end{align}
Parameter $z$ is called the redshift and is sometimes used as a time scale of the Universe, since it corresponds directly to the scale factor.

Redshift can be known from observations -- for example, if we observe a star light with the spectral lines corresponding to known elements, but shifted towards lower energies, we can calculate $z$. It can be further used to determine the Hubble parameter, by using the Hubble law:
$$d_lH_0=z+\frac{1}{2}(1-q_0)z^2+\ldots$$
Here $d_l$ denotes the luminosity distance from the light source, defined as the distance in a static Universe that would lead to such changes in luminosity of a given object as observed. Absolute luminosity of a given observed object can often be theoretically predicted (for example if it is a supernova of known type), therefore $d_l$ is easy to measure. The acceleration parameter $q_0$ is defined as
$$q_0\equiv-\frac{\ddot R_0R_0}{\dot R_0^2}.$$
			\begin{figure}[H]
			\begin{center}
				\includegraphics[width=.5\textwidth]{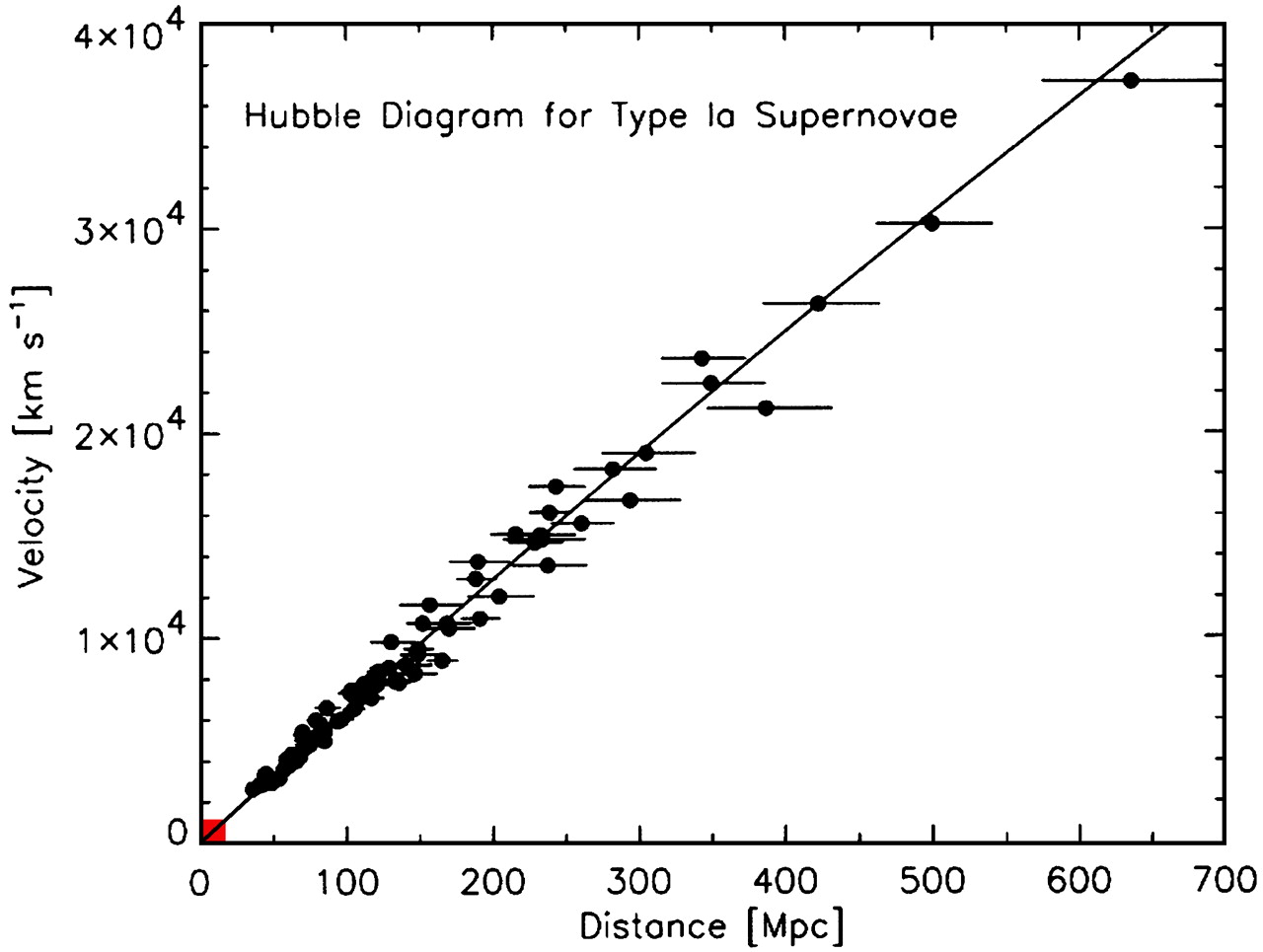}
				\caption{The Hubble diagram for Type Ia Supernovae. In the non-relativistic regime velocity is proportional to the redshift. Source: \protect\url{http://www.pnas.org/content/101/1/8/F3.large.jpg}}
			\end{center}
			\end{figure}

\end{subsection}
\end{section}
\begin{section}{The Friedmann equations}
Having the FLRW metric and the momentum-energy tensor, one can combine them via the Einstein equation:
	$$R_{\mu\nu}-\frac{1}{2}\Ric g_{\mu\nu}=8\pi GT_{\mu\nu}+\Lambda g_{\mu\nu}.$$
As already mentioned in section \ref{sec:cosmo}, we work under an assumption that in large scales the contents of the Universe can be approximately treated as a perfect fluid, whose energy-momentum tensor is
	$$T^\mu_{\;\nu}=\left[\begin{array}{cccc}\rho&0&0&0\\0&-\pres&0&0\\0&0&-\pres&0\\0&0&0&-\pres\\\end{array}\right],$$
	where $\rho$ stands for density and $p$ for pressure. Hence
	$$T_{\mu\nu}=g_{\mu\alpha}T^\alpha_{\;\nu}=\left[
	\begin{array}{cccc}\rho&0&0&0\\0&\frac{R^2}{1-kr^2}\pres&0&0\\0&0&R^2r^2\pres&0\\0&0&0&R^2r^2\pres\sin^2\theta\end{array}
	\right].$$
The Ricci tensor for the FLRW metric reads
	$$R_{\mu\nu}=\left[\begin{array}{cccc}
	-3\frac{\ddot R}{R}&0&0&0\\0&\frac{2k+2\dot R^2+R\ddot R}{1-kr^2}&0&0\\
	0&0&(2k+2\dot R^2+R\ddot R)r^2&0\\0&0&0&(2k+2\dot R^2+R\ddot R)r^2\sin^2\theta
	\end{array}\right],$$
and the Ricci scalar is
	$$\Ric=-6\frac{k+\dot R^2+R\ddot R}{R^2}.$$
The $t-t$ component of the Einstein equation (after division by 3) is so called \emph{Friedmann evolution equation} (or just \emph{Friedmann equation}):
	\begin{align}\label{eq:fried}
	H^2+\frac{k}{R^2}=\frac{8\pi G}{3}\rho+\frac{\Lambda}{3},
	\end{align}
where $H=\frac{\dot R}{R}$ is the Hubble parameter.

The sum of the $t-t$ component (multiplied by $-\frac{1}{6}$) and the $r-r$ component (multiplied by $-\frac{1}{2}\frac{1-kr^2}{R^2}$) \kom{\xxx{sprawdzić po raz pięćdziesiąty}} we call the \emph{Friedmann acceleration equation} (or just  \emph{acceleration equation}):
\begin{align}\label{eq:acc}
\dot H+H^2=-\frac{4\pi G}{3}(\rho+3\pres)+\frac{\Lambda}{3}.
\end{align}
Taking the time derivative of (\ref{eq:fried}) and combining it with (\ref{eq:fried}) and (\ref{eq:acc}), we obtain the \emph{continuity equation}:
	\begin{align}\label{eq:cont}
	\dot\rho=-3H(\rho+\pres).
	\end{align}
To simplify the form of the Friedmann equations we can now introduce ,,density'' and ,,pressure'' functions corresponding to the cosmological constant $\Lambda$ and the curvature factor $k$:
	\begin{align*}
	\rho_\Lambda&\equiv\frac{\Lambda}{8\pi G},&\rho_k&\equiv-\frac{3k}{8\pi G}R^{-2},\\
	\pres_\Lambda&\equiv-\rho_\Lambda,&\pres_k&\equiv-\frac{1}{3}\rho_k.\\
	\end{align*}
We assume that the contents of the Universe can be approximately treated as a mixture of two kinds of objects:
	\begin{itemize}
	\item radiation (ultra-relativistic particles), which satisfies $p_\rad=\frac{1}{3}\rho_\rad$, 
	\item matter (non-relativistic objects; sometimes called \emph{dust}), which satisfies $p_\mat=0$.
	\end{itemize}
Of course, there also exist particles that are relativistic (so the kinetic energy is comparable to mass, what leads to non-zero pressure), but not ultra-relativistic (so $\rho=3\pres$ relation is not satisfied). Nevertheless, it is a common assumption in cosmology to treat contents of the Universe as a mixture of radiation and dust (see e.g. \cite{bib:dm}).

The Friedmann equations now read:
	\begin{gather}\label{eq:fried2}
		\begin{cases}
		H^2=\frac{8\pi G}{3}\rho\\
		\dot H+H^2=-\frac{4\pi G}{3}(\rho+3\pres)\\
		\dot\rho=-3H(\rho+\pres)
		\end{cases},
	\end{gather}
where
	\begin{align}\begin{aligned}\label{eq:rhop}
	\rho&=\rho_\rad+\rho_\mat+\rho_\Lambda+\rho_k,\\
	\pres&=\pres_\rad+\pres_\mat+\pres_\Lambda+\pres_k.
	\end{aligned}\end{align}
In case of matter, the energy density should evolve like $R^{-3}$ because it is proportional to the total amount of matter divided by the unit volume of the Universe.

For radiation, the density should evolve like $R^{-4}$. Apart the volume-increasing effect, also the wavelength of any radiation increases proportionally to $R$, according to the redshift equation:
$$1+z=\frac{R_0}{R_e},$$
lowering the total energy\footnote{One should understand that the assumption about constant total amount of matter or radiation is quite na\"\i ve, since e.g. matter can annihilate into photons, changing their number. This assumption works more properly if we neglect such interactions (because of expansion of the Universe or matter-antimatter asymmetry) or assume that annihilation and creation processes are in equilibrium. Interactions between particles of matter that produce particles of similar mass are allowed, since due to energy conservation they do not change the total energy density.}.\\
Hence:
	\begin{align}\begin{aligned}\label{eq:rhoR}
	\rho_\mat&=\rho^0_\mat\left(\frac{R}{R_0}\right)^{-3}, & \rho_\rad&=\rho^0_\rad\left(\frac{R}{R_0}\right)^{-4},\\
	\rho_\Lambda&=\rho^0_\Lambda, & \rho_k&=\rho^0_k\left(\frac{R}{R_0}\right)^{-2},
	\end{aligned}\end{align}
where
	\begin{align*}
	\rho^0_\Lambda&\equiv\frac{\Lambda}{8\pi G} & \rho^0_k&\equiv-\frac{3k}{8\pi GR_0^2}.
	\end{align*}
We define the critical density $\rho_c$ as a solution of the flat Universe ($k=0$) and no cosmological constant case:
$$\rho_c\equiv\frac{3H^2}{8\pi G}.$$
Because the Universe consists of many components, it is convenient to define the density parameter, which is a ratio of certain component's density and the critical one:
$$\Omega_\alpha\equiv\frac{\rho_\alpha}{\rho_c},$$
where $\alpha$ denotes given component.

Since experiments, like Planck \cite{bib:planck} and WMAP \cite{bib:wmap}, show that the Universe is flat or nearly flat, the density of our Universe is assumed to be the critical one.

\begin{subsection}{Solution of the Friedmann equation}
We are going to solve the Friedmann evolution equation:
$$\left(\frac{\dot R}{R}\right)^2=\frac{8\pi G}{3}\rho.$$
Taking square root\footnote{The Universe is expanding, so $\dot R>0$, therefore we take the square root with ,,+'' sign. Solution with negative sign (i.e. collapsing of the Universe) is possible only in the closed Universe case, ie. $k>0$.} and integrating over $dt$, we obtain
$$\int_0^t\frac{dR}{R\rho^\frac{1}{2}}=\sqrt{\frac{8\pi G}{3}}t.$$
Assuming the Universe can be treated as a mixture of dust (non-relativistic matter) and radiation (ultra-relativistic particles), in the presence of the curvature $k$ and the cosmological constant $\Lambda$ the equation reads
	\begin{align}\label{eq:rtgen}
	\sqrt{\frac{8\pi G}{3}}t=\int_0^{R(t)}\frac{dR}{\left[\rho^0_\rad R_0^4/R^2+\rho^0_\mat R_0^3/R+\rho^0_k R_0^2+\rho^0_\Lambda R^2\right]^\frac{1}{2}}.
	\end{align}
This integral cannot be calculated analytically. The usual assumption is, in the early Universe the scale factor $R$ is so small, that the radiation term is the only that matters. Let us now discuss it.\\
\begin{subsubsection}{Radiation domination epoch}
Radiation dominates when
	\begin{align*}
	\rho^0_\rad\left(\frac{R_0}{R}\right)^4&\gg\rho^0_\mat\left(\frac{R_0}{R}\right)^3
	&\Longleftrightarrow\qquad\qquad R&\ll R_0\frac{\rho^0_\rad}{\rho^0_\mat},\\
	\rho^0_\rad\left(\frac{R_0}{R}\right)^4&\gg\rho^0_k\left(\frac{R_0}{R}\right)^2
	&\Longleftrightarrow\qquad\qquad R&\ll R_0\left(\frac{\rho^0_\rad}{\rho^0_k}\right)^\frac{1}{2},\\
	\rho^0_\rad\left(\frac{R_0}{R}\right)^4&\gg\rho^0_\Lambda
	&\Longleftrightarrow\qquad\qquad R&\ll R_0\left(\frac{\rho^0_\rad}{\rho^0_\Lambda}\right)^\frac{1}{4}.
	\end{align*}
Present ratios of densities, appearing in these conditions, are (from \cite{bib:planck})
	$$\frac{\rho^0_\rad}{\rho^0_\mat}\approx 2.9\e{-4},\qquad
	\left(\frac{\rho^0_\rad}{\rho^0_k}\right)^\frac{1}{2}\approx \infty,\qquad
	\left(\frac{\rho^0_\rad}{\rho^0_\Lambda}\right)^\frac{1}{4}\approx 1.0\e{-1}.$$
According to \cite{bib:planck}, $\rho^0_k$ is too small to be comparable with other components' densities.

Hence, the most restricting condition is
	$$R\ll R_0\frac{\rho^0_\rad}{\rho^0_\mat}\qquad\Longleftrightarrow\qquad \frac{R}{R_0}\ll 2.9\e{-4}.$$
During the radiation domination epoch, equation (\ref{eq:rtgen}) simplifies to
	\begin{align}\notag
	\sqrt{\frac{8\pi G}{3}}t=\int_0^{R(t)}\frac{RdR}{\left[\rho^0_\rad\right]^\frac{1}{2}R_0^2}
	=\frac{R^2}{2\left(\rho^0_\rad\right)^\frac{1}{2}R_0^2}\\
	\label{eq:rt}
	\Longrightarrow\qquad t=\frac{1}{2}\sqrt{\frac{3}{8\pi G\rho^0_\rad}}\left(\frac{R}{R_0}\right)^2=
	\frac{1}{2H_0}\frac{1}{\sqrt{\Omega^0_\rad}}\left(\frac{R}{R_0}\right)^2.
	\end{align}
Now we can check for what times the scale factor condition is satisfied:
	$$\frac{R}{R_0}\ll 2.9\e{-4}\qquad\Longleftrightarrow\qquad
	t\ll\frac{1}{2H_0}\frac{8.41\e{-8}}{\sqrt{\Omega^0_\rad}}\equiv t_\rad.$$
Using the $\Omega^0_\rad$ and $H_0$ values from \cite{bib:planck}, we obtain the upper time boundary of the radiation dominated epoch:
	\begin{align}\label{eq:trad}
	t\ll t_\rad\approx 63.6\text{ ky}.
	\end{align}
In the next section we use eq. (\ref{eq:rt}) to find the corresponding bound in terms of temperature.
\end{subsubsection}
\end{subsection}
\end{section}
\begin{section}{Thermodynamics of the Universe}
In this chapter we investigate Universe as a thermodynamical system. We consider such quantities as temperature, entropy density or chemical potentials of particles involved.
\begin{subsection}{Temperature dependence of the thermodynamical parameters}\label{sec:thermodep}
Here we show how do the energy density, pressure and entropy density depend on temperature of the Universe\footnote{Precise meaning of this term is explained in section \ref {sec:thermeq}}.\\

For any gas, its number density, energy density and pressure can be expressed in terms of the distribution function $f(\vec\mom)$ as
\begin{align*}
n&\equiv\frac{g}{(2\pi)^3}\int f(\vec\mom)d^3\mom,\\
\rho&\equiv\frac{g}{(2\pi)^3}\int f(\vec\mom)Ed^3\mom,\\
\pres&\equiv\frac{g}{(2\pi)^3}\int f(\vec\mom)\frac{\vec\mom^2}{3E}d^3\mom,
\end{align*}
where $g$ stands for the number of internal degrees of freedom and energy is defined, as usually, as $E^2=\vec\mom^2+m^2$ for the particles of given mass $m$.
Assuming equilibrium distribution (Bose-Einstein for bosons, Fermi-Dirac for fermions) and therefore using $f(\vec\mom)=f(E)=\left[\exp\left(\frac{E-\mu}{T}\right)\pm 1\right]^{-1}$, one obtains
	\begin{align*}
	n&\equiv\frac{g}{2\pi^2}\int_m^\infty\frac{\sqrt{E^2-m^2}}{e^{\frac{E-\mu}{T}}\pm 1}EdE
	=\frac{gT^3}{2\pi^2}\int_\frac{m}{T}^\infty\frac{\sqrt{u^2-\left(\frac{m}{T}\right)^2}}{e^{u-\frac{\mu}{T}}\pm 1}udu,\\
	\rho&\equiv\frac{g}{2\pi^2}\int_m^\infty\frac{\sqrt{E^2-m^2}}{e^{\frac{E-\mu}{T}}\pm 1}E^2dE
	=\frac{gT^4}{2\pi^2}\int_\frac{m}{T}^\infty\frac{\sqrt{u^2-\left(\frac{m}{T}\right)^2}}{e^{u-\frac{\mu}{T}}\pm 1}u^2du,\\
	\pres&\equiv\frac{g}{6\pi^2}\int_m^\infty\frac{(E^2-m^2)^\frac{3}{2}}{e^{\frac{E-\mu}{T}}\pm 1}dE
	=\frac{gT^4}{6\pi^2}\int_\frac{m}{T}^\infty\frac{\left(u^2-\left(\frac{m}{T}\right)^2\right)^\frac{3}{2}}
	{e^{u-\frac{\mu}{T}}\pm 1}du,
	\end{align*}
where in the denominator one should use $+1$ for fermions and $-1$ for bosons. In the last column we introduce an auxiliary variable $u=\frac{E}{T}$.

Let us consider now the ultra-relativistic and the non-relativistic limit of these formulae.
\begin{itemize}
\item In the ultra-relativistic limit ($T\gg m, \mu$), density and pressure take the form
	\begin{align}\label{eq:rel}\begin{aligned}
	n&\approx\frac{gT^3}{2\pi^2}\int_0^\infty\frac{1}{e^u\pm 1}u^2du=\frac{\zeta(3)}{\pi^2}gT^3\cdot
		\begin{cases}
		1&\text{for bosons}\\
		\frac{3}{4}&\text{for fermions}
		\end{cases},\\
	\rho&\approx\frac{gT^4}{2\pi^2}\int_0^\infty\frac{1}{e^u\pm 1}u^3du=\frac{\pi^2}{30}gT^4\cdot
		\begin{cases}
		1&\text{for bosons}\\
		\frac{7}{8}&\text{for fermions}
		\end{cases},\\
	\pres&=\frac{1}{3}\rho.
	\end{aligned}\end{align}
To express the total radiation energy density, one can introduce the \emph{total number of effectively massless degrees of freedom} $g_*$, defined as
	$$g_*\equiv\sum_{b\in\text{bosons}}g_b\left(\frac{T_b}{T}\right)^4
	+\frac{7}{8}\sum_{f\in\text{fermions}}g_f\left(\frac{T_f}{T}\right)^4.$$
In principle, temperatures of various components of the Universe do not have to be equal, hence $T_b$ ($T_f$) denotes temperature of a given kind of bosonic (fermionic) particles.

The total energy density of ultra-relativistic particles reads now
	$$\rho_\rad=\frac{\pi^2}{30}g_*T^4.$$
To calculate the entropy density, i.e. entropy per unit volume, let us use the first law of thermodynamics (with chemical potential neglected\footnote{A short discussion of neglecting the chemical potential is provided in section \ref{sec:chemPot}.}):
	$$dU=TdS-pdV,$$
where $U$ is the energy, $S$ is the entropy and $V$ is the volume of the considered portion of fluid. Dividing\footnote{Of course, this ,,division'' is only a mnemotechnic, actually we act with differential 1-form $dU$ on the vector $\pd_V$ from the space tangent to abstract manifold equipped with coordinates $S$ and $V$.} by $dV$ we obtain
\begin{align}\label{eq:thermolaw}
\rho=\frac{dU}{dV}=T\frac{dS}{dV}-\pres=Ts-\pres,
\end{align}
where $s$ is the entropy density. Using $p=\frac{1}{3}\rho$, we can obtain a formula for the total radiation entropy density (analogically as for $\rho_\rad$):
	$$s_\rad=\frac{2\pi^2}{45}g_{*s}T^3,$$
where
	$$g_{*s}\equiv\sum_{b\in\text{bosons}}g_b\left(\frac{T_b}{T}\right)^3
	+\frac{7}{8}\sum_{f\in\text{fermions}}g_f\left(\frac{T_f}{T}\right)^3.$$
\item In the non-relativistic limit ($T\ll m$) we also assume that the chemical potential is much smaller than the mass ($\mu\ll m$, see section \ref{sec:chemPot}). The exponential term $e^\frac{E-\mu}{T}\approx e^\frac{E}{T}$ is much bigger than one, hence there is no difference between bosons and fermions. We obtain
	\begin{align}\label{eq:nonrelBessel}\begin{aligned}
	n&\approx\frac{gT^3}{2\pi^2}\int_{\frac{m}{T}}^\infty\sqrt{u^2-\left(\frac{m}{T}\right)^2}e^{-u+\frac{\mu}{T}}udu
	=\frac{g}{2\pi^2}m^2TK_2\left(\frac{m}{T}\right)e^{\frac{\mu}{T}},\\
	\rho&\approx\frac{gT^4}{2\pi^2}\int_{\frac{m}{T}}^\infty\sqrt{u^2-\left(\frac{m}{T}\right)^2}e^{-u+\frac{\mu}{T}}u^2du
	=\frac{g}{2\pi^2}m^2T\left[mK_1\left(\frac{m}{T}\right)+3TK_2\left(\frac{m}{T}\right)\right]e^{\frac{\mu}{T}},\\
	\pres&\approx\frac{gT^4}{6\pi^2}\int_{\frac{m}{T}}^\infty\left(u^2-\left(\frac{m}{T}\right)^2\right)^{\frac{3}{2}}
	e^{-u+\frac{\mu}{T}}du
	=\frac{g}{2\pi^2}m^2T^2K_2\left(\frac{m}{T}\right)e^{\frac{\mu}{T}},
	\end{aligned}\end{align}
where $K_i$ is the $i$-th modified Bessel function of the second kind. For big $x$, the Bessel functions can be well approximated by $K_i(x)\approx\sqrt{\frac{\pi}{2x}}e^{-x}$. Therefore, if $x\equiv\frac{m}{T}\gg 1$, we can approximate appropriate functions by
	\begin{align}\label{eq:nonrel}\begin{aligned}
	n&\approx g\left(\frac{mT}{2\pi}\right)^\frac{3}{2}e^{-\frac{m}{T}}e^{\frac{\mu}{T}},\\
	\rho&\approx gm\left(\frac{mT}{2\pi}\right)^\frac{3}{2}e^{-\frac{m}{T}}e^{\frac{\mu}{T}}=mn,\\
	\pres&\approx gT\left(\frac{mT}{2\pi}\right)^\frac{3}{2}e^{-\frac{m}{T}}e^{\frac{\mu}{T}}=Tn\ll\rho.
	\end{aligned}\end{align}
Note that density and pressure of the non-relativistic particles are suppressed by the $e^{-\frac{m}{T}}$ factor, hence the main contribution is radiation.
\end{itemize}
On the other hand, we can use eq. (\ref{eq:rhoR}):
	\begin{align*}
	\rho_\rad=\rho^0_\rad\left(\frac{R_0}{R}\right)^4,\\
	\Rightarrow \rho^0_\rad\left(\frac{R_0}{R}\right)^4 \approx\frac{\pi^2}{30}g_*T^4.
	\end{align*}
Using equation (\ref{eq:rt}) we obtain the time-temperature relation:
	\begin{align}\label{eq:tT}
	t=\sqrt{\frac{45}{16\pi^3}}\frac{\mpl}{T^2\sqrt{g_*}}\approx 0.301\frac{\mpl}{T^2\sqrt{g_*}},
	\end{align}
where $\mpl\equiv G^{-\frac{1}{2}}$ is the Planck mass.

We can now express the radiation domination condition (\ref{eq:trad}) in terms of temperature:
	\begin{align}\label{eq:tempRD}
	T\gg T_\rad=\left[\sqrt{\frac{45}{16\pi^3}}\frac{\mpl}{\sqrt{g_*}}t_\rad^{-1}\right]^\frac{1}{2}\approx 8.1\e{-10}\gev\approx 9.4\e{3}\text{ K}.
	\end{align}

The numbers of effective degrees of freedom, $g_*$ and $g_{*s}$, can be calculated theoretically at given temperature, knowing masses of all the Standard Model particles. Their behaviour is shown in fig. \ref{fig:ggs}.
	\begin{figure}[H]
	\begin{center}
		\includegraphics[width=\textwidth]{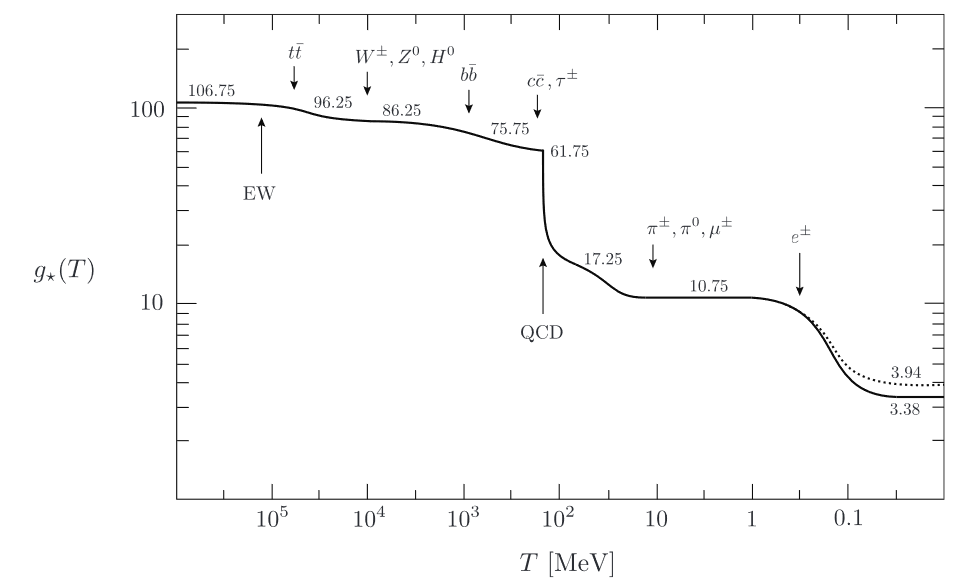}
		\caption{Effective number of degrees of freedom: $g_*$ (solid) and $g_{*s}$ (dotted). Source: \cite{bib:baumann}}
		\label{fig:ggs}
	\end{center}
	\end{figure}
\end{subsection}
\begin{subsection}{Entropy conservation}
In this section we will derive conservation of the entropy of the Universe. Derivation is quite detailed since in most of the cosmology textbooks the chemical potential is neglected at the very beginning, here it is present all the time. We discuss neglection of the chemical potential in section \ref{sec:chemPot}.\\

Let us recall the first law of thermodynamics:
	\begin{align}\label{eq:dU}
	dU=TdS-\pres dV+\mu dN.
	\end{align}
We would like to express it using $T$ and $V$ as variables:
	\begin{align*}
	dU=T\frac{\pd S}{\pd T}dT+T\frac{\pd S}{\pd V}dV-\pres dV+\mu ndV+\mu Vdn=\\
	=\left(T\frac{\pd S}{\pd T}+\mu V\frac{dn}{dT}\right)dT+\left(T\frac{\pd S}{\pd V}-\pres+\mu n\right)dV.
	\end{align*}
Having this form, we can easily obtain the partial derivatives of entropy with respect to $T$ and $V$:
	\begin{align*}
	\frac{\pd U}{\pd T}&=T\frac{\pd S}{\pd T}+\mu V\frac{dn}{dT}\qquad\Rightarrow & \frac{\pd S}{\pd T}&=\frac{1}{T}\left(\frac{\pd U}{\pd T}-\mu V\frac{dn}{dT}\right),\\
	\rho=\frac{\pd U}{\pd V}&=T\frac{\pd S}{\pd V}-\pres+\mu n\qquad\Rightarrow & \frac{\pd S}{\pd V}&=\frac{\rho+\pres-\mu n}{T}.
	\end{align*}
The second order mixed derivatives must be equal to each other:
	\begin{align*}
	\frac{\pd^2S}{\pd V\pd T}&=\frac{\pd^2S}{\pd T\pd V},\\
	\frac{1}{T}\frac{\pd \rho}{\pd T}-\frac{\mu}{T}\frac{dn}{dT}&=\frac{1}{T}\frac{\pd(\rho+\pres-\mu n)}{\pd T}-\frac{\rho+\pres-\mu n}{T^2},\\
	\Rightarrow \frac{\pd\pres}{\pd T}&=\frac{\rho+\pres-\mu n}{T}+n\frac{d\mu}{dT}.
	\end{align*}
Since pressure and chemical potential are functions of temperature only, it is equivalent to
	\begin{align}\label{eq:dpdmu}
	d\pres=\frac{\rho+\pres-\mu n}{T}dT+nd\mu.
	\end{align}
Therefore, using eq. (\ref{eq:dU}) and eq. (\ref{eq:dpdmu}), we obtain
	\begin{align*}
	dS&=\frac{1}{T}\left[d(V\rho)+\pres dV-\mu d(Vn)\right]
	=\frac{1}{T}\left[d\left(V(\rho+\pres-\mu n)\right)-Vd\pres+Vnd\mu\right]=\\
	&=\frac{1}{T}\left[d\left(V(\rho+\pres-\mu n)\right)-V\frac{\rho+\pres-\mu n}{T}dT\right]=
	d\left[\frac{V(\rho+\pres-\mu n)}{T}\right],
	\end{align*}
hence the time derivative reads
	\begin{align*}
	\frac{dS}{dt}&=3HV\frac{\rho+\pres-\mu n}{T}+V\frac{\dot\rho+\dot\pres-\frac{d}{dt}(\mu n)}{T}-V\frac{\rho+\pres-\mu n}{T^2}\dot T=\\
	&=\frac{V}{T}\left[\dot\rho+3H(\rho+\pres)\right]-\frac{\mu V}{T}(3Hn+\dot n)+\frac{V}{T}\left[\dot\pres-n\dot\mu-(\rho+\pres-\mu n)\frac{\dot T}{T}\right]=\\
	&=-\frac{\mu V}{T}(3Hn+\dot n).
	\end{align*}
The first term has been cancelled due to the continuity equation (\ref{eq:fried2}). The last one vanishes according to eq. (\ref{eq:dpdmu}). Note that when $\mu=0$, entropy is conserved.\\
Using entropy density defined as $s=\frac{S}{V}=\frac{S}{R^3}$, we obtain
	\begin{align}\notag
	\dot s+3Hs=-\frac{\mu}{T}(3Hn+\dot n),\\
	\frac{\dot s}{s}=-3H-\frac{\mu}{sT}(3Hn+\dot n).\label{eq:dsdt}
	\end{align}
This result will be used in derivation of the Boltzmann equation in chapter \ref{sec:BEq}.
\end{subsection}
\begin{subsection}{Thermodynamic equilibrium}\label{sec:thermeq}
Before we go further in our considerations, we should precisely state, what do we mean by \emph{thermodynamic equilibrium}.\\

We distinguish two kinds of thermodynamic equilibrium:
\begin{itemize}
\item \emph{kinetic} (or \emph{thermal}) \emph{equilibrium},
\item \emph{chemical equilibrium}.
\end{itemize}

Throughout the thesis, whenever any kind of particles is said to be just \emph{in equilibrium} (thermal or chemical), it should be understood as being \emph{in equilibrium with photons}.\\

Chemical equilibrium between two kinds of particles means that processes transforming one of the kinds into the second one are balanced by oppositely directed processes, what is connected to the chemical potentials (we discuss it in section \ref{sec:chemPot}). As argued in section \ref{sec:chemPot}, chemical potential of photons is equal to 0. Because of large effectiveness of electromagnetic processes, all the electrically charged particles (and Higgs particle, which is strongly coupled to the massive charged bosons and quarks) -- that is, all the SM particles with possible exception of neutrinos -- are assumed to be in chemical equilibrium with photons.

Let us consider some kind of dark (i.e. not interacting electromagnetically) particles from beyond the Standard Model. In some moment, because the Universe expands, the particle gas can become too dilute to effectively annihilate into SM (convenient criterion is that the annihilation rate, $\Gamma_\text{ann}=n\vev{\sigma v}_\text{ann}$, becomes comparable to the Hubble parameter, that is, the rate of expansion of the Universe), what ends chemical equilibrium. This moment is called the \emph{freeze-out} or \emph{chemical decoupling}\footnote{In the one-component dark matter case chemical decoupling is equivalent to freeze-out. For multi-component dark matter we distinguish chemical decoupling, which is the end of equilibrium with photons, from freeze-out, which is stabilization of the yield of a given kind of particles.}. For the cold dark matter particles of mass $m$, the freeze-out occurs usually when $\frac{m}{T}\approx 20-30$, what is satisfied also in the model considered here.\\

Thermal equilibrium between two kinds of particles means that they share common temperature (that is, their energies are described by Fermi-Dirac or Bose-Einstein distributions parametrized with the same temperature). It is obtained mainly through the elastic scattering processes. Thermal equilibrium with photons for the SM particles\footnote{With possible exception of neutrinos, see e.g. \cite{bib:neutr}.} is maintained all the time, since they are electrically charged -- or, like Higgs particles, their interactions with electrically charged matter are strong.

Whenever the  temperature of the Universe is mentioned, it should be understood as the temperature of photons. Since the Standard Model is thermally coupled to photons, it is also temperature of all the SM particles.

In analogy to chemical equilibrium, thermal equilibrium of a given kind of DM particles ends when DM-SM scattering rate\footnote{Note that here we use $n_\text{scat}$, which is number density of SM particles being the scattering partners for DM. $\Gamma_\text{scat}^{-1}=\tau_\text{scat}$ is the mean time between two consecutive scatterings of a given DM particle moving with mean relative velocity $v$ through a cloud of SM particles with number density $n_\text{scat}$.}, $\Gamma_\text{scat}=n_\text{scat}\vev{\sigma v}_\text{scat}$, becomes comparable to the Hubble parameter. In our calculations we assume that DM particles remain in thermal equilibrium long after chemical decoupling. To justify this assumption, let us compare the annihilation rate with the scattering rate:
	\begin{align*}
	\frac{\Gamma_\text{scat}}{\Gamma_\text{ann}}=
	\frac{n_\text{scat}\vev{\sigma v}_\text{scat}}{n\vev{\sigma v}_\text{ann}}.
	\end{align*}
Let us note that:
	\begin{itemize}
	\item non-relativistic DM can annihilate only into lighter states, while the scattering particle can have any mass -- therefore, for scattering more channels are allowed,
	\item if the scattering partner is lighter than the given DM particle (in particular, if it is any of states that DM can annihilate into), $n_\text{scat}$ is much higher than $n$ due to exponential term $e^{-\frac{m}{T}}$ in eq. (\ref{eq:nonrel}).
	\end{itemize}
If allowed, the cross section for annihilation into pair of given SM particles should be roughly the same as cross section for scattering on one of these particles, i.e.
	$$\vev{\sigma v}_{\chi\chi\to a\bar a}\sim\vev{\sigma v}_{\chi a\to\chi a},$$
where $\chi$ denotes DM particle and $a$ is the SM scattering partner. Hence, $\Gamma_\text{scat}$ contains all the terms (and also a term for every additional channel allowed) present in $\Gamma_\text{ann}$, but multiplied by $n_\text{scat}\gg n$ instead of $n$. For light scattering partners (i.e. still relativistic during the chemical decoupling of DM particles, what means $m_\text{scat}<\frac{1}{20}m$), near the temperature of freeze-out ($T\sim\frac{m}{20}$) we can use the relativistic approximation (\ref{eq:rel}) for $n_\text{scat}$, obtaining
	$$\frac{n_\text{scat}}{n}\approx \left(\frac{T}{m}\right)^\frac{3}{2}e^{\frac{m}{T}}\sim 10^6.$$
Chemical decoupling takes place when $\Gamma_\text{ann}\sim H$, so $\Gamma_\text{scat}\gg\Gamma_\text{ann}$ means $\Gamma_\text{scat}\gg H$ -- hence, thermal equilibrium is still maintained long after the chemical decoupling.

There exists, however, a case, described in the next section, when the annihilation cross-section is enhanced in comparison to scattering and the thermal decoupling can occur even before the chemical one.
\begin{subsubsection}{Resonant annihilation model and the early kinetic decoupling}\label{sec:ekd}
Let us assume that dark matter interacts with the SM only through a mediator resonant particle $R$ (fig. \ref{fig:resDiags})
	\begin{figure}[H]\begin{center}
		\includegraphics[width=0.5\textwidth]{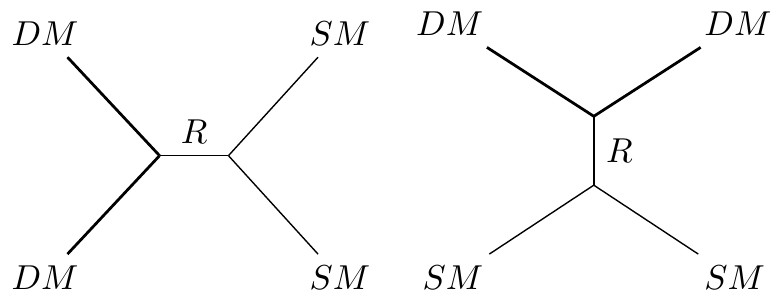}
		\caption{DM-SM interaction diagrams in the resonant annihilation model: annihilation (left) and scattering (right)}
		\label{fig:resDiags}
	\end{center}\end{figure}
Let $\Gamma$ be the total mediator decay rate, $M$ be the mass of the mediator, $\Gamma \ll M$. According to \cite{bib:mdbg}, the cross-section for the annihilation satisfies
$$\sigma_\text{ann}\sim\frac{1}{(s-M^2)^2+\Gamma^2M^2}\stackrel{v_\text{rel}\ll 1}{\approx}\frac{1}{\Gamma^2M^2},$$
while the scattering cross-section
$$\sigma_\text{scat}\sim\frac{1}{(t-M^2)^2+\Gamma^2M^2}\stackrel{v_\text{rel}\ll 1}{\approx}\frac{1}{M^4},$$
where $s$ and $t$ are the Mandelstam variables and we assume that the relative velocity $v_\text{rel}$ of the incident particles is small since usually we consider non-relativistic dark matter. Since $\Gamma\ll M$, the scattering cross-section should be much smaller than the annihilation cross-section. If kinetic decoupling of the dark matter particles occurs early, we should assume that their temperature changes like $R^{-2}$ (for non-relativistic dust, see eq. (\ref{eq:RTnonrel})), not like $R^{-1}$ (for particles coupled to the Standard Model dominated by radiation, see eq. (\ref{eq:RTrel})). Therefore, evolution of the density of the early kinetic decoupled DM is different than in the standard late decoupling case.

To obtain the correct value of the DM relic abundance (i.e. $\Omega h^2\approx 0.12$), the scattering cross-section has to be additionally suppressed, giving as a result temperature of the kinetic decoupling of the same order as temperature of the chemical decoupling (further discussion can be found in \cite{bib:mdbg, bib:decoWIMP}).

In this thesis we assume that because of presence of scattering processes in $s$-channel, thermal decoupling happens much later than the freeze-out.
\end{subsubsection}
\end{subsection}
\begin{subsection}{Particles decoupled from the thermal bath}
After kinetic decoupling from the thermal bath, temperature of a given kind of particles is not necessarily the temperature of photons. What is behaviour of this parameter then?

First, note that the phase space element is not affected by expansion of the Universe. Let $d^3\mom\,d^3x$ denote the phase space element at temperature $T$ and let $d^3\mom'd^3x'$ correspond to $T'$. Indeed, since momentum of a free particle changes like $R^{-1}$ and distance scales like $R$, we obtain
	\begin{align}\label{eq:phasespace}
	d^3\mom'd^3x'=\left(\frac{R}{R'}\right)^3d^3\mom\left(\frac{R'}{R}\right)^3d^3x=d^3\mom\,d^3x.
	\end{align}
Let us now consider particles decoupled from the thermal bath, that are contained in a given phase space element. Their number at temperature $T$ should be equal to the number of particles in corresponding phase space element at temperature $T'$ (for both temperatures smaller than the temperature of decoupling),
	$$f(\vec\mom)d^3\mom\,d^3x\equiv dN\equiv f'(\vec\mom')d^3\mom'd^3x'\stackrel{(\ref{eq:phasespace})}{=}f'(\vec \mom')d^3\mom\,d^3x,$$
where $f(\vec\mom)$ is the appropriate distribution function. Hence
	$$\frac{1}{\exp\left(\frac{E-\mu}{T}\right)\pm 1}=f(\vec\mom)
	\equiv f'(\vec\mom')=\frac{1}{\exp\left(\frac{E'-\mu'}{T'}\right)\pm 1}.$$
For particles that are relativistic at time of decoupling, we use $E\approx\mom$ to obtain
	$$\exp\left(\frac{\mom-\mu}{T}\right)\equiv
	\exp\left(\frac{\mom'-\mu'}{T'}\right)=\exp\left(\frac{\frac{R}{R'}\mom-\mu'}{T'}\right).$$
This equality should hold for every $\mom$, therefore
	\begin{align}\begin{aligned}\label{eq:RTrel}
	T'=\frac{R}{R'}T,\qquad\mu'=\frac{R}{R'}\mu,
	\end{aligned}\end{align}
that is, temperature and chemical potential of particles relativistic during the kinetic decoupling scale later as $R^{-1}$.

If particles are non-relativistic when decoupling, energy and momentum satisfy $E\approx m+\frac{\mom^2}{2m}$, therefore for every $\mom$
	$$\exp\left(\frac{\frac{\mom^2}{2m}+m-\mu}{T}\right)\equiv\exp\left(\frac{\frac{(\mom')^2}{2m}+m-\mu'}{T'}\right)
	=\exp\left(\frac{\left(\frac{R}{R'}\right)^2\frac{\mom^2}{2m}+m-\mu'}{T'}\right),$$
hence
	\begin{align}\begin{aligned}\label{eq:RTnonrel}
	T'=\left(\frac{R}{R'}\right)^2T,\qquad\mu'=\left(\frac{R}{R'}\right)^2\mu,
	\end{aligned}\end{align}
that is, temperature and chemical potential of particles non-relativistic during the kinetic decoupling scale later as $R^{-2}$.
\end{subsection}
\begin{subsection}{The chemical potential}\label{sec:chemPot}
Usually in cosmology textbooks, the chemical potential is neglected without any or with little comment. In this section we shortly investigate the reasons that allow us to do so.
\begin{subsubsection}{Photons}
There are few arguments, clearly listed in \cite{bib:chemPot}, proving that the chemical potential of photons, $\mu_\gamma$, has to be equal to 0. The simplest is, that during the scattering of electrically charged particles an arbitrary number of photons can be produced, even if equilibrium is maintained. It means that the sum of the chemical potential of scattering particles has to be equal itself plus the chemical potential of photons, therefore $\mu_\gamma=0$.
\end{subsubsection}
\begin{subsubsection}{Other particles}
For any two species $a$ and $b$ in chemical equilibrium obtained through the annihilation processes, the following condition must be satisfied:
$$\mu_a+\mu_{\bar a}=\mu_b+\mu_{\bar b}.$$
So, for any particle $a$ in equilibrium with photons, its chemical potential has to satisfy
	\begin{align}\label{eq:mumu}
	\mu_a+\mu_{\bar a}=0,
	\end{align}
where $\bar a$ is the corresponding antiparticle.

Chemical potential can be connected with the matter-antimatter ratio. Since particles and antiparticles differ only by charge, not by mass nor the number of degrees of freedom, the difference between their number densities (in the non-relativistic limit) is, according to eq. (\ref{eq:nonrel}),
$$n_a-n_{\bar{a}}\approx 2\sinh\left(\frac{\mu_a}{T}\right)g\left(\frac{mT}{2\pi}\right)^\frac{3}{2}e^{-\frac{m}{T}}.$$
We assume that, due to large degree of symmetry between matter and antimatter in the early Universe, chemical potential can be neglected in comparison to temperature.

Following directly from eq. (\ref{eq:mumu}), for any particle $a$ being its own antiparticle $\mu_a=0$ without any approximations and assumptions other than equilibrium with photons (note that the particles in equilibrium with photons do not necessarily need to be charged since the equilibrium can be achieved indirectly in several steps of reaction).
\end{subsubsection}

After chemical decoupling, chemical equilibrium is no longer maintained and therefore potential can no longer be neglected in comparison to temperature, nevertheless we will assume that even after the decoupling the chemical potential is still negligible in comparison to mass. This assumption will be used in the section \ref{sec:BEq} and in the non-relativistic approximation of thermodynamical quantities (\ref{eq:nonrel}).
\end{subsection}
\begin{subsection}{Quantum degeneracies}\label{sec:qdeg}
The important question is when the quantum degeneracies, i.e. Bose-Einstein condensation or fermionic degeneracy, can become relevant for dark matter interactions.

To find the answer, we can compare the mean de Broglie wavelength,
$$\vev{\lambda}=\frac{2\pi}{\vev{E_k}}, \qquad \vev{E_k}\sim T,$$
with the mean distance between the particles:
$$\vev{d}=n^{-1/3}.$$
Hence (using the non-relativistic approximation of the number density):
$$\frac{\vev{d}}{\lambda}\sim \frac{1}{2\pi}Tn^{-1/3}=\frac{1}{2\pi}T\left[g\left(\frac{mT}{2\pi}\right)^{3/2}e^\frac{\mu-m}{T}\right]^{-1/3}=
\frac{1}{\sqrt{2\pi}}\sqrt{\frac{T}{m}}g^{-1/3}e^\frac{-\mu+m}{3T},$$
where $m$ is mass, $T$ is temperature, $\mu$ is the chemical potential and $g$ is the number of internal degrees of freedom of a given kind of particles. We assume that the chemical potential can be neglected in comparison to mass.

For $\frac{m}{T}=x>1.5$ this function is fast growing with $x$. For $x\approx 13$, value of $\frac{\vev{d}}{\lambda}$ is around ten, therefore it is justified to neglect the quantum degeneracies around the moment of freeze-out, which occurs at $x\approx 20-30$.
\end{subsection}
\end{section}
\begin{section}{The Boltzmann equation}\label{sec:BEq}
The Boltzmann transport equation describes the change of the number of particles in a given phase space region, caused by collisions (or, in general, any interactions) between them. The purpose of this section is to provide a step-by-step derivation of the Boltzmann equation, used in this thesis to calculate abundances of DM particles, including explicitly listed assumptions. We follow derivation provided in \cite{bib:kt}, with additional detailed discussion when needed.\\

The most general form of the equation reads:
$$\hat Lf=\hat Cf,$$
where $\hat L$ is called the Liouville operator, $\hat C$ is called the collision term and ${f=f(x^\mu,\mom^\mu)}$ is the distribution function.\\
The Liouville operator is defined in curved space-time as
$$\hat L\equiv\mom^\mu\pd_\mu-\Gamma^\mu_{\,\alpha\beta}\mom^\alpha\mom^\beta\frac{\pd}{\pd\mom^\mu}.$$
For the \FLRW metric and under the assumption that the distribution function is homogeneous and isotropic (i.e. $f(x^\mu,\mom^\mu)=f(t,E)$), the Liouville operator takes the following form:
$$\hat L=E\pd_t-H|\vec\mom|^2\pd_E,$$
where $H$ is the Hubble parameter, defined using the scale factor $R$ as $H=\frac{\dot R}{R}$. The Boltzmann equations now reads:
	\begin{align}\label{eq:beq0}
	E\pd_tf-H|\vec\mom|^2\pd_Ef=\hat Cf.
	\end{align}
Let us consider the Boltzmann equation describing particles denoted as $X$. We would like to investigate the time dependence of the number density of the $X$ particles, i.e. number of particles per unit comoving volume. Formally, in terms of the distribution function $f$, it is given by:
$$n_X(t)\equiv\frac{g_X}{(2\pi)^3}\int fd^3\mom_X,$$
where $g_X$ stands for the number of internal spin degrees of freedom of the particle (1 for scalars, 2 for massless bosons and spin-$\frac{1}{2}$ fermions, 3 for massive bosons).
Therefore, we can multiply the Boltzmann equation (\ref{eq:beq0}) by $\frac{1}{E_X}\frac{g_X}{(2\pi)^3}$ and integrate over the momentum of $X$ particles to obtain
\begin{align}\label{eq:parts}
\dot n_X-H\frac{g_X}{(2\pi)^3}\int\frac{|\vec\mom_X|^2}{E_X}\pd_Efd^3\mom_X=\frac{g_X}{(2\pi)^3}\int\frac{\hat Cf}{E_X}d^3\mom_X.
\end{align}
We integrate the LHS by parts. The boundary term can be neglected assuming that the distribution function vanishes in the limit of infinite energy faster than $E^{-3}$. After integration we obtain
$$\LHS=\dot n_X+3Hn_X.$$
The collision operator is connected with a given process\footnote{To obtain the full collision term one has to sum over all the processes that change the number of considered particles.} $Xab\ldots\leftrightarrow ij\ldots$ in the following way:
	\begin{align}\begin{aligned}\label{eq:collision}
	\frac{g_X}{(2\pi)^3}\int\frac{\hat C f_X}{E_X}d^3\mom_X = -\int d\Phi_Xd\Phi_ad\Phi_b\ldots
	d\Phi_id\Phi_j\ldots(2\pi)^4\delta^{(4)}(\mom_X+\mom_a+\mom_b+\ldots-\mom_i-\mom_j-\ldots)\cdot\\
	\cdot\left[|\bar\M|^2_{Xab\ldots\rightarrow ij\ldots}f_Xf_af_b\ldots(1\pm f_i)(1\pm f_j)\ldots
	-|\bar\M|^2_{ij\ldots\rightarrow Xab\ldots}f_if_j\ldots(1\pm f_X)(1\pm f_a)(1\pm f_b)\ldots\right],
	\end{aligned}\end{align}
where $d\Phi_\alpha\equiv\frac{g_\alpha}{(2\pi)^3}\frac{d^3\mom_\alpha}{2E_\alpha}$ denotes the phase space element for particles denoted by $\alpha$, $|\bar M|^2$ stands for the spin-averaged squared matrix element for the process and the quantum statistics term $1\pm f_\alpha$ reads:
$$1\pm f_\alpha=\begin{cases}1+f_\alpha&\text{for bosons}\\1-f_\alpha&\text{for fermions}\end{cases}.$$
In the absence of quantum degeneracy, i.e. Bose-Einstein condensation or fermionic degeneracy (what is a good assumption since the gas of particles is usually very dilute, see section \ref{sec:qdeg}), the quantum statistics terms can be well approximated by 1. We also assume $T$ (or, equivalently\footnote{Since $CPT$ is assumed to be the exact symmetry of any Lorentz-invariant quantum field theory with a Hermitian Hamiltonian; for details see \cite{bib:cpt1, bib:cpt2}.}, $CP$\footnote{So far, the only observations of violation of the $CP$ symmetry are $K^0-\bar K^0$ system \cite{bib:cp-k} and $B$ meson decays \cite{bib:cp-b}. Hence, $CP$ symmetry is well-motivated assumption.}) invariance of all the processes considered, so
$$|\bar\M|^2_{Xab\ldots\rightarrow ij\ldots}=|\bar\M|^2_{ij\ldots\rightarrow Xab\ldots}\equiv|\bar\M|^2_{Xab\ldots ij\ldots}.$$
Now we can rewrite the Boltzmann equation as
	\begin{align}\label{eq:beq1}
	\dot n_X+3Hn_X= -\sum_{ab\ldots ij\ldots}\int d\Phi_Xd\Phi_ad\Phi_b\ldots d\Phi_id\Phi_j\ldots(2\pi)^4\delta(\mom)\cdot|\bar\M|^2_{Xab\ldots ij\ldots}(f_Xf_af_b\ldots-f_if_j\ldots)
	\end{align}
(we use $\delta(\mom)$ as an abbreviation of $\delta^{(4)}(\mom_X+\mom_a+\mom_b+\ldots-\mom_i-\mom_j-\ldots)$).
If a given kind of particles is in thermodynamical equilibrium of the temperature $T$ and chemical potential $\mu$, it will have thermal distribution function\footnote{We approximate the Bose-Einstein and the Dirac-Fermi statistics by the Maxwell-Boltzmann distribution since, as mentioned previously, we neglect quantum degeneracies}:
$$f\equiv\exp\left(-\frac{E-\mu}{T}\right).$$

Before chemical decoupling, chemical potential of the DM particles is assumed to be equal to 0 (what is discussed in section \ref{sec:chemPot}). Therefore $\mu=0$ means chemical equilibrium with Standard Model.

The chemical potential part can be factored-out from the distribution function:
\begin{align}\label{eq:f}
f=\exp\left(\frac{\mu(T)}{T}\right)\exp\left(-\frac{E}{T}\right).
\end{align}
Let us denote as $\bar f$ the distribution function without the chemical potential part (so called \emph{equilibrium distribution}):
$$\bar f(E)\equiv\exp\left(-\frac{E}{T}\right).$$
We introduce the \emph{equilibrium number density} which corresponds to $\mu=0$:
$$\bar n_X(t)\equiv\frac{g_X}{(2\pi)^3}\int\bar fd^3\mom_X\stackrel{(\ref{eq:nonrelBessel})}{=}\frac{g_X}{2\pi^2}T^3x^2K_2(x),$$
where $x\equiv\frac{m}{T}$.

The distribution function is connected with the equilibrium one by
	\begin{align}\label{eq:ffeq}
	f_\alpha=\exp\left(\frac{\mu_X(T)}{T}\right)\bar f_\alpha=\frac{n_\alpha(T)}{\bar n_\alpha(T)}\bar f_\alpha.
	\end{align}
As mentioned in chapter \ref{sec:thermeq}, dark matter is assumed to be in thermal equilibrium with photons for a long time after chemical decoupling. Therefore in this thesis, temperature of DM particles is always assumed to be equal to the Standard Model temperature, so equation (\ref{eq:ffeq}) can be used without changing the value of $T$. Using this assumption we can transform the Boltzmann equation (\ref{eq:beq1}) into the following form:
	\begin{align}\label{eq:beq2}
	\dot n_X+3Hn_X= -\sum_{ab\ldots ij\ldots}<\sigma_{Xab\ldots ij\ldots} v>\left(n_Xn_an_b\ldots-\bar n_X\bar n_a\bar n_b\ldots\frac{n_in_j\ldots}{\bar n_i\bar n_j\ldots}\right),
	\end{align}
where we used the thermally averaged cross-section $<\sigma_{Xab\ldots ij\ldots} v>$, defined as
	$$<\sigma_{Xab\ldots ij\ldots} v>\equiv\frac{1}{\bar n_X\bar n_a\bar n_b\ldots}\int d\Phi_Xd\Phi_ad\Phi_b\ldots
	d\Phi_id\Phi_j\ldots(2\pi)^4\delta(p)\cdot|\bar\M|^2_{Xab\ldots ij\ldots}\bar f_X\bar f_a\bar f_b\ldots$$
In order to simplify the equation, we can change variables from $n$ to $Y=\frac{n}{s}$, where $s$ is the total entropy density of the Universe, introduced in (\ref{eq:thermolaw}) (as discussed in section \ref{sec:thermodep}, non-relativistic particles have negligible contribution to the entropy density, so we do not need to take non-relativistic dark matter into account). The time derivative of $Y$ is
$$\dot Y=\frac{1}{s}(\dot n-n\frac{\dot s}{s})\stackrel{(\ref{eq:dsdt})}{=}\frac{1}{s}\left(1+\frac{\mu}{T}Y\right)(\dot n+3Hn).$$

In terms of $Y$ eq. (\ref{eq:beq2}) reads
\begin{align*}
\dot Y = -\frac{1}{s}\left(1+\frac{\mu}{T}Y\right)\sum_{ab\ldots ij\ldots}<\sigma_{Xab\ldots ij\ldots} v>s^N\left(Y_XY_aY_b\ldots-\bar Y_X\bar Y_a\bar Y_b\ldots\frac{Y_iY_j\ldots}{\bar Y_i\bar Y_j\ldots}\right),
\end{align*}
where $N$ is the number of particles in the initial state of reaction ($X$, $a$, $b$, $\ldots$).\\
It is convenient to express the Boltzmann equation in terms of temperature rather than time\footnote{Since dark matter is assumed to be cold around the moment of decoupling, we can assume that in considered range of temperatures it has no significant influence on the time-temperature relation}. Instead of time derivative, we will consider $\frac{dY}{dx}$, $x\equiv\frac{m}{T}$:
	\begin{align*}
	\frac{dY}{dx} = \frac{1}{\dot x}\dot Y=x^{-2}\left(1+\frac{\mu}{T}Y\right)\frac{m}{\dot T}\sum_{ab\ldots ij\ldots}s^{N-1}<\sigma_{Xab\ldots ij\ldots} v>\left(Y_XY_aY_b\ldots-\bar Y_X\bar Y_a\bar Y_b\ldots\frac{Y_iY_j\ldots}{\bar Y_i\bar Y_j\ldots}\right).
	\end{align*}
We will assume that $1\gg \frac{\mu}{T}Y=\frac{n\mu}{sT}$. As discussed in section \ref{sec:chemPot}, $\frac{\mu}{T}$ cannot be treated as small after the freeze-out, but $\frac{n}{s}$ factor suppresses it to be much smaller than 1. Usually freeze-out occurs at $x\approx 20-30$, so we can use the non-relativistic approximation:
$$\frac{n\mu}{sT}=\frac{45g}{g_{*s}}\sqrt{\frac{\pi}{2}}\left(\frac{m}{T}\right)^\frac{3}{2}\frac{\mu}{T}e^{-\frac{m-\mu}{T}}.$$
We see that as long as the chemical potential is much smaller than the mass, the $\frac{\mu}{T}Y=\frac{n\mu}{sT}$ term can be neglected because of the exponential part $e^{-x}$.\\
Now, the Boltzmann equation takes the following form:
	\begin{align*}
	\frac{dY}{dx} = \frac{1}{\dot x}\dot Y=x^{-2}\frac{m}{\dot T}\sum_{ab\ldots ij\ldots}s^{N-1}<\sigma_{Xab\ldots ij\ldots} v>\left(Y_XY_aY_b\ldots-\bar Y_X\bar Y_a\bar Y_b\ldots\frac{Y_iY_j\ldots}{\bar Y_i\bar Y_j\ldots}\right).
	\end{align*}
If we consider $2\rightarrow n$ processes only ($N=2$), we obtain
	\begin{align*}
	\frac{dY}{dx}=-L(x)\sum_{a,ij\ldots}\vev{\sigma_{Xa,ij\ldots} v}\left(Y_XY_a-\bar Y_X\bar Y_a\frac{Y_iY_j\ldots}{\bar Y_i\bar Y_j\ldots}\right),
	\end{align*}
where
	\begin{align*}
	L(x)\equiv-\frac{sm}{x^2\dot T}.
	\end{align*}
From eq. (\ref{eq:tT}) we obtain (under an assumption that the $g_*$ function is slowly-changing\footnote{Chemical decoupling of dark matter occurs long before QCD transition threshold shown in fig. \ref{fig:ggs}.} and therefore $t\frac{\dot g_*}{\sqrt{g_*}}\ll 1$)
	$$\dot T=-\sqrt{\frac{4\pi^3}{45}}\frac{\sqrt{g_*}}{\mpl}T^3$$
and hence
	\begin{align}\label{eq:l}
	L(x)=\sqrt\frac{\pi}{45}m\mpl x^{-2}\frac{g_{*s}}{\sqrt{g_*}}.
	\end{align}
For decays, $N=1$ and the corresponding part of the Boltzmann equation is
\begin{align*}
\frac{dY}{dx}=-\tilde L(x)\sum_{ij\ldots}\Gamma_{X\rightarrow ij\ldots}\left(Y_X-\bar Y_X\frac{Y_iY_j\ldots}{\bar Y_i\bar Y_j\ldots}\right),
\end{align*}
where
\begin{align}\label{eq:ll}
\tilde L(x)\equiv-\frac{m}{x^2\dot T}=\sqrt{\frac{45}{4\pi^3}}\frac{\mpl}{m^2}\frac{x}{\sqrt{g_*}}.
\end{align}
Of course, $g_*$ and $g_{*s}$ are the temperature dependent functions. Since we assume that dark matter particles are non-relativistic around the moment of freeze-out, they do not contribute neither to radiation energy density, nor to radiation entropy density. Hence we can assume that existence of dark matter does not influence behaviour of $g_*$ and $g_{*s}$ and therefore their values are the same as shown in fig. \ref{fig:ggs}.\\

Let us briefly summarize the assumptions that have been made throughout the derivation of the Boltzmann equation:
	\begin{itemize}
	\item Isotropy and homogeneity of the Universe lead to isotropy of the distribution function: $f(\mom,t)=f(E,t)$.
	\item The distribution function $f(E)$ has to vanish with energy faster than $E^{-3}$ to cancel the boundary term appearing during integration the LHS of eq. (\ref{eq:parts}) by parts.
	\item The quantum degeneracies are neglected, according to section \ref{sec:qdeg}. Therefore all the statistical factors $1\pm f$ appearing in eq. (\ref{eq:collision}) are substituted by 1. Also the Fermi-Dirac and Bose-Einstein distribution functions are approximated by Maxwell-Boltzmann distribution.
	\item Due to $T$ (equivalently: $CP$) invariance, the squared matrix elements of a given process and the inverted process are assumed to be equal, i.e.
	$$|\bar\M|^2_{Xab\ldots\rightarrow ij\ldots}=|\bar\M|^2_{ij\ldots\rightarrow Xab\ldots}\equiv|\bar\M|^2_{Xab\ldots ij\ldots}$$
	\item Dark matter is assumed to be in thermal equilibrium with photons for a long time after chemical decoupling (freeze-out), see section \ref{sec:thermeq}.
	\item Dark matter is assumed to be non-relativistic around the point of freeze-out (it appears that indeed $\frac{m}{T}\oddo{\text{fo}}{}\approx 20-30$). Hence it has no significant influence on the expansion rate of the Universe, the total entropy density, the total energy density and the time-temperature relation during the radiation domination epoch.
	\item Long before the chemical decoupling, the chemical potential of DM particles is assumed to be negligible in comparison to temperature. All the time it is assumed to be negligible in comparison to mass of the particles (see section \ref{sec:chemPot}).
	\item The $g_*$ function is slowly changing with time, therefore its time dependence can be neglected during calculating the time derivative of the temperature of the Universe.
	\end{itemize}
\end{section}
\begin{section}{The model}\label{sec:vfdm}
The vector-fermion dark matter model considered here is described in full detail in \cite{bib:vfdm}.\\

We consider a simple extension of the Standard Model gauge group with the simplest continuous gauge group, i.e. $U(1)$, which does not act on the Standard Model particles:
$$\G=\underbrace{SU(3)_c\times SU(2)_L\times U(1)_Y}_{\text{Standard Model gauge group}}\times U(1)_X.$$
We add to the SM the gauge field $X_\mu$, connected with $U(1)_X$. We also introduce one complex scalar $S$ and one fermionic field $\chi$, with following charges:
	$$S:(1,1,0,1), \qquad \chi:(1,1,0,\frac{1}{2}).$$
Note that the newly introduced particles transform trivially under $\G_\text{SM}$. As it should be for dark matter candidates, $X_\mu$, $S$ and $\chi$ are electrically neutral.

Our Lagrangian has the following form:
	$$\Lag=\Lag_{SM}+\Lag_{DM}+\Lag_\text{portal},$$
where $\Lag_{SM}$ is the Standard Model Lagrangian, $\Lag_{DM}$ is the dark sector Lagrangian and $\Lag_\text{portal}$ is responsible for the DM-SM portal coupling:
	$$\Lag_{DM}=-\frac{1}{2}\F_{\mu\nu}\F^{\mu\nu}+\left(\D_\mu S\right)^*\D^\mu
	S+\mu_S^2|S|^2-\lambda_S|S|^4+\bar\chi(i\slashed D-m_D)\chi-\frac{1}{\sqrt{2}}(y_xS^*\chi^T\C\chi+h.c.),$$
	$$\Lag_\text{portal}=-\kappa|S|^2|H|^2.$$
Here $\F_{\mu\nu}\equiv\pd_\mu X_\nu-\pd_\nu X_\mu$ is the gauge field tensor of $X_\mu$, $\C\equiv-i\gamma_2\gamma_0$ denotes charge conjugation operator and $D_\mu$ is the covariant derivative connected with $U(1)_X$ group:
	$$D_\mu\equiv\pd_\mu+ig_xq_xX_\mu,$$
where $g_x$ is the coupling constant and $q_x$ denotes the charge with respect to $U(1)_X$, which is 0 for SM particles, 1 for $S$ and $\frac{1}{2}$ for $\chi$, as defined above.

The Lagrangian is invariant under the charge conjugation that acts on the dark sector particles in the following way:
	\begin{align}\label{eq:conj}
	\chi\rightarrow\chi^\C&\equiv-i\gamma_2\chi^*, & S\rightarrow S^\C&\equiv S^*,
	& X_\mu\rightarrow X_\mu^\C\equiv-X_\mu.
	\end{align}
The role of this symmetry will be discussed in section \ref{sec:sym}.

To find the physical scalar states, we have to investigate the scalar potential part of the Lagrangian:
	$$V(H,S)=-\mu_H^2|H|^2+\lambda_H|H|^4-\mu_S^2|S|^2+\lambda_S|S|^4+\kappa|H|^2|S|^2.$$
To construct the theory, a stable minimal-potential vacuum state is needed, in other words, the potential cannot bend down in any direction. Therefore its parameters have to satisfy the following conditions:
	$$\lambda_H>0, \qquad \lambda_S>0, \qquad \kappa > -2\sqrt{\lambda_H\lambda_S}.$$
As a vacuum state we can choose any $H$ and $S$ values that correspond to the minimum of the potential. Since $V$ depends on the absolute values only (so it is invariant under $U(1)$ rotations of the fields), the vacuum states can be chosen as
	$$\vev{H}=\Vd{0}{\frac{v}{\sqrt{2}}}, \qquad \vev{S}=\frac{v_x}{\sqrt{2}},$$
where $v$ and $v_x$ are real parameters. The chosen states of the vacuum are called the \emph{vacuum expectation values} (often abbreviated to \emph{vevs}) of the fields.

To minimize the potential, $v$ and $v_x$ have to satisfy:
	$$0=\frac{\pd V}{\pd H}\oddo{H=\vev{H},\,S=\vev{S}}{}=\left(2\lambda_H v^2-2\mu_H^2+\kappa v_x^2\right)\frac{v}{\sqrt{2}},$$
	$$0=\frac{\pd V}{\pd S}\oddo{H=\vev{H},\,S=\vev{S}}{}=\left(2\lambda_S v_x^2-2\mu_S^2+\kappa v^2\right)\frac{v_x}{\sqrt{2}}.$$
The vacuum expectation values $v$ and $v_x$ are responsible for masses of the gauge bosons (bosons of the SM and $X_\mu$, respectively). We want them to be massive, therefore vevs should not be zeroes. Hence, the potential minimalization condition in terms of $v$ and $v_x$ reads:
	\begin{align*}
	v^2=\frac{2\kappa\mu_S^2-4\lambda_S\mu_H^2}{\kappa^2-4\lambda_H\lambda_S},\\
	v_x^2=\frac{2\kappa\mu_H^2-4\lambda_H\mu_S^2}{\kappa^2-4\lambda_H\lambda_S}.
	\end{align*}
We can represent $H$ and $S$ as
	\begin{align}
	\begin{split}\label{eq:ssb}
	H&=\vev{H}+\Vd{\pi^+}{\frac{h+i\pi^0}{\sqrt{2}}},\\
	S&=\vev{S}+\frac{\phi+i\sigma}{\sqrt{2}}.
	\end{split}
	\end{align}
Here $\pi^+$, $\pi^0$ and $\sigma$ are the Goldstone bosons. Note that the potential is not invariant under rotations of the physical fields -- the initial gauge symmetry has been spontaneously broken (SSB).

We would like to find the mass eigenstates. The squared-mass matrix for the real parts of scalars' fluctuations $(h,\phi)$ reads
	$$\M^2=\left.\Md{\frac{\pd^2V}{\pd h^2}}{\frac{\pd^2V}{\pd h\pd\phi}}{\frac{\pd^2V}{\pd h\pd\phi}}
	{\frac{\pd^2V}{\pd \phi^2}}\right|_{H=\vev{H},\,S=\vev{S}}{}
	=\Md{2\lambda_Hv^2}{\kappa vv_x}{\kappa vv_x}{2\lambda_Sv_x^2}.$$
In order to diagonalize it, we have to use linear combinations of the fields, $(h_1,h_2)$, instead of $(h,\phi)$:
	\begin{align}\label{eq:rot}
	\Vd{h_1}{h_2}=\R^{-1}\Vd{h}{\phi},
	\end{align}
where
	$$\R=\Md{\cos\alpha}{-\sin\alpha}{\sin\alpha}{\cos\alpha}$$
and $\alpha$ is the scalars mixing angle. The squared mass matrix will be diagonalized when
	$$\tan\alpha=\frac{\kappa vv_x}{\lambda_Hv^2-\lambda_Sv_x^2}$$
and masses of $h_1$ and $h_2$ states become
	$$m_1^2=\lambda_H v^2(1+\sec 2\alpha)+\lambda_S v_x^2(1-\sec 2\alpha),$$
	$$m_2^2=\lambda_H v^2(1-\sec 2\alpha)+\lambda_S v_x^2(1+\sec 2\alpha).$$
Hereafter we will always assume that $h_1$ is the observed Higgs particle, i.e. $m_1=125\gev$. The second mass, $m_2$, can be in principle higher as well as lower than $m_1$. Also $v$ is assumed to be the Standard Model Higgs vev, $v=246\gev$.

To find the mass of $X_\mu$, let us check what happens to $(D_\mu S)^*D^\mu S$ after SSB:
	$$(D_\mu S)^*D^\mu S\;\ni\;(ig_xX_\mu S)^*ig_xX^\mu S=g_x^2|X|^2|S|^2
	\;\stackrel{SSB}{\longrightarrow}\;\frac{g_x^2}{2}X^\mu X_\mu(v_x^2+2v_x\phi+\phi^2).$$
Hence $m_X=g_xv_x$.

Now let us focus on the fermionic part of the dark sector Lagrangian:
	$$\Lag_{DF}=\bar\chi(i\slashed D-m_D)\chi-\frac{1}{\sqrt{2}}(y_xS^*\chi^T\C\chi+h.c.).$$
Using the charge conjugation symmetry defined in eq. (\ref{eq:conj}), one can transform it into the following form:
	\begin{align*}
	\Lag_{DF}=&\frac{i}{2}\left(\bar\chi\gamma^\mu\pd_\mu\chi+\bar{\chi^\C}\gamma^\mu\pd_\mu\chi^\C\right)
	-\frac{m_D}{2}\left(\bar\chi\chi+\bar{\chi^\C}\chi^\C\right)
	-\frac{g_x}{4}\left(\bar\chi\gamma^\mu\chi-\bar{\chi^\C}\gamma^\mu\chi^\C\right)X_\mu\\
	&-\frac{y_xv_x}{2}\left(\bar{\chi^\C}\chi+\bar\chi\chi^\C\right)
	-\frac{y_x}{2}\left(\bar{\chi^\C}\chi+\bar\chi\chi^\C\right)\phi.
	\end{align*}
The Lagrangian can be expressed through mass eigenstates $\psi_\pm$, which are Majorana states:
$$\psi_+\equiv\psi^\C_+=\frac{1}{\sqrt{2}}(\chi+\chi^\C), \qquad \psi_-\equiv\psi^\C_-=\frac{1}{i\sqrt{2}}(\chi-\chi^\C)$$
with masses $m_\pm=m_D\pm yv_x$:
	\begin{align*}
	\Lag_{DF}=&\frac{i}{2}(\bar\psi_+\gamma^\mu\pd_\mu\psi_+
	+\bar\psi_-\gamma^\mu\pd_\mu\psi_-)
	-\frac{1}{2}m_+\bar\psi_+\psi_+-\frac{1}{2}m_-\bar\psi_-\psi_-\\
	&-\frac{i}{4}g_x(\bar\psi_+\gamma^\mu\psi_--\bar\psi_-\gamma^\mu\psi_+)X_\mu
	-\frac{y_x}{2}(\bar\psi_+\psi_++\bar\psi_-\psi_-)\phi.
	\end{align*}
Note that both $X_\mu$ and $\psi_\pm$ particles are massive and neutral and hence -- if they are stable -- can be dark matter candidates. The considered model of vector-fermion dark matter is a gauged version of the model considered by S. Weinberg in \cite{bib:wein}.\\

Let us now explicitly write down the DM interaction part of the Lagrangian:
$$\Lag_\text{int}=v_xg_x^2X^\mu X_\mu\phi+\frac{g_x^2}{2}X^\mu X_\mu\phi^2
-\frac{i}{4}g_x(\bar\psi_+\gamma^\mu\psi_--\bar\psi_-\gamma^\mu\psi_+)X_\mu
-\frac{y_x}{2}(\bar\psi_+\psi_++\bar\psi_-\psi_-)\phi,$$
where
$$\phi=\cos\alpha\;h_2-\sin\alpha\;h_1.$$
Basing on this Lagrangian, we can list the DM interaction Feynman vertices, which are shown in fig. \ref{fig:intDiags} ($\R$ denotes the mixing matrix defined in eq. (\ref{eq:rot})).
	\begin{figure}[H]
	\begin{center}
		\includegraphics[width=\textwidth]{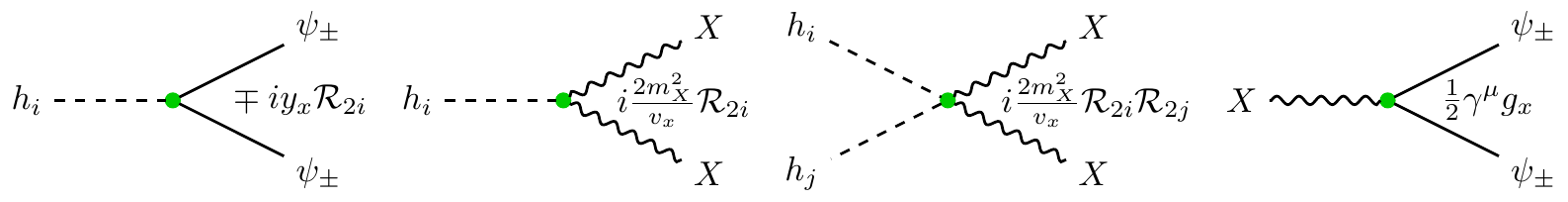}
		\caption{Feynman vertices of interactions involving the dark sector particles.}
		\label{fig:intDiags}
	\end{center}
	\end{figure}
Diagrams for the lowest order 2-2 processes involving dark particles are shown in fig. \ref{fig:2-2}.
	\begin{figure}
	\begin{center}
		\includegraphics[width=\textwidth]{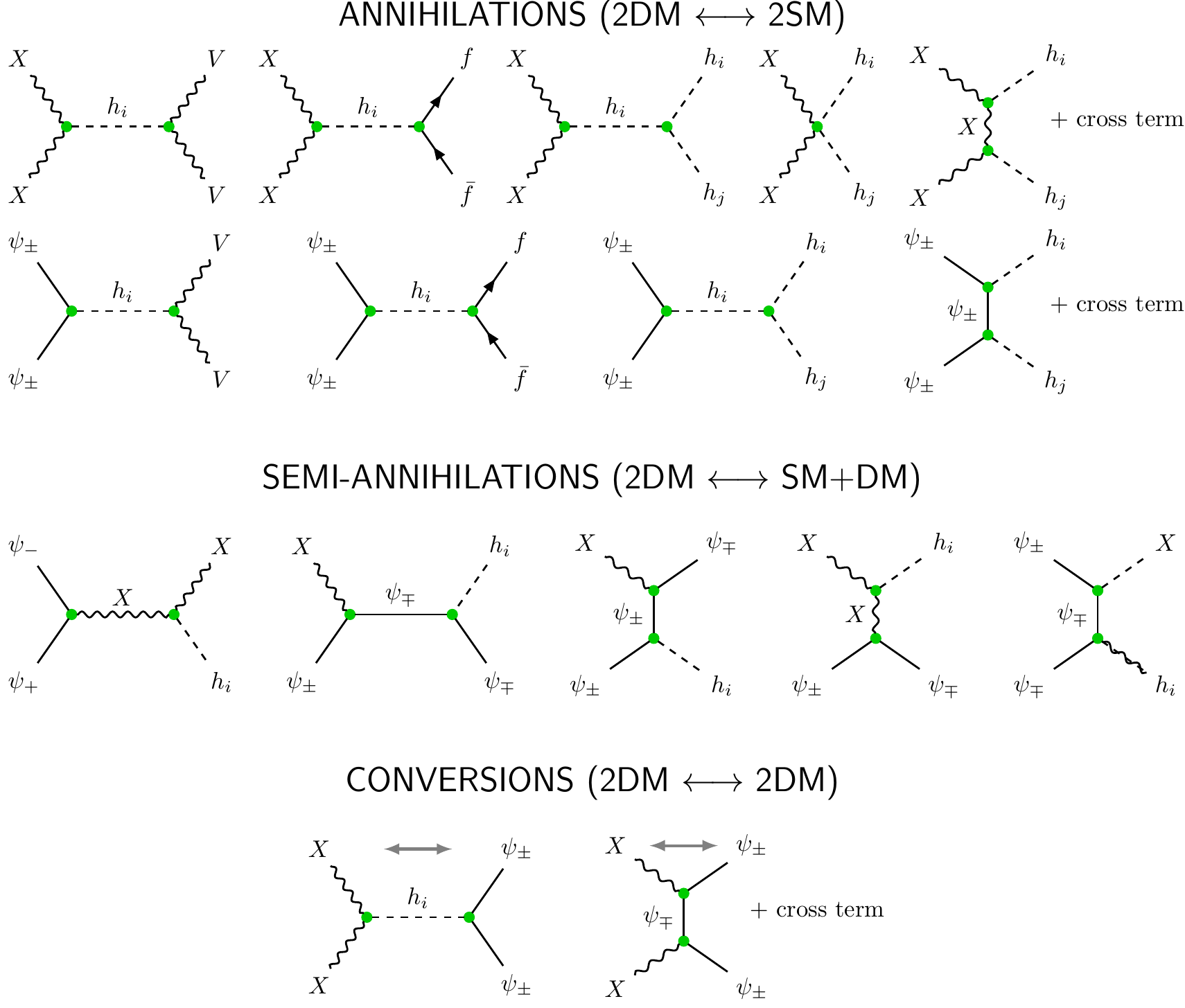}
	\end{center}
	\caption{The lowest order 2-2 processes Feynman diagrams}
	\label{fig:2-2}
	\end{figure}
In general, there are three kinds of these processes:
	\begin{itemize}
	\item annihilation -- transformation of 2 DM particles to 2 SM particles or vice-versa,
	\item semi-annihilation -- transformation of 2 DM particles to 1 SM particle and 1 DM particle or vice-versa,
	\item conversion -- annihilation within the dark sector.
	\end{itemize}
The collision terms in the Boltzmann equations contain contributions from these processes and decays (if they are allowed).
\begin{subsection}{Discrete symmetries of the Lagrangian and stability of the dark matter particles}\label{sec:sym}
The DM interaction Lagrangian has following discrete symmetries that are relevant for stability of the dark sector particles:
	\begin{center}
	\begin{tabular}{|c|c|c|c|c|}
	\hline
\rowcolor{col:gray}
	Symmetry &$X_\mu$ & $\psi_+$&$\psi_-$ & $\phi$\\
	\hline
	$\Z_2$ &$-$& $+$ &$-$ &$+$\\
\rowcolor{col:lightgray}
	$\Z^{'}_2$ &$-$ & $-$ &$+$ &$+$\\
	$\Z^{''}_2$ &$+$ & $-$ &$-$ &$+$\\
	\hline
	\end{tabular}
	\end{center}
Note that $\Z_2^{''}$ symmetry is a combination of $\Z_2$ and $\Z_2^{'}$. $\Z_2$ is actually the charge conjugation symmetry $\C$ expressed in terms of gauge field $X_\mu$ and fermionic mass eigenstates $\psi_\pm$.

The charge of each symmetry must be conserved during any process, therefore it is clear that the only dark matter decay vertex consists of all three dark fields, i.e. $X_\mu$ and $\psi_\pm$. The decay can be allowed or not, depending on masses of the particles. In order to decay, a given kind of DM particles has to be heavier than sum of the masses of two others. Since $m_\pm=m_D\pm y_xv_x$, $\psi_-$ is always lighter than $\psi_+$\footnote{Without loss of generality we assume $y_x>0$, $v_x>0$.}, so it has to be stable. We have therefore three possibilities:
	\begin{itemize}
	\item $m_+>m_-+m_X \Rightarrow$ $\psi_+$ is unstable,
	\item $m_X>m_-+m_+ \Rightarrow$ $X_\mu$ is unstable,
	\item other cases $\Rightarrow$ all three particles are stable.
	\end{itemize}
Hence, depending on the parameters, our model can be either 2- or 3-component dark matter model.
\end{subsection}
\begin{subsection}{Parameters of the theory}
The model is equipped with six free parameters, which can be chosen in several ways. A convenient and intuitive choice includes masses of the extra particles ($X_\mu$, $\psi_\pm$, $h_2$), the Higgs sector mixing angle and the $U(1)_X$ coupling constant:
\begin{align*}
m_X,\; m_+,\; m_-,\; m_2,\; \sin\alpha,\; g_x.
\end{align*}
Other parameters appearing in the Lagrangian can be expressed as follows:
	\begin{align*}
	v_x&=\frac{m_X}{g_x}, &
	\kappa&=\frac{(m_1^2 -m_2^2)}{2vm_X}\sin(2\alpha)g_x,\\
	\lambda_H&=\frac{m_1^2\cos^2\alpha+m_2^2\sin^2\alpha}{2v^2}, &
	\lambda_S&=\frac{m_1^2\sin^2\alpha+m_2^2\cos^2\alpha}{2m_X^2}g_x^2,\\
	y_x&=\frac{m_+-m_-}{2m_X}g_x, &
	m_D&=\frac{m_++m_-}{2}.
	\end{align*}
\end{subsection}
\begin{subsection}{The Boltzmann equations for the considered model}
The Boltzmann equations for the three new particles introduced in the model, $X_\mu$ and $\psi_\pm$, take the following form:
\begin{align*}
\frac{\mref}{m_X}\frac{dY_X}{dx}&=-L(x)\left[\vev{\sigma^{XX\alpha\alpha^\p}_v}\left(Y_X^2-\bar Y_X^2\right)+
\vev{\sigma^{X\psi_+\psi_-h_i}_v}\left(Y_XY_{\psi_+}-\bar Y_X\bar Y_{\psi_+}\frac{Y_{\psi_-}}{\bar Y_{\psi_-}}\right)\right.+\\
&\vev{\sigma^{X\psi_-\psi_+ h_i}_v}\left(Y_XY_{\psi_-}-\bar Y_X \bar Y_{\psi_-}\frac{ Y_{\psi_+}}{\bar Y_{\psi_+}}\right)+
\vev{\sigma^{Xh_i\psi_+\psi_- }_v}\bar Y_{h_i}\left(Y_X-\bar Y_X \frac{ Y_{\psi_+}Y_{\psi_-}}{\bar Y_{\psi_+}\bar Y_{\psi_-}}\right)+\\
&\vev{\sigma^{XX\psi_+\psi_+}_v}\left(Y_X^2- \bar Y_X^2\frac{Y_{\psi_+}^2}{\bar Y_{\psi_+}^2}\right)+
\left.\vev{\sigma^{XX\psi_-\psi_-}_v}\left(Y_X^2- \bar Y_X^2\frac{Y_{\psi_-}^2}{\bar Y_{\psi_-}^2}\right)\right]\\
&+\tilde L(x)\Gamma_{\psi_+\to X\psi_-}\left(Y_{\psi_+}-  \bar Y_{\psi_+}\frac{Y_X}{\bar Y_X}\frac{Y_{\psi_-}}{\bar Y_{\psi_-}}\right),\\
\frac{\mref}{m_-}\frac{dY_{\psi_-}}{dx}&=-L(x)\left[\vev{\sigma^{\psi_-\psi_-\alpha\alpha^\p}_v}\left(Y_{\psi_-}^2-\bar Y_{\psi_-}^2\right)\right.+
\vev{\sigma^{\psi_-\psi_+ X h_i}_v}\left(Y_{\psi_-}Y_{\psi_+}-\bar Y_{\psi_-} \bar Y_{\psi_+}\frac{ Y_X}{\bar Y_X}\right)+\\
&\vev{\sigma^{X\psi_-\psi_+ h_i}_v}\left(Y_XY_{\psi_-}-\bar Y_X \bar Y_{\psi_-}\frac{ Y_{\psi_+}}{\bar Y_{\psi_+}}\right)+
\vev{\sigma^{\psi_- h_iX\psi_+}_v} \bar Y_{h_i} \left(Y_{\psi_-}- \bar Y_{\psi_-}\frac{Y_{\psi_+}}{\bar Y_{\psi_+}}\frac{ Y_X}{\bar Y_X}\right)+\\
&\vev{\sigma^{\psi_-\psi_-XX}_v}\left(Y_{\psi_-}^2-\bar Y_{\psi_-}^2 \frac{Y_X^2}{\bar Y_X^2}\right)+
\left.\vev{\sigma^{\psi_-\psi_-\psi_+\psi_+}_v}\left(Y_{\psi_-}^2- \bar Y_{\psi_-}^2\frac{Y_{\psi_+}^2}{\bar Y_{\psi_+}^2}\right)\right]\\
&+\tilde L(x)\Gamma_{\psi_+\to X\psi_-} \left(Y_{\psi_+}- \bar Y_{\psi_+}\frac{Y_{\psi_-}}{\bar Y_{\psi_-}}\frac{ Y_X}{\bar Y_X}\right),\\
\frac{\mref}{m_+}\frac{dY_{\psi_+}}{dx}&=-L(x)\left[\vev{\sigma^{\psi_+\psi_+\alpha\alpha^\p}_v}\left(Y_{\psi_+}^2-\bar Y_{\psi_+}^2\right)\right.+
\vev{\sigma^{\psi_+\psi_- X h_i}_v}\left(Y_{\psi_+}Y_{\psi_-}-\bar Y_{\psi_+} \bar Y_{\psi_-}\frac{ Y_X}{\bar Y_X}\right)+\\
&\vev{\sigma^{X\psi_+\psi_- h_i}_v}\left(Y_XY_{\psi_+}-\bar Y_X \bar Y_{\psi_+}\frac{ Y_{\psi_-}}{\bar Y_{\psi_-}}\right)+
\vev{\sigma^{\psi_+ h_iX\psi_-}_v}\bar Y_{h_i}\left(Y_{\psi_+}- \bar Y_{\psi_+}\frac{Y_{\psi_-}}{\bar Y_{\psi_-}}\frac{ Y_X}{\bar Y_X}\right)+\\
&\vev{\sigma^{\psi_+\psi_+XX}_v}\left(Y_{\psi_+}^2-\bar Y_{\psi_+}^2 \frac{Y_X^2}{\bar Y_X^2}\right)+
\left.\vev{\sigma^{\psi_+\psi_+\psi_-\psi_-}_v}\left(Y_{\psi_+}^2- \bar Y_{\psi_+}^2\frac{Y_{\psi_-}^2}{\bar Y_{\psi_-}^2}\right)\right]\\
&-\tilde L(x)\Gamma_{\psi_+\to X\psi_-} \left(Y_{\psi_+}- \bar Y_{\psi_+}\frac{Y_{\psi_-}}{\bar Y_{\psi_-}}\frac{ Y_X}{\bar Y_X}\right),
\end{align*}
where $\mref$ is the mass used to define $x=\frac{\mref}{T}$, and $L(x)$, $\tilde L(x)$ are defined via eq. (\ref{eq:l}) \& (\ref{eq:ll}):
\begin{align*}
L(x)&=\sqrt\frac{\pi}{45}\mref\mpl x^{-2}\frac{g_{*s}}{\sqrt{g_*}},\\
\tilde L(x)&=\sqrt{\frac{45}{4\pi^3}}\frac{\mpl}{\mref^2}\frac{x}{\sqrt{g_*}}.
\end{align*}
In the equations, $\alpha$ and $\alpha^{'}$ denote any Standard Model particles (including $h_1$ and $h_2$, which are both assumed to be always in thermal and chemical equilibrium with photons) that can be produced in the annihilation process, and $h_i$ denotes $h_1$ or $h_2$.
\end{subsection}
\end{section}
\begin{section}{Results}
We solve the Boltzmann equation for our model using our own dedicated \protect\url{C++} code. It uses the \protect\url{calcHEP} package (see \cite{bib:calchep}) to calculate necessary spin-averaged matrix elements. There exists also \protect\url{micrOMEGAs} -- a code popular among dark matter physicists, allowing to analyze various properties of dark matter in a given model (see \cite{bib:micromegas}). However, developing our own code was necessary to deal with the case of 3 stable components, as \protect\url{micrOMEGAs} deals well with 1-2 component dark matter models only.\\

Data from LHC (see \cite{bib:sina}) provide an upper bound for the absolute value of the mixing angle: $|\sin\alpha|\lsim 0.3$, which we take into account while choosing parameters for the plots.\\

To make comparison of the results easier, for all the plots $\mref$, mass used to define variable $x\equiv\frac{\mref}{T}$, is set to the same value, that is $100$ GeV. Tables under the plots show first two non-zero coefficients of the thermally averaged cross-sections (given in picobarns) expanded in the powers of $x^{-1}$. $N$ denotes the first power of the expansion with non-zero coefficient, i.e. $\vev{\sigma v}=a_Nx^{-N}+a_{N+1}x^{-(N+1)}+\ldots$. For the decays, the decay rate (in GeV) is provided. The values of $Y(x=100)$ are collected in the tables at the botom of each figure.\\

To plot solutions of the Boltzmann equations, obtained with our code and \protect\url{micrOMEGAs}, we used \protect\url{gnuplot} (see \cite{bib:gnuplot}).\\

Following plots present the basic features of solutions of the Boltzmann equations in our model. More detailed analysis can be found in \cite{bib:vfdm}.
\begin{subsection}{Comparison of our code with micrOMEGAs}
To be sure that the code works properly, we compared it with the \micromegas results. An example of comparison is presented in Fig. \ref{fig:micro}, where results of our code are shown as lines, while points are \micromegas' output. The plots show very good agreement in the 2 component case, while for the 3 component case there is a large discrepancy.

	\begin{figure}[H]
	\makebox[\linewidth][c]{\centering\begin{tabular}{c}
		\begin{minipage}{.6\textwidth}\centering
			\includegraphics[width=\textwidth]{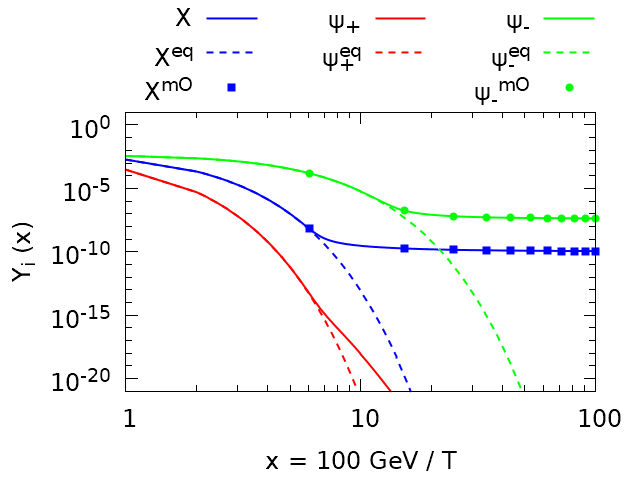}
\scalebox{.85}{
			\begin{tabular}{|c|c|c|c|}
				\hline
				process & $a_N$ & $a_{N+1}$ & $N$ \\
				\hline
				$XX \rightarrow \text{SM}$ & $8.5\cdot 10^{-3}$ & $-8.2\cdot 10^{-3}$ & $0$ \\
				$\psi_+\psi_+ \rightarrow \text{SM}$ & $2.4\cdot 10^{-4}$ & $-2.8\cdot 10^{-4}$ & $1$ \\
				$\psi_-\psi_- \rightarrow \text{SM}$ & $1.9\cdot 10^{-3}$ & $-2.6\cdot 10^{-2}$ & $1$ \\
				\hline
				$\Psi_+\Psi_+ \rightarrow XX$ & $8.6\cdot 10^{-4}$ & $1\cdot 10^{-3}$ & $0$ \\
				$XX \rightarrow \Psi_-\Psi_-$ & $2.3\cdot 10^{-4}$ & $-2\cdot 10^{-4}$ & $0$ \\
				$\Psi_+,\Psi_+ \rightarrow \Psi_-\Psi_-$ & $3.6\cdot 10^{-3}$ & $-2.7\cdot 10^{-4}$ & $0$ \\
				\hline
				$\Psi_+\Psi_- \rightarrow Xh_1$ & $1.8\cdot 10^{-5}$ & $-1.5\cdot 10^{-5}$ & $0$ \\
				$\Psi_+\Psi_- \rightarrow Xh_2$ & $2.4\cdot 10^{-3}$ & $-3.3\cdot 10^{-3}$ & $0$ \\
				$\Psi_+h_1 \rightarrow X\Psi_-$ & $3.8\cdot 10^{-5}$ & $6.4\cdot 10^{-4}$ & $0$ \\
				$\Psi_+h_2 \rightarrow X\Psi_-$ & $4.1\cdot 10^{-3}$ & $4.3\cdot 10^{-2}$ & $0$ \\
				$X\Psi_+ \rightarrow \Psi_-h_1$ & $6.2\cdot 10^{-5}$ & $-3.5\cdot 10^{-5}$ & $0$ \\
				$X\Psi_+ \rightarrow \Psi_-h_2$ & $6.1\cdot 10^{-3}$ & $-3.5\cdot 10^{-3}$ & $0$ \\
				\hline
				$\Psi_+ \rightarrow X\Psi_-$ & \multicolumn{3}{|c|}{$2.3\cdot 10^{-2}$} \\
				\hline
			\end{tabular}
}
		\end{minipage}
		\begin{minipage}{.6\textwidth}\centering
			\includegraphics[width=\textwidth]{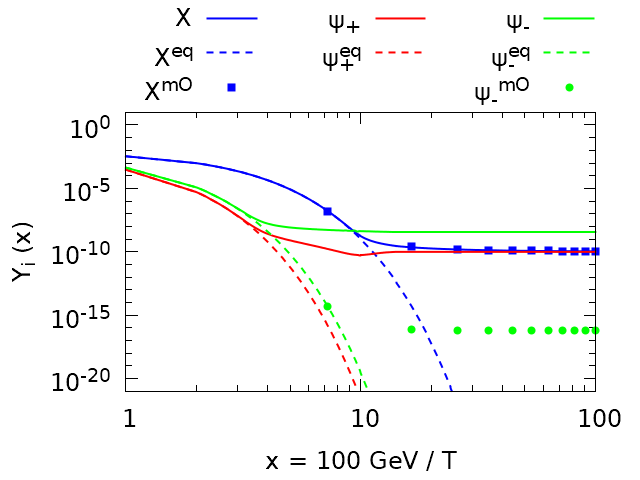}
\scalebox{.85}{
			\begin{tabular}{|c|c|c|c|}
				\hline
				process & $a_N$ & $a_{N+1}$ & $N$ \\
				\hline
				$XX \rightarrow \text{SM}$ & $1.4\cdot 10^{-2}$ & $-1.8\cdot 10^{-2}$ & $0$ \\
				$\psi_+\psi_+ \rightarrow \text{SM}$ & $4.6\cdot 10^{-8}$ & $-4.4\cdot 10^{-8}$ & $1$ \\
				$\psi_-\psi_- \rightarrow \text{SM}$ & $1.7\cdot 10^{-6}$ & $-2.4\cdot 10^{-6}$ & $1$ \\
				\hline
				$\Psi_+\Psi_+ \rightarrow XX$ & $2.2\cdot 10^{-4}$ & $3.5\cdot 10^{-5}$ & $0$ \\
				$\Psi_-\Psi_- \rightarrow XX$ & $1.7\cdot 10^{-4}$ & $5.5\cdot 10^{-5}$ & $0$ \\
				$\Psi_+,\Psi_+ \rightarrow \Psi_-\Psi_-$ & $3.7\cdot 10^{-3}$ & $4.5\cdot 10^{-3}$ & $0$ \\
				\hline
				$\Psi_+\Psi_- \rightarrow Xh_1$ & $2.8\cdot 10^{-6}$ & $-2.3\cdot 10^{-6}$ & $0$ \\
				$\Psi_+\Psi_- \rightarrow Xh_2$ & $2.7\cdot 10^{-4}$ & $-2.2\cdot 10^{-4}$ & $0$ \\
				$X\Psi_- \rightarrow \Psi_+h_1$ & $8.2\cdot 10^{-7}$ & $9.9\cdot 10^{-5}$ & $0$ \\
				$\Psi_+h_2 \rightarrow X\Psi_-$ & $6.26\cdot 10^{0}$ & $-1.06\cdot 10^{4}$ & $2$ \\
				$X\Psi_- \rightarrow \Psi_+h_2$ & $1.59\cdot 10^{0}$ & $-2.69\cdot 10^{3}$ & $2$ \\
				$X\Psi_+ \rightarrow \Psi_-h_1$ & $1.3\cdot 10^{-5}$ & $9.7\cdot 10^{-5}$ & $0$ \\
				$X\Psi_+ \rightarrow \Psi_-h_2$ & $1.1\cdot 10^{-3}$ & $9.9\cdot 10^{-3}$ & $0$ \\
				\hline
			\end{tabular}
}
		\end{minipage}\\~\\
			\begin{tabular}{|c|c|c|}
			\hline
			$Y(x=100)$&left&right\\
			\hline
			$Y_X$&$1.1\e{-10}$&$1.1\e{-10}$\\
			$Y_+$&$6.5\e{-62}$&$9.7\e{-11}$\\
			$Y_-$&$4.3\e{-8}$&$3.6\e{-9}$\\
			\hline
			\end{tabular}\\
	\end{tabular}}
		\caption{Comparison of our code with micrOMEGAs. The case of 2 stable components agrees, while in the case of 3 stable components there is strong discrepancy. Parameters of the plots are: $m_2=150\gev$, $\sin\alpha=0.1$, $g_x=0.1$. Masses of the dark sector particles are: $m_X=300\gev$, $m_+=500\gev$, $m_-=100\gev$ (left), $m_X=200\gev$, $m_+=500\gev$, $m_-=450\gev$ (right)}
		\label{fig:micro}
	\end{figure}

\end{subsection}
\begin{subsection}{Influence of $m_X$ and $m_\pm$}
In this section we investigate how do various configurations of masses lead to different behaviour of yield functions. Throughout whole section, $m_2=150\gev$, $\sin\alpha=0.1=g_x$.\\
\begin{subsubsection}{Two stable components}
Let us first focus on the case of two stable components. Fig. \ref{fig:2-1dm1-3} and \ref{fig:2-1dm4-6} show results for one stable particle ($X$ and $\psi_-$, respectively) much lighter than the second one.

In the first case, increasing the mass of $X$ leads to increase of abundances of both fermions. The coupling between fermions and the Higgs particles is proportional to $y_x=\frac{m_+-m_-}{2m_X}g_x$, therefore if $m_X$ is larger, fermions decouple earlier and their yields are larger. Note that behaviour of $X$ yield is nontrivial; first it rises with $m_X$ but than it decreases. Since mass of the second Higgs particle is $m_2=150\gev$, after exceeding this threshold new channel of annihilation appears.

In the second case, mass of $\psi_-$ does not influence behaviour of $Y_X$, while $Y_+$ is increasing and $Y_-$ is decreasing.\\

	\begin{figure}[H]
	\makebox[\linewidth][c]{\centering\begin{tabular}{c}
		\begin{minipage}{.45\textwidth}\centering
			\includegraphics[width=\textwidth]{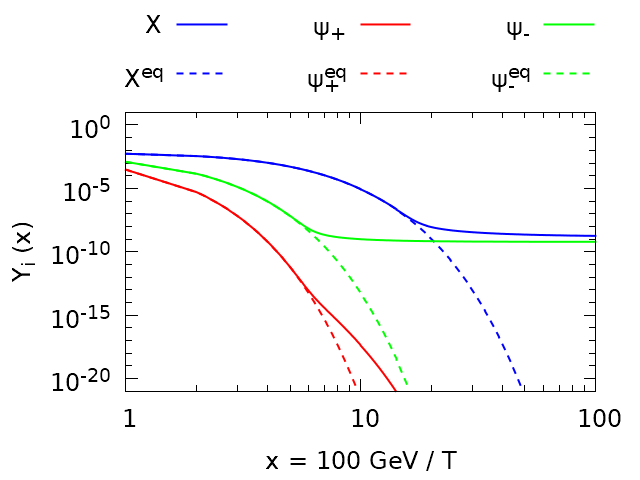}
\scalebox{.85}{
			\begin{tabular}{|c|c|c|c|}
				\hline
				process & $a_N$ & $a_{N+1}$ & $N$ \\
				\hline
				$XX \rightarrow \text{SM}$ & $1.9\cdot 10^{-3}$ & $-1.5\cdot 10^{-2}$ & $0$ \\
				$\psi_+\psi_+ \rightarrow \text{SM}$ & $9.9\cdot 10^{-4}$ & $-1.1\cdot 10^{-3}$ & $1$ \\
				$\psi_-\psi_- \rightarrow \text{SM}$ & $1.3\cdot 10^{-2}$ & $-2.7\cdot 10^{-2}$ & $1$ \\
				\hline
				$\Psi_+\Psi_+ \rightarrow XX$ & $4.2\cdot 10^{-4}$ & $2\cdot 10^{-3}$ & $0$ \\
				$\Psi_-\Psi_- \rightarrow XX$ & $1.3\cdot 10^{-4}$ & $3.9\cdot 10^{-3}$ & $0$ \\
				$\Psi_+,\Psi_+ \rightarrow \Psi_-\Psi_-$ & $2.8\cdot 10^{-2}$ & $1.1\cdot 10^{-2}$ & $0$ \\
				\hline
				$\Psi_+\Psi_- \rightarrow Xh_1$ & $3.2\cdot 10^{-5}$ & $-3.5\cdot 10^{-5}$ & $0$ \\
				$\Psi_+\Psi_- \rightarrow Xh_2$ & $3.1\cdot 10^{-3}$ & $-3.5\cdot 10^{-3}$ & $0$ \\
				$\Psi_+h_1 \rightarrow X\Psi_-$ & $8.4\cdot 10^{-4}$ & $2\cdot 10^{-3}$ & $0$ \\
				$\Psi_+h_2 \rightarrow X\Psi_-$ & $3.1\cdot 10^{-2}$ & $3.1\cdot 10^{-1}$ & $0$ \\
				$X\Psi_+ \rightarrow \Psi_-h_1$ & $2.4\cdot 10^{-3}$ & $-6.2\cdot 10^{-3}$ & $0$ \\
				$X\Psi_+ \rightarrow \Psi_-h_2$ & $3\cdot 10^{-1}$ & $-8.1\cdot 10^{-1}$ & $0$ \\
				\hline
				$\Psi_+ \rightarrow X\Psi_-$ & \multicolumn{3}{|c|}{$1.1\cdot 10^{-1}$} \\
				\hline
			\end{tabular}
}
		\end{minipage}
		\begin{minipage}{.45\textwidth}\centering
		\includegraphics[width=\textwidth]{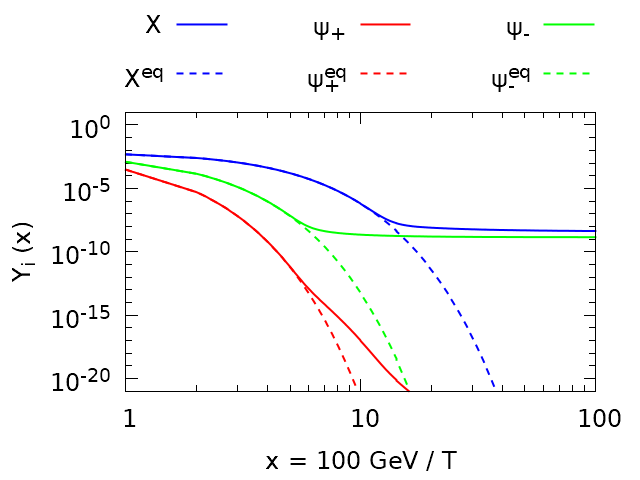}
\scalebox{.85}{
			\begin{tabular}{|c|c|c|c|}
				\hline
				process & $a_N$ & $a_{N+1}$ & $N$ \\
				\hline
				$XX \rightarrow \text{SM}$ & $3.9\cdot 10^{-4}$ & $-9.8\cdot 10^{-4}$ & $0$ \\
				$\psi_+\psi_+ \rightarrow \text{SM}$ & $3.5\cdot 10^{-4}$ & $-3.8\cdot 10^{-4}$ & $1$ \\
				$\psi_-\psi_- \rightarrow \text{SM}$ & $4.4\cdot 10^{-3}$ & $-9.2\cdot 10^{-3}$ & $1$ \\
				\hline
				$\Psi_+\Psi_+ \rightarrow XX$ & $4.2\cdot 10^{-4}$ & $8.6\cdot 10^{-4}$ & $0$ \\
				$\Psi_-\Psi_- \rightarrow XX$ & $1.2\cdot 10^{-4}$ & $1.2\cdot 10^{-3}$ & $0$ \\
				$\Psi_+,\Psi_+ \rightarrow \Psi_-\Psi_-$ & $1.3\cdot 10^{-2}$ & $5.8\cdot 10^{-3}$ & $0$ \\
				\hline
				$\Psi_+\Psi_- \rightarrow Xh_1$ & $6.5\cdot 10^{-6}$ & $-6\cdot 10^{-6}$ & $0$ \\
				$\Psi_+\Psi_- \rightarrow Xh_2$ & $6.5\cdot 10^{-4}$ & $-5.8\cdot 10^{-4}$ & $0$ \\
				$\Psi_+h_1 \rightarrow X\Psi_-$ & $1\cdot 10^{-4}$ & $2.6\cdot 10^{-3}$ & $0$ \\
				$\Psi_+h_2 \rightarrow X\Psi_-$ & $8.5\cdot 10^{-3}$ & $2\cdot 10^{-1}$ & $0$ \\
				$X\Psi_+ \rightarrow \Psi_-h_1$ & $6.5\cdot 10^{-4}$ & $-1.1\cdot 10^{-3}$ & $0$ \\
				$X\Psi_+ \rightarrow \Psi_-h_2$ & $7\cdot 10^{-2}$ & $-1.3\cdot 10^{-1}$ & $0$ \\
				\hline
				$\Psi_+ \rightarrow X\Psi_-$ & \multicolumn{3}{|c|}{$4.4\cdot 10^{-2}$} \\
				\hline
			\end{tabular}
}
		\end{minipage}
		\begin{minipage}{.45\textwidth}\centering
		\includegraphics[width=\textwidth]{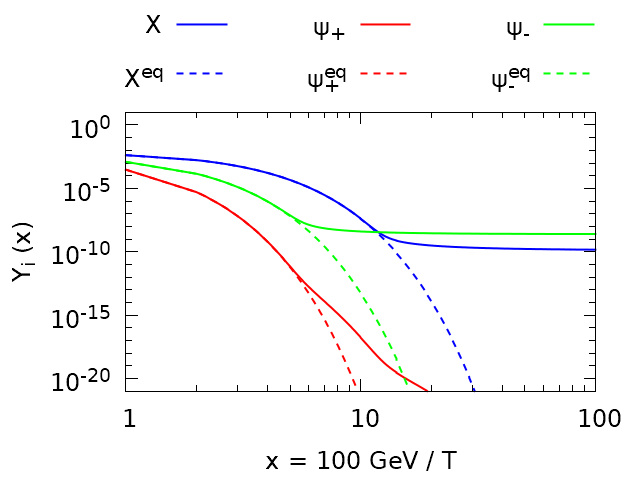}
\scalebox{.85}{
			\begin{tabular}{|c|c|c|c|}
				\hline
				process & $a_N$ & $a_{N+1}$ & $N$ \\
				\hline
				$XX \rightarrow \text{SM}$ & $1.3\cdot 10^{-2}$ & $-8\cdot 10^{-4}$ & $0$ \\
				$\psi_+\psi_+ \rightarrow \text{SM}$ & $1.5\cdot 10^{-4}$ & $-1.7\cdot 10^{-4}$ & $1$ \\
				$\psi_-\psi_- \rightarrow \text{SM}$ & $1.9\cdot 10^{-3}$ & $-4\cdot 10^{-3}$ & $1$ \\
				\hline
				$\Psi_+\Psi_+ \rightarrow XX$ & $4.2\cdot 10^{-4}$ & $4.7\cdot 10^{-4}$ & $0$ \\
				$\Psi_-\Psi_- \rightarrow XX$ & $1.1\cdot 10^{-4}$ & $4.6\cdot 10^{-4}$ & $0$ \\
				$\Psi_+,\Psi_+ \rightarrow \Psi_-\Psi_-$ & $7.9\cdot 10^{-3}$ & $3.7\cdot 10^{-3}$ & $0$ \\
				\hline
				$\Psi_+\Psi_- \rightarrow Xh_1$ & $1.1\cdot 10^{-6}$ & $1.2\cdot 10^{-7}$ & $0$ \\
				$\Psi_+\Psi_- \rightarrow Xh_2$ & $1.1\cdot 10^{-4}$ & $2.1\cdot 10^{-5}$ & $0$ \\
				$\Psi_+h_1 \rightarrow X\Psi_-$ & $7.3\cdot 10^{-5}$ & $1.5\cdot 10^{-3}$ & $0$ \\
				$\Psi_+h_2 \rightarrow X\Psi_-$ & $1\cdot 10^{-2}$ & $1.1\cdot 10^{-1}$ & $0$ \\
				$X\Psi_+ \rightarrow \Psi_-h_1$ & $2.6\cdot 10^{-4}$ & $-2.7\cdot 10^{-4}$ & $0$ \\
				$X\Psi_+ \rightarrow \Psi_-h_2$ & $2.6\cdot 10^{-2}$ & $-3\cdot 10^{-2}$ & $0$ \\
				\hline
				$\Psi_+ \rightarrow X\Psi_-$ & \multicolumn{3}{|c|}{$1.5\cdot 10^{-2}$} \\
				\hline
			\end{tabular}
}
		\end{minipage}\\~\\
		\begin{tabular}{|c|c|c|c|}
		\hline
		$Y(x=100)$&left&middle&right\\
		\hline
		$Y_X$&$1.8\e{-9}$&$4.3\e{-9}$&$1.4\e{-10}$\\
		$Y_+$&$1.5\e{-62}$&$5.8\e{-49}$&$2.7\e{-37}$\\
		$Y_-$&$6.0\e{-10}$&$1.4\e{-9}$&$2.5\e{-9}$\\
		\hline
		\end{tabular}
	\end{tabular}}
		\caption{The case of two stable components with one of them much lighter. Parameters of the plots are: $m_+=500\gev$, $m_-=300\gev$, $m_2=150\gev$, $\sin\alpha=0.1$, $g_x=0.1$, $m_X$ is $100\gev$ (left), $130\gev$ (middle), $160\gev$ (right).}
		\label{fig:2-1dm1-3}
	\end{figure}

	\begin{figure}[H]
	\makebox[\linewidth][c]{\centering\begin{tabular}{c}
		\begin{minipage}{.45\textwidth}\centering
			\includegraphics[width=\textwidth]{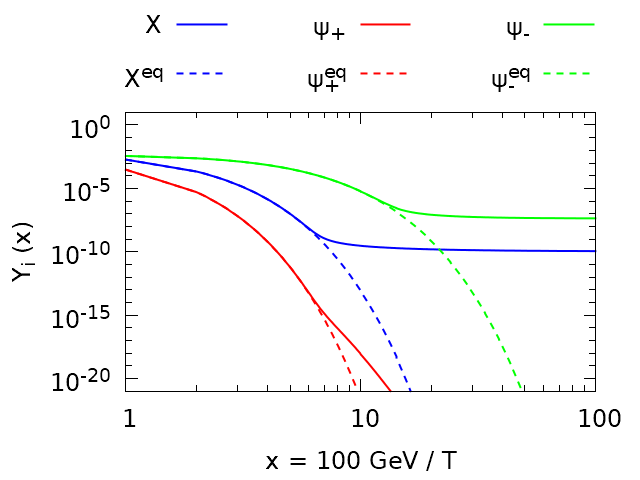}
\scalebox{.85}{
			\begin{tabular}{|c|c|c|c|}
				\hline
				process & $a_N$ & $a_{N+1}$ & $N$ \\
				\hline
				$XX \rightarrow \text{SM}$ & $8.5\cdot 10^{-3}$ & $-8.2\cdot 10^{-3}$ & $0$ \\
				$\psi_+\psi_+ \rightarrow \text{SM}$ & $2.4\cdot 10^{-4}$ & $-2.8\cdot 10^{-4}$ & $1$ \\
				$\psi_-\psi_- \rightarrow \text{SM}$ & $1.9\cdot 10^{-3}$ & $-2.6\cdot 10^{-2}$ & $1$ \\
				\hline
				$\Psi_+\Psi_+ \rightarrow XX$ & $8.6\cdot 10^{-4}$ & $1\cdot 10^{-3}$ & $0$ \\
				$XX \rightarrow \Psi_-\Psi_-$ & $2.3\cdot 10^{-4}$ & $-2\cdot 10^{-4}$ & $0$ \\
				$\Psi_+,\Psi_+ \rightarrow \Psi_-\Psi_-$ & $3.6\cdot 10^{-3}$ & $-2.7\cdot 10^{-4}$ & $0$ \\
				\hline
				$\Psi_+\Psi_- \rightarrow Xh_1$ & $1.8\cdot 10^{-5}$ & $-1.5\cdot 10^{-5}$ & $0$ \\
				$\Psi_+\Psi_- \rightarrow Xh_2$ & $2.4\cdot 10^{-3}$ & $-3.3\cdot 10^{-3}$ & $0$ \\
				$\Psi_+h_1 \rightarrow X\Psi_-$ & $3.8\cdot 10^{-5}$ & $6.4\cdot 10^{-4}$ & $0$ \\
				$\Psi_+h_2 \rightarrow X\Psi_-$ & $4.1\cdot 10^{-3}$ & $4.3\cdot 10^{-2}$ & $0$ \\
				$X\Psi_+ \rightarrow \Psi_-h_1$ & $6.2\cdot 10^{-5}$ & $-3.5\cdot 10^{-5}$ & $0$ \\
				$X\Psi_+ \rightarrow \Psi_-h_2$ & $6.1\cdot 10^{-3}$ & $-3.5\cdot 10^{-3}$ & $0$ \\
				\hline
				$\Psi_+ \rightarrow X\Psi_-$ & \multicolumn{3}{|c|}{$2.3\cdot 10^{-2}$} \\
				\hline
			\end{tabular}
}
		\end{minipage}
		\begin{minipage}{.45\textwidth}\centering
		\includegraphics[width=\textwidth]{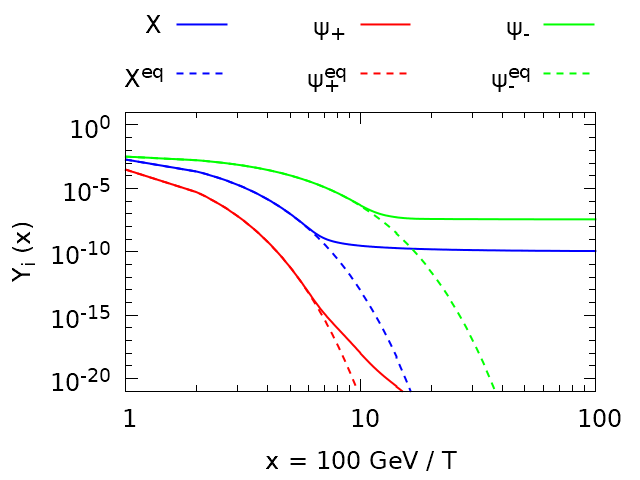}
\scalebox{.85}{
			\begin{tabular}{|c|c|c|c|}
				\hline
				process & $a_N$ & $a_{N+1}$ & $N$ \\
				\hline
				$XX \rightarrow \text{SM}$ & $8.5\cdot 10^{-3}$ & $-8.2\cdot 10^{-3}$ & $0$ \\
				$\psi_+\psi_+ \rightarrow \text{SM}$ & $1.7\cdot 10^{-4}$ & $-1.9\cdot 10^{-4}$ & $1$ \\
				$\psi_-\psi_- \rightarrow \text{SM}$ & $2.2\cdot 10^{-4}$ & $-1.2\cdot 10^{-3}$ & $1$ \\
				\hline
				$\Psi_+\Psi_+ \rightarrow XX$ & $7.9\cdot 10^{-4}$ & $8.8\cdot 10^{-4}$ & $0$ \\
				$XX \rightarrow \Psi_-\Psi_-$ & $1.8\cdot 10^{-4}$ & $-1.4\cdot 10^{-4}$ & $0$ \\
				$\Psi_+,\Psi_+ \rightarrow \Psi_-\Psi_-$ & $3.3\cdot 10^{-3}$ & $-5.8\cdot 10^{-5}$ & $0$ \\
				\hline
				$\Psi_+\Psi_- \rightarrow Xh_1$ & $3\cdot 10^{-6}$ & $9.5\cdot 10^{-6}$ & $0$ \\
				$\Psi_+\Psi_- \rightarrow Xh_2$ & $3.5\cdot 10^{-4}$ & $1\cdot 10^{-3}$ & $0$ \\
				$\Psi_+h_1 \rightarrow X\Psi_-$ & $3.3\cdot 10^{-5}$ & $5.6\cdot 10^{-4}$ & $0$ \\
				$\Psi_+h_2 \rightarrow X\Psi_-$ & $3.6\cdot 10^{-3}$ & $3.8\cdot 10^{-2}$ & $0$ \\
				$X\Psi_+ \rightarrow \Psi_-h_1$ & $5.7\cdot 10^{-5}$ & $-2.9\cdot 10^{-5}$ & $0$ \\
				$X\Psi_+ \rightarrow \Psi_-h_2$ & $5.6\cdot 10^{-3}$ & $-3.1\cdot 10^{-3}$ & $0$ \\
				\hline
				$\Psi_+ \rightarrow X\Psi_-$ & \multicolumn{3}{|c|}{$1.4\cdot 10^{-2}$} \\
				\hline
			\end{tabular}
}
		\end{minipage}
		\begin{minipage}{.45\textwidth}\centering
		\includegraphics[width=\textwidth]{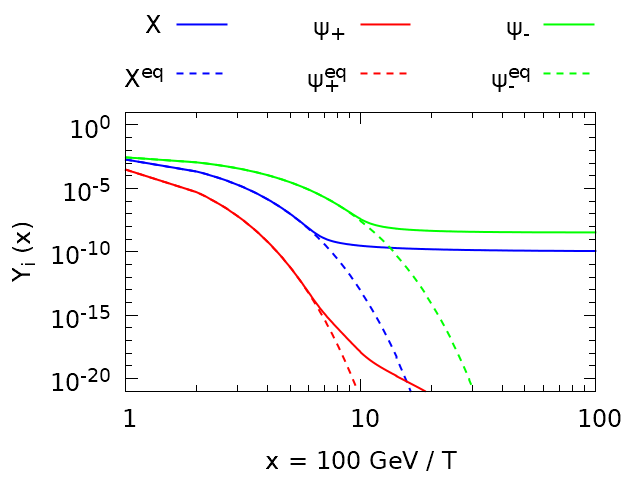}
\scalebox{.85}{
			\begin{tabular}{|c|c|c|c|}
				\hline
				process & $a_N$ & $a_{N+1}$ & $N$ \\
				\hline
				$XX \rightarrow \text{SM}$ & $8.5\cdot 10^{-3}$ & $-8.2\cdot 10^{-3}$ & $0$ \\
				$\psi_+\psi_+ \rightarrow \text{SM}$ & $1.2\cdot 10^{-4}$ & $-1.4\cdot 10^{-4}$ & $1$ \\
				$\psi_-\psi_- \rightarrow \text{SM}$ & $9.2\cdot 10^{-3}$ & $-1.2\cdot 10^{-2}$ & $1$ \\
				\hline
				$\Psi_+\Psi_+ \rightarrow XX$ & $7.2\cdot 10^{-4}$ & $7.4\cdot 10^{-4}$ & $0$ \\
				$XX \rightarrow \Psi_-\Psi_-$ & $1.4\cdot 10^{-4}$ & $-8\cdot 10^{-5}$ & $0$ \\
				$\Psi_+,\Psi_+ \rightarrow \Psi_-\Psi_-$ & $3\cdot 10^{-3}$ & $1.4\cdot 10^{-4}$ & $0$ \\
				\hline
				$\Psi_+\Psi_- \rightarrow Xh_1$ & $1\cdot 10^{-6}$ & $5.7\cdot 10^{-6}$ & $0$ \\
				$\Psi_+\Psi_- \rightarrow Xh_2$ & $1\cdot 10^{-4}$ & $6.6\cdot 10^{-4}$ & $0$ \\
				$\Psi_+h_1 \rightarrow X\Psi_-$ & $3.1\cdot 10^{-5}$ & $4.4\cdot 10^{-4}$ & $0$ \\
				$\Psi_+h_2 \rightarrow X\Psi_-$ & $3.5\cdot 10^{-3}$ & $3.1\cdot 10^{-2}$ & $0$ \\
				$X\Psi_+ \rightarrow \Psi_-h_1$ & $5.1\cdot 10^{-5}$ & $-2.3\cdot 10^{-5}$ & $0$ \\
				$X\Psi_+ \rightarrow \Psi_-h_2$ & $5.1\cdot 10^{-3}$ & $-2.3\cdot 10^{-3}$ & $0$ \\
				\hline
				$\Psi_+ \rightarrow X\Psi_-$ & \multicolumn{3}{|c|}{$6.6\cdot 10^{-3}$} \\
				\hline
			\end{tabular}
}
		\end{minipage}\\~\\
		\begin{tabular}{|c|c|c|c|}\hline
		$Y(x=100)$&left&middle&right\\
		\hline
		$Y_X$&$1.1\e{-10}$&$1.1\e{-10}$&$1.1\e{-10}$\\
		$Y_+$&$6.4\e{-62}$&$4.0\e{-49}$&$2.9\e{-37}$\\
		$Y_-$&$4.3\e{-8}$&$3.6\e{-8}$&$3.3\e{-9}$\\
		\hline\end{tabular}
	\end{tabular}}
		\caption{The case of 2 stable components with one of them much lighter. Parameters of the plots are: $m_X=300\gev$ $m_+=500\gev$, $m_2=150\gev$, $\sin\alpha=0.1$, $g_x=0.1$, $m_-$ is $100\gev$ (left), $130\gev$ (middle), $160\gev$ (right).}
		\label{fig:2-1dm4-6}
	\end{figure}

Figs. \ref{fig:2dm1-3}, \ref{fig:2dm4-6} and \ref{fig:2dm7-9} show results for the case when masses of the stable particles are relatively close (or even equal) to each other. In fig. \ref{fig:2dm1-3}, mass of $X$ increases, what leads to decreasing value of $Y_X$, since before the decoupling the value of $Y$ is proportional to $e^{-x}=\exp(-m_X/\mref)$. However, as $m_X$ rises, the cross-section for annihilation of $\psi_-$ into the SM decreases as the $\psi_-\psi_-\rightarrow h_i$ vertex is proportional to $y_x=\frac{m_+-m_-}{2m_X}g_x$. Therefore, the decoupling of $\psi_-$ takes place earlier, when $Y_+$ is larger. Conversion processes from heavier $\psi_+$ fermions into lighter $X$ bosons increas the value of $Y_X$. Note that even when $m_X=m_-$, final yields are not the same: $Y_X<Y_-$. The cross-section for annihilation of $X$ particles is constant in the leading order, while the cross-section for annihilation of $\psi_-$ behaves like $x^{-1}$, therefore is much smaller. Hence the boson decouples later, when its yield is smaller.

In fig. \ref{fig:2dm4-6}, behaviour of $Y_X$ remains unchanged while $m_-$ rises. This is because the $X$ coupling to the Higgs particles is proportional to $m_X$ and remains constant, much bigger than the conversion cross-section between $X$ and $\psi_-$. Lowering the $Y_-$ value is caused by lower equilibrium value due to mass increasing.

In fig. \ref{fig:2dm7-9}, yields of all three particles rise when $m_-$ increases. Due to smaller difference of masses, and therefore smaller value of $y_x$, the cross-section for $\psi_+$ and $\psi_-$ fermions annihilation into SM decreases, what leads to earlier decoupling.

	\begin{figure}[H]
	\makebox[\linewidth][c]{\centering\begin{tabular}{c}
		\begin{minipage}{.45\textwidth}\centering
			\includegraphics[width=\textwidth]{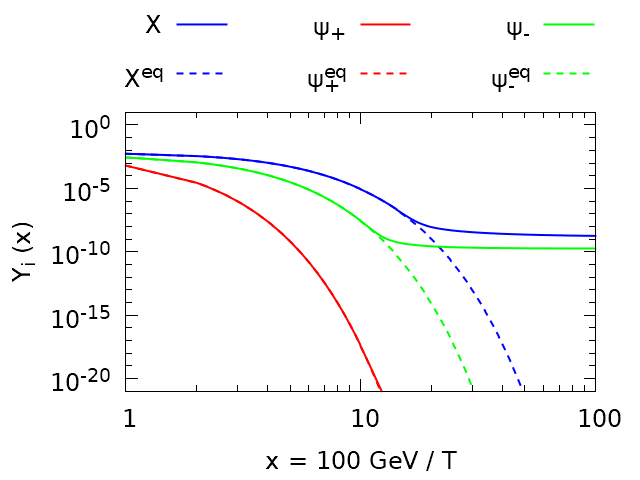}
\scalebox{.85}{
			\begin{tabular}{|c|c|c|c|}
				\hline
				process & $a_N$ & $a_{N+1}$ & $N$ \\
				\hline
				$XX \rightarrow \text{SM}$ & $1.2\cdot 10^{-3}$ & $-8.6\cdot 10^{-3}$ & $0$ \\
				$\psi_+\psi_+ \rightarrow \text{SM}$ & $3.4\cdot 10^{-3}$ & $-4.8\cdot 10^{-3}$ & $1$ \\
				$\psi_-\psi_- \rightarrow \text{SM}$ & $1.8\cdot 10^{-1}$ & $-1.9\cdot 10^{-1}$ & $1$ \\
				\hline
				$\Psi_+\Psi_+ \rightarrow XX$ & $9.1\cdot 10^{-4}$ & $7.1\cdot 10^{-3}$ & $0$ \\
				$\Psi_-\Psi_- \rightarrow XX$ & $7.2\cdot 10^{-5}$ & $2\cdot 10^{-2}$ & $0$ \\
				$\Psi_+,\Psi_+ \rightarrow \Psi_-\Psi_-$ & $3.7\cdot 10^{-2}$ & $5.8\cdot 10^{-4}$ & $0$ \\
				\hline
				$\Psi_+\Psi_- \rightarrow Xh_1$ & $1.9\cdot 10^{-4}$ & $-3.5\cdot 10^{-4}$ & $0$ \\
				$\Psi_+\Psi_- \rightarrow Xh_2$ & $1.9\cdot 10^{-2}$ & $-3.6\cdot 10^{-2}$ & $0$ \\
				$\Psi_+h_1 \rightarrow X\Psi_-$ & $1.2\cdot 10^{-3}$ & $1.6\cdot 10^{-3}$ & $0$ \\
				$\Psi_+h_2 \rightarrow X\Psi_-$ & $5.2\cdot 10^{-2}$ & $2.9\cdot 10^{-1}$ & $0$ \\
				$X\Psi_+ \rightarrow \Psi_-h_1$ & $2.2\cdot 10^{-3}$ & $-5.6\cdot 10^{-3}$ & $0$ \\
				$X\Psi_+ \rightarrow \Psi_-h_2$ & $2.6\cdot 10^{-1}$ & $-7.1\cdot 10^{-1}$ & $0$ \\
				\hline
				$\Psi_+ \rightarrow X\Psi_-$ & \multicolumn{3}{|c|}{$1.3\cdot 10^{-1}$} \\
				\hline
			\end{tabular}
}
		\end{minipage}
		\begin{minipage}{.45\textwidth}\centering
		\includegraphics[width=\textwidth]{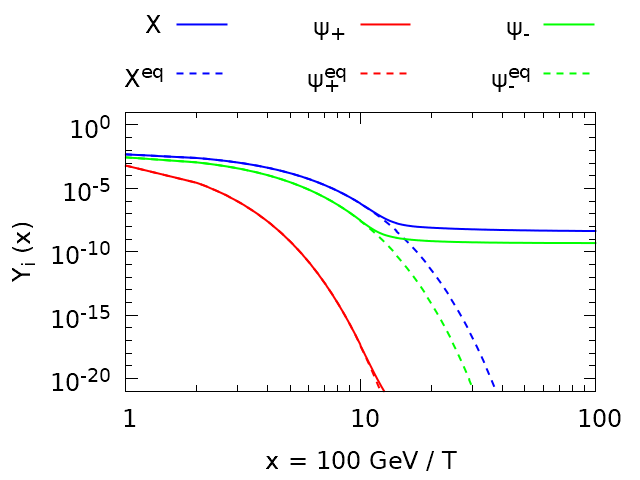}
\scalebox{.85}{
			\begin{tabular}{|c|c|c|c|}
				\hline
				process & $a_N$ & $a_{N+1}$ & $N$ \\
				\hline
				$XX \rightarrow \text{SM}$ & $3.9\cdot 10^{-4}$ & $-9.8\cdot 10^{-4}$ & $0$ \\
				$\psi_+\psi_+ \rightarrow \text{SM}$ & $1.4\cdot 10^{-3}$ & $-2\cdot 10^{-3}$ & $1$ \\
				$\psi_-\psi_- \rightarrow \text{SM}$ & $7.5\cdot 10^{-2}$ & $-8.2\cdot 10^{-2}$ & $1$ \\
				\hline
				$\Psi_+\Psi_+ \rightarrow XX$ & $9.2\cdot 10^{-4}$ & $3.4\cdot 10^{-3}$ & $0$ \\
				$\Psi_-\Psi_- \rightarrow XX$ & $3.5\cdot 10^{-5}$ & $5.9\cdot 10^{-3}$ & $0$ \\
				$\Psi_+,\Psi_+ \rightarrow \Psi_-\Psi_-$ & $2\cdot 10^{-2}$ & $9.7\cdot 10^{-4}$ & $0$ \\
				\hline
				$\Psi_+\Psi_- \rightarrow Xh_1$ & $6.2\cdot 10^{-5}$ & $-1.1\cdot 10^{-4}$ & $0$ \\
				$\Psi_+\Psi_- \rightarrow Xh_2$ & $6.5\cdot 10^{-3}$ & $-1.1\cdot 10^{-2}$ & $0$ \\
				$\Psi_+h_1 \rightarrow X\Psi_-$ & $3.2\cdot 10^{-4}$ & $2.5\cdot 10^{-3}$ & $0$ \\
				$\Psi_+h_2 \rightarrow X\Psi_-$ & $2\cdot 10^{-2}$ & $2.1\cdot 10^{-1}$ & $0$ \\
				$X\Psi_+ \rightarrow \Psi_-h_1$ & $7.9\cdot 10^{-4}$ & $-1.5\cdot 10^{-3}$ & $0$ \\
				$X\Psi_+ \rightarrow \Psi_-h_2$ & $8.4\cdot 10^{-2}$ & $-1.7\cdot 10^{-1}$ & $0$ \\
				\hline
				$\Psi_+ \rightarrow X\Psi_-$ & \multicolumn{3}{|c|}{$7.1\cdot 10^{-2}$} \\
				\hline
			\end{tabular}
}
		\end{minipage}
		\begin{minipage}{.45\textwidth}\centering
		\includegraphics[width=\textwidth]{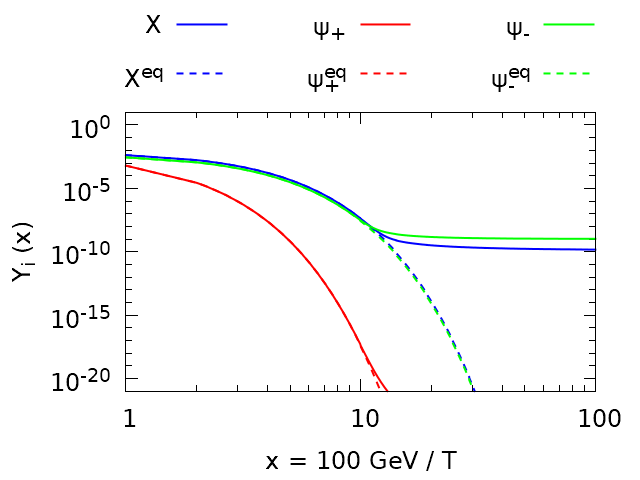}
\scalebox{.85}{
			\begin{tabular}{|c|c|c|c|}
				\hline
				process & $a_N$ & $a_{N+1}$ & $N$ \\
				\hline
				$XX \rightarrow \text{SM}$ & $1.3\cdot 10^{-2}$ & $-8\cdot 10^{-4}$ & $0$ \\
				$\psi_+\psi_+ \rightarrow \text{SM}$ & $6.2\cdot 10^{-4}$ & $-8.3\cdot 10^{-4}$ & $1$ \\
				$\psi_-\psi_- \rightarrow \text{SM}$ & $3.3\cdot 10^{-2}$ & $-3.6\cdot 10^{-2}$ & $1$ \\
				\hline
				$\Psi_+\Psi_+ \rightarrow XX$ & $9.3\cdot 10^{-4}$ & $1.9\cdot 10^{-3}$ & $0$ \\
				$XX \rightarrow \Psi_-\Psi_-$ & $9.4\cdot 10^{-2}$ & $-1.59\cdot 10^{2}$ & $2$ \\
				$\Psi_-\Psi_- \rightarrow XX$ & $2.1\cdot 10^{-1}$ & $-3.58\cdot 10^{2}$ & $2$ \\
				$\Psi_+,\Psi_+ \rightarrow \Psi_-\Psi_-$ & $1.1\cdot 10^{-2}$ & $9.7\cdot 10^{-4}$ & $0$ \\
				\hline
				$\Psi_+\Psi_- \rightarrow Xh_1$ & $1.6\cdot 10^{-5}$ & $-1.9\cdot 10^{-5}$ & $0$ \\
				$\Psi_+\Psi_- \rightarrow Xh_2$ & $1.7\cdot 10^{-3}$ & $-1.8\cdot 10^{-3}$ & $0$ \\
				$\Psi_+h_1 \rightarrow X\Psi_-$ & $1.1\cdot 10^{-4}$ & $1.8\cdot 10^{-3}$ & $0$ \\
				$\Psi_+h_2 \rightarrow X\Psi_-$ & $1.2\cdot 10^{-2}$ & $1.3\cdot 10^{-1}$ & $0$ \\
				$X\Psi_+ \rightarrow \Psi_-h_1$ & $3.2\cdot 10^{-4}$ & $-4.1\cdot 10^{-4}$ & $0$ \\
				$X\Psi_+ \rightarrow \Psi_-h_2$ & $3.2\cdot 10^{-2}$ & $-4.4\cdot 10^{-2}$ & $0$ \\
				\hline
				$\Psi_+ \rightarrow X\Psi_-$ & \multicolumn{3}{|c|}{$3.4\cdot 10^{-2}$} \\
				\hline
			\end{tabular}
}
		\end{minipage}\\~\\
		\begin{tabular}{|c|c|c|c|}\hline
		$Y(x=100)$&left&middle&right\\
		\hline
		$Y_X$&$1.8\e{-9}$&$4.3\e{-9}$&$1.5\e{-10}$\\
		$Y_+$&$3.3\e{-80}$&$1.6\e{-66}$&$9.1\e{-55}$\\
		$Y_-$&$1.8\e{-10}$&$4.8\e{-10}$&$1.0\e{-9}$\\
		\hline\end{tabular}
	\end{tabular}}
		\caption{The case of 2 stable components with similar masses. Parameters of the plots are: $m_+=400\gev$, $m_-=160\gev$, $m_2=150\gev$, $\sin\alpha=0.1$, $g_x=0.1$, $m_X$ is $100\gev$ (left), $130\gev$ (middle), $160\gev$ (right).}
		\label{fig:2dm1-3}
	\end{figure}

	\begin{figure}[H]
	\makebox[\linewidth][c]{\centering\begin{tabular}{c}
		\begin{minipage}{.45\textwidth}\centering
			\includegraphics[width=\textwidth]{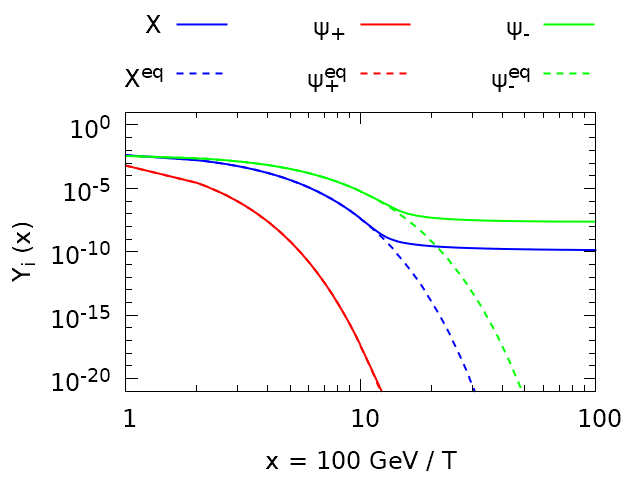}
\scalebox{.85}{
			\begin{tabular}{|c|c|c|c|}
				\hline
				process & $a_N$ & $a_{N+1}$ & $N$ \\
				\hline
				$XX \rightarrow \text{SM}$ & $1.3\cdot 10^{-2}$ & $-8\cdot 10^{-4}$ & $0$ \\
				$\psi_+\psi_+ \rightarrow \text{SM}$ & $1.7\cdot 10^{-3}$ & $-2.3\cdot 10^{-3}$ & $1$ \\
				$\psi_-\psi_- \rightarrow \text{SM}$ & $3.7\cdot 10^{-3}$ & $-5\cdot 10^{-2}$ & $1$ \\
				\hline
				$\Psi_+\Psi_+ \rightarrow XX$ & $1.1\cdot 10^{-3}$ & $3.8\cdot 10^{-3}$ & $0$ \\
				$XX \rightarrow \Psi_-\Psi_-$ & $1.9\cdot 10^{-3}$ & $-3.3\cdot 10^{-3}$ & $0$ \\
				$\Psi_+,\Psi_+ \rightarrow \Psi_-\Psi_-$ & $1.5\cdot 10^{-2}$ & $-2.1\cdot 10^{-3}$ & $0$ \\
				\hline
				$\Psi_+\Psi_- \rightarrow Xh_1$ & $2\cdot 10^{-4}$ & $-5.3\cdot 10^{-4}$ & $0$ \\
				$\Psi_+\Psi_- \rightarrow Xh_2$ & $2.5\cdot 10^{-2}$ & $-7.1\cdot 10^{-2}$ & $0$ \\
				$\Psi_+h_1 \rightarrow X\Psi_-$ & $2.7\cdot 10^{-4}$ & $1.8\cdot 10^{-3}$ & $0$ \\
				$\Psi_+h_2 \rightarrow X\Psi_-$ & $1.9\cdot 10^{-2}$ & $1.4\cdot 10^{-1}$ & $0$ \\
				$X\Psi_+ \rightarrow \Psi_-h_1$ & $4.3\cdot 10^{-4}$ & $-6.2\cdot 10^{-4}$ & $0$ \\
				$X\Psi_+ \rightarrow \Psi_-h_2$ & $4.4\cdot 10^{-2}$ & $-6.8\cdot 10^{-2}$ & $0$ \\
				\hline
				$\Psi_+ \rightarrow X\Psi_-$ & \multicolumn{3}{|c|}{$7.1\cdot 10^{-2}$} \\
				\hline
			\end{tabular}
}
		\end{minipage}
		\begin{minipage}{.45\textwidth}\centering
		\includegraphics[width=\textwidth]{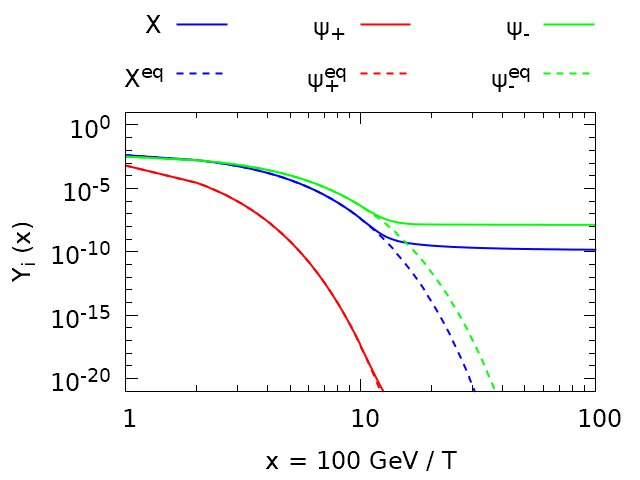}
\scalebox{.85}{
			\begin{tabular}{|c|c|c|c|}
				\hline
				process & $a_N$ & $a_{N+1}$ & $N$ \\
				\hline
				$XX \rightarrow \text{SM}$ & $1.3\cdot 10^{-2}$ & $-8\cdot 10^{-4}$ & $0$ \\
				$\psi_+\psi_+ \rightarrow \text{SM}$ & $1\cdot 10^{-3}$ & $-1.5\cdot 10^{-3}$ & $1$ \\
				$\psi_-\psi_- \rightarrow \text{SM}$ & $4\cdot 10^{-4}$ & $-2.3\cdot 10^{-3}$ & $1$ \\
				\hline
				$\Psi_+\Psi_+ \rightarrow XX$ & $1\cdot 10^{-3}$ & $2.7\cdot 10^{-3}$ & $0$ \\
				$XX \rightarrow \Psi_-\Psi_-$ & $5.9\cdot 10^{-4}$ & $7.8\cdot 10^{-5}$ & $0$ \\
				$\Psi_+,\Psi_+ \rightarrow \Psi_-\Psi_-$ & $1.3\cdot 10^{-2}$ & $-5.8\cdot 10^{-4}$ & $0$ \\
				\hline
				$\Psi_+\Psi_- \rightarrow Xh_1$ & $5.6\cdot 10^{-5}$ & $-1\cdot 10^{-4}$ & $0$ \\
				$\Psi_+\Psi_- \rightarrow Xh_2$ & $6.2\cdot 10^{-3}$ & $-1.1\cdot 10^{-2}$ & $0$ \\
				$\Psi_+h_1 \rightarrow X\Psi_-$ & $1.7\cdot 10^{-4}$ & $1.9\cdot 10^{-3}$ & $0$ \\
				$\Psi_+h_2 \rightarrow X\Psi_-$ & $1.4\cdot 10^{-2}$ & $1.4\cdot 10^{-1}$ & $0$ \\
				$X\Psi_+ \rightarrow \Psi_-h_1$ & $3.7\cdot 10^{-4}$ & $-5.3\cdot 10^{-4}$ & $0$ \\
				$X\Psi_+ \rightarrow \Psi_-h_2$ & $3.8\cdot 10^{-2}$ & $-5.6\cdot 10^{-2}$ & $0$ \\
				\hline
				$\Psi_+ \rightarrow X\Psi_-$ & \multicolumn{3}{|c|}{$5.2\cdot 10^{-2}$} \\
				\hline
			\end{tabular}
}
		\end{minipage}
		\begin{minipage}{.45\textwidth}\centering
		\includegraphics[width=\textwidth]{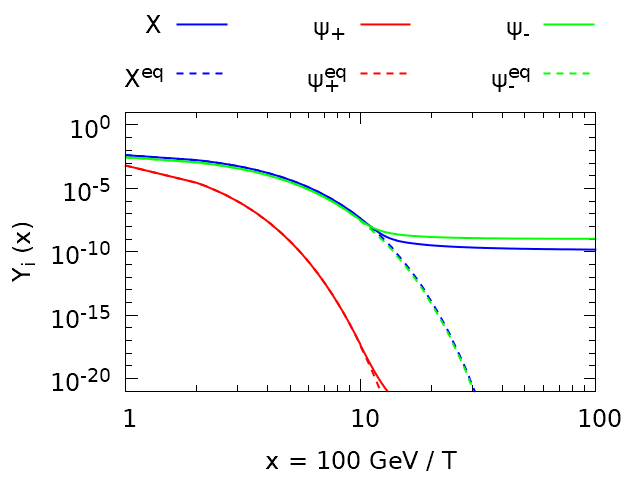}
\scalebox{.85}{
			\begin{tabular}{|c|c|c|c|}
				\hline
				process & $a_N$ & $a_{N+1}$ & $N$ \\
				\hline
				$XX \rightarrow \text{SM}$ & $1.3\cdot 10^{-2}$ & $-8\cdot 10^{-4}$ & $0$ \\
				$\psi_+\psi_+ \rightarrow \text{SM}$ & $6.2\cdot 10^{-4}$ & $-8.3\cdot 10^{-4}$ & $1$ \\
				$\psi_-\psi_- \rightarrow \text{SM}$ & $3.3\cdot 10^{-2}$ & $-3.6\cdot 10^{-2}$ & $1$ \\
				\hline
				$\Psi_+\Psi_+ \rightarrow XX$ & $9.3\cdot 10^{-4}$ & $1.9\cdot 10^{-3}$ & $0$ \\
				$XX \rightarrow \Psi_-\Psi_-$ & $9.4\cdot 10^{-2}$ & $-1.59\cdot 10^{2}$ & $2$ \\
				$\Psi_-\Psi_- \rightarrow XX$ & $2.1\cdot 10^{-1}$ & $-3.58\cdot 10^{2}$ & $2$ \\
				$\Psi_+,\Psi_+ \rightarrow \Psi_-\Psi_-$ & $1.1\cdot 10^{-2}$ & $9.7\cdot 10^{-4}$ & $0$ \\
				\hline
				$\Psi_+\Psi_- \rightarrow Xh_1$ & $1.6\cdot 10^{-5}$ & $-1.9\cdot 10^{-5}$ & $0$ \\
				$\Psi_+\Psi_- \rightarrow Xh_2$ & $1.7\cdot 10^{-3}$ & $-1.8\cdot 10^{-3}$ & $0$ \\
				$\Psi_+h_1 \rightarrow X\Psi_-$ & $1.1\cdot 10^{-4}$ & $1.8\cdot 10^{-3}$ & $0$ \\
				$\Psi_+h_2 \rightarrow X\Psi_-$ & $1.2\cdot 10^{-2}$ & $1.3\cdot 10^{-1}$ & $0$ \\
				$X\Psi_+ \rightarrow \Psi_-h_1$ & $3.2\cdot 10^{-4}$ & $-4.1\cdot 10^{-4}$ & $0$ \\
				$X\Psi_+ \rightarrow \Psi_-h_2$ & $3.2\cdot 10^{-2}$ & $-4.4\cdot 10^{-2}$ & $0$ \\
				\hline
				$\Psi_+ \rightarrow X\Psi_-$ & \multicolumn{3}{|c|}{$3.4\cdot 10^{-2}$} \\
				\hline
			\end{tabular}
}
		\end{minipage}\\~\\
		\begin{tabular}{|c|c|c|c|}\hline
		$Y(x=100)$&left&middle&right\\
		\hline
		$Y_X$&$1.4\e{-10}$&$1.4\e{-10}$&$1.5\e{-10}$\\
		$Y_+$&$3.3\e{-79}$&$1.5\e{-66}$&$9.1\e{-55}$\\
		$Y_-$&$2.3\e{-8}$&$1.3\e{-8}$&$1.0\e{-9}$\\
		\hline\end{tabular}
	\end{tabular}}
		\caption{The case of 2 stable components with similar masses. Parameters of the plots are: $m_X=160\gev$, $m_+=400\gev$, $m_2=150\gev$, $\sin\alpha=0.1$, $g_x=0.1$, $m_-$ is $100\gev$ (left), $130\gev$ (middle), $160\gev$ (right).}
		\label{fig:2dm4-6}
	\end{figure}

	\begin{figure}[H]
	\makebox[\linewidth][c]{\centering\begin{tabular}{c}
		\begin{minipage}{.45\textwidth}\centering
			\includegraphics[width=\textwidth]{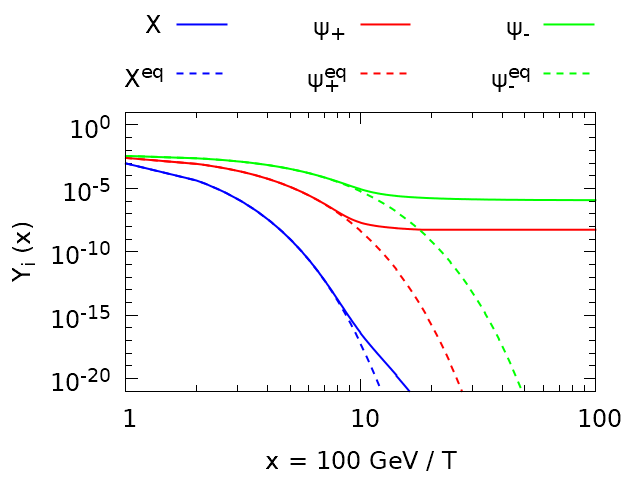}
\scalebox{.85}{
			\begin{tabular}{|c|c|c|c|}
				\hline
				process & $a_N$ & $a_{N+1}$ & $N$ \\
				\hline
				$XX \rightarrow \text{SM}$ & $5.3\cdot 10^{-3}$ & $-4\cdot 10^{-3}$ & $0$ \\
				$\psi_+\psi_+ \rightarrow \text{SM}$ & $1\cdot 10^{-6}$ & $-3.8\cdot 10^{-6}$ & $1$ \\
				$\psi_-\psi_- \rightarrow \text{SM}$ & $4.3\cdot 10^{-5}$ & $-5.7\cdot 10^{-4}$ & $1$ \\
				\hline
				$XX \rightarrow \Psi_+\Psi_+$ & $1.2\cdot 10^{-3}$ & $-1.7\cdot 10^{-3}$ & $0$ \\
				$XX \rightarrow \Psi_-\Psi_-$ & $8.2\cdot 10^{-4}$ & $-9.7\cdot 10^{-4}$ & $0$ \\
				$\Psi_+,\Psi_+ \rightarrow \Psi_-\Psi_-$ & $3.4\cdot 10^{-4}$ & $6.6\cdot 10^{-4}$ & $0$ \\
				\hline
				$Xh_1 \rightarrow \Psi_+\Psi_-$ & $3\cdot 10^{-4}$ & $-1.1\cdot 10^{-3}$ & $0$ \\
				$Xh_2 \rightarrow \Psi_+\Psi_-$ & $1.8\cdot 10^{-2}$ & $-5.4\cdot 10^{-2}$ & $0$ \\
				$X\Psi_- \rightarrow \Psi_+h_1$ & $9.6\cdot 10^{-6}$ & $3.9\cdot 10^{-6}$ & $0$ \\
				$X\Psi_- \rightarrow \Psi_+h_2$ & $9.2\cdot 10^{-4}$ & $1.4\cdot 10^{-4}$ & $0$ \\
				$X\Psi_+ \rightarrow \Psi_-h_1$ & $1.1\cdot 10^{-5}$ & $7.2\cdot 10^{-6}$ & $0$ \\
				$X\Psi_+ \rightarrow \Psi_-h_2$ & $1.1\cdot 10^{-3}$ & $7.8\cdot 10^{-4}$ & $0$ \\
				\hline
				$X \rightarrow \Psi_+\Psi_-$ & \multicolumn{3}{|c|}{$2.2\cdot 10^{-2}$} \\
				\hline
			\end{tabular}
}
		\end{minipage}
		\begin{minipage}{.45\textwidth}\centering
		\includegraphics[width=\textwidth]{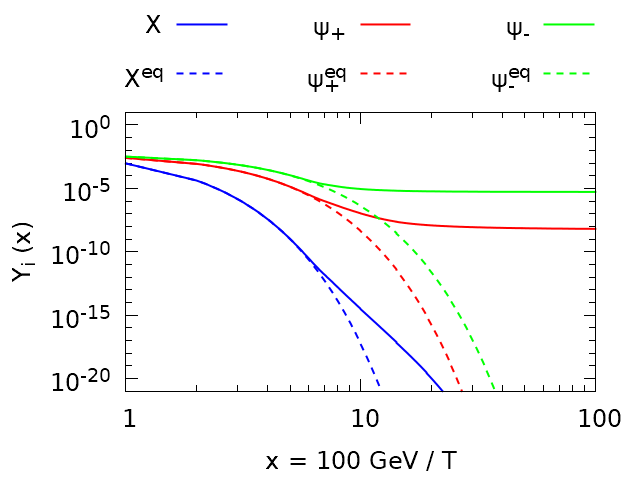}
\scalebox{.85}{
			\begin{tabular}{|c|c|c|c|}
				\hline
				process & $a_N$ & $a_{N+1}$ & $N$ \\
				\hline
				$XX \rightarrow \text{SM}$ & $5.3\cdot 10^{-3}$ & $-4\cdot 10^{-3}$ & $0$ \\
				$\psi_+\psi_+ \rightarrow \text{SM}$ & $6.2\cdot 10^{-7}$ & $-1.9\cdot 10^{-6}$ & $1$ \\
				$\psi_-\psi_- \rightarrow \text{SM}$ & $2.4\cdot 10^{-6}$ & $-1.2\cdot 10^{-5}$ & $1$ \\
				\hline
				$XX \rightarrow \Psi_+\Psi_+$ & $1.1\cdot 10^{-3}$ & $-1.5\cdot 10^{-3}$ & $0$ \\
				$XX \rightarrow \Psi_-\Psi_-$ & $8.5\cdot 10^{-4}$ & $-1.1\cdot 10^{-3}$ & $0$ \\
				$\Psi_+,\Psi_+ \rightarrow \Psi_-\Psi_-$ & $2.3\cdot 10^{-4}$ & $5.4\cdot 10^{-4}$ & $0$ \\
				\hline
				$Xh_1 \rightarrow \Psi_+\Psi_-$ & $3.1\cdot 10^{-4}$ & $-1.1\cdot 10^{-3}$ & $0$ \\
				$Xh_2 \rightarrow \Psi_+\Psi_-$ & $1.9\cdot 10^{-2}$ & $-5.6\cdot 10^{-2}$ & $0$ \\
				$X\Psi_- \rightarrow \Psi_+h_1$ & $8\cdot 10^{-6}$ & $1\cdot 10^{-5}$ & $0$ \\
				$X\Psi_- \rightarrow \Psi_+h_2$ & $7.4\cdot 10^{-4}$ & $1\cdot 10^{-3}$ & $0$ \\
				$X\Psi_+ \rightarrow \Psi_-h_1$ & $1\cdot 10^{-5}$ & $9\cdot 10^{-6}$ & $0$ \\
				$X\Psi_+ \rightarrow \Psi_-h_2$ & $9.4\cdot 10^{-4}$ & $9.3\cdot 10^{-4}$ & $0$ \\
				\hline
				$X \rightarrow \Psi_+\Psi_-$ & \multicolumn{3}{|c|}{$2.1\cdot 10^{-2}$} \\
				\hline
			\end{tabular}
}
		\end{minipage}
		\begin{minipage}{.45\textwidth}\centering
		\includegraphics[width=\textwidth]{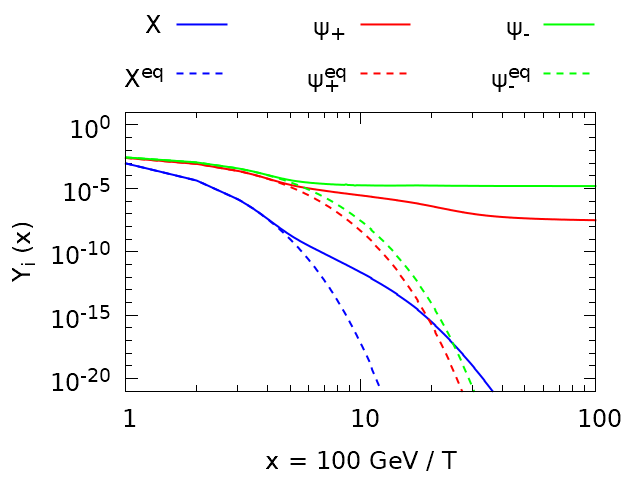}
\scalebox{.85}{
			\begin{tabular}{|c|c|c|c|}
				\hline
				process & $a_N$ & $a_{N+1}$ & $N$ \\
				\hline
				$XX \rightarrow \text{SM}$ & $5.3\cdot 10^{-3}$ & $-4\cdot 10^{-3}$ & $0$ \\
				$\psi_+\psi_+ \rightarrow \text{SM}$ & $2.1\cdot 10^{-7}$ & $-6.5\cdot 10^{-7}$ & $1$ \\
				$\psi_-\psi_- \rightarrow \text{SM}$ & $6.7\cdot 10^{-7}$ & $-6.3\cdot 10^{-7}$ & $1$ \\
				\hline
				$XX \rightarrow \Psi_+\Psi_+$ & $1\cdot 10^{-3}$ & $-1.3\cdot 10^{-3}$ & $0$ \\
				$XX \rightarrow \Psi_-\Psi_-$ & $8.9\cdot 10^{-4}$ & $-1.1\cdot 10^{-3}$ & $0$ \\
				$\Psi_+,\Psi_+ \rightarrow \Psi_-\Psi_-$ & $1.3\cdot 10^{-4}$ & $4\cdot 10^{-4}$ & $0$ \\
				\hline
				$Xh_1 \rightarrow \Psi_+\Psi_-$ & $3.1\cdot 10^{-4}$ & $-1.1\cdot 10^{-3}$ & $0$ \\
				$Xh_2 \rightarrow \Psi_+\Psi_-$ & $1.9\cdot 10^{-2}$ & $-5.6\cdot 10^{-2}$ & $0$ \\
				$X\Psi_- \rightarrow \Psi_+h_1$ & $7.7\cdot 10^{-6}$ & $1.1\cdot 10^{-5}$ & $0$ \\
				$X\Psi_- \rightarrow \Psi_+h_2$ & $7.1\cdot 10^{-4}$ & $1.1\cdot 10^{-3}$ & $0$ \\
				$X\Psi_+ \rightarrow \Psi_-h_1$ & $8.6\cdot 10^{-6}$ & $1\cdot 10^{-5}$ & $0$ \\
				$X\Psi_+ \rightarrow \Psi_-h_2$ & $8.1\cdot 10^{-4}$ & $1.1\cdot 10^{-3}$ & $0$ \\
				\hline
				$X \rightarrow \Psi_+\Psi_-$ & \multicolumn{3}{|c|}{$1.9\cdot 10^{-2}$} \\
				\hline
			\end{tabular}
}
		\end{minipage}\\~\\
		\begin{tabular}{|c|c|c|c|}\hline
		$Y(x=100)$&left&middle&right\\
		\hline
		$Y_X$&$6.1\e{-67}$&$2.3\e{-53}$&$2.5\e{-39}$\\
		$Y_+$&$5.4\e{-9}$&$6.3\e{-9}$&$3.1\e{-8}$\\
		$Y_-$&$1.2\e{-6}$&$5.2\e{-6}$&$1.5\e{-5}$\\
		\hline\end{tabular}
	\end{tabular}}
		\caption{The case of 2 stable components with similar masses. Parameters of the plots are: $m_X=400\gev$, $m_+=180\gev$, $m_2=150\gev$, $\sin\alpha=0.1$, $g_x=0.1$, $m_-$ is $100\gev$ (left), $130\gev$ (middle), $160\gev$ (right).}
		\label{fig:2dm7-9}
	\end{figure}
\end{subsubsection}
\begin{subsubsection}{Three stable components}
In this subsection we discuss the case with three stable components of dark matter.\\

Figs. \ref{fig:3-1dm1-3} and \ref{fig:3-1dm4-6} show the case of one component much lighter than two others. As previously, here we can see effects of exceeding the threshold of annihilation into $h_2$ when mass of a given particle becomes larger than $m_2$. Note the non trivial behaviour of $Y_+$ in fig. \ref{fig:3-1dm1-3} -- after decoupling from equilibrium with the Standard Model, $\psi_+$ particles are still coupled to $X$ through the conversion processes, therefore $Y_+$ stabilizes only after freeze-out of $X$.

	\begin{figure}[H]
	\makebox[\linewidth][c]{\centering\begin{tabular}{c}
		\begin{minipage}{.45\textwidth}\centering
			\includegraphics[width=\textwidth]{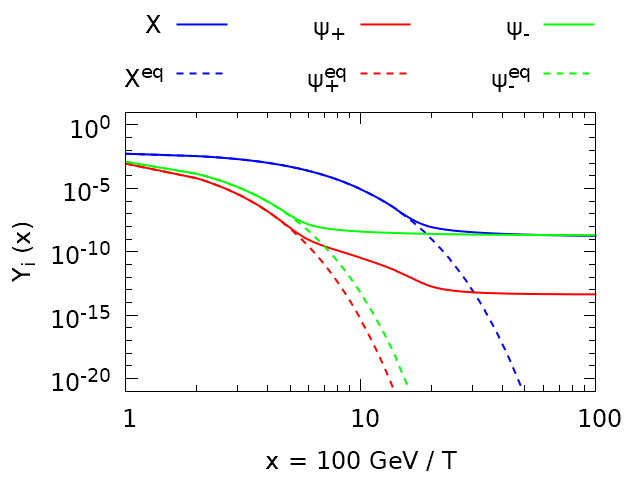}
\scalebox{.85}{
			\begin{tabular}{|c|c|c|c|}
				\hline
				process & $a_N$ & $a_{N+1}$ & $N$ \\
				\hline
				$XX \rightarrow \text{SM}$ & $1.9\cdot 10^{-3}$ & $-1.5\cdot 10^{-2}$ & $0$ \\
				$\psi_+\psi_+ \rightarrow \text{SM}$ & $1.3\cdot 10^{-6}$ & $-2.8\cdot 10^{-6}$ & $1$ \\
				$\psi_-\psi_- \rightarrow \text{SM}$ & $1.4\cdot 10^{-4}$ & $-3.1\cdot 10^{-4}$ & $1$ \\
				\hline
				$\Psi_+\Psi_+ \rightarrow XX$ & $5.1\cdot 10^{-4}$ & $9.7\cdot 10^{-5}$ & $0$ \\
				$\Psi_-\Psi_- \rightarrow XX$ & $3.6\cdot 10^{-4}$ & $1.4\cdot 10^{-4}$ & $0$ \\
				$\Psi_+,\Psi_+ \rightarrow \Psi_-\Psi_-$ & $1.3\cdot 10^{-2}$ & $3.4\cdot 10^{-2}$ & $0$ \\
				\hline
				$\Psi_+\Psi_- \rightarrow Xh_1$ & $3.1\cdot 10^{-6}$ & $-2.9\cdot 10^{-6}$ & $0$ \\
				$\Psi_+\Psi_- \rightarrow Xh_2$ & $2.9\cdot 10^{-4}$ & $-2.6\cdot 10^{-4}$ & $0$ \\
				$\Psi_+h_1 \rightarrow X\Psi_-$ & $8.4\cdot 10^{-4}$ & $-2.7\cdot 10^{-4}$ & $0$ \\
				$\Psi_+h_2 \rightarrow X\Psi_-$ & $8.3\cdot 10^{-2}$ & $-1.2\cdot 10^{-2}$ & $0$ \\
				$X\Psi_+ \rightarrow \Psi_-h_1$ & $7.2\cdot 10^{-5}$ & $7.3\cdot 10^{-4}$ & $0$ \\
				$\Psi_-h_2 \rightarrow X\Psi_+$ & $4.67\cdot 10^{1}$ & $-8\cdot 10^{4}$ & $2$ \\
				$X\Psi_+ \rightarrow \Psi_-h_2$ & $2.27\cdot 10^{1}$ & $-3.89\cdot 10^{4}$ & $2$ \\
				\hline
			\end{tabular}
}
		\end{minipage}
		\begin{minipage}{.45\textwidth}\centering
		\includegraphics[width=\textwidth]{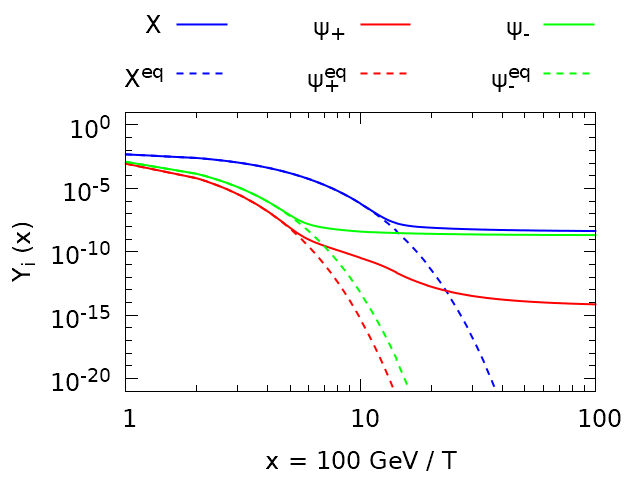}
\scalebox{.85}{
			\begin{tabular}{|c|c|c|c|}
				\hline
				process & $a_N$ & $a_{N+1}$ & $N$ \\
				\hline
				$XX \rightarrow \text{SM}$ & $3.9\cdot 10^{-4}$ & $-9.8\cdot 10^{-4}$ & $0$ \\
				$\psi_+\psi_+ \rightarrow \text{SM}$ & $5\cdot 10^{-7}$ & $-1.1\cdot 10^{-6}$ & $1$ \\
				$\psi_-\psi_- \rightarrow \text{SM}$ & $4.9\cdot 10^{-5}$ & $-1\cdot 10^{-4}$ & $1$ \\
				\hline
				$\Psi_+\Psi_+ \rightarrow XX$ & $5\cdot 10^{-4}$ & $9.7\cdot 10^{-5}$ & $0$ \\
				$\Psi_-\Psi_- \rightarrow XX$ & $3.3\cdot 10^{-4}$ & $1.8\cdot 10^{-4}$ & $0$ \\
				$\Psi_+,\Psi_+ \rightarrow \Psi_-\Psi_-$ & $8.2\cdot 10^{-3}$ & $1.7\cdot 10^{-2}$ & $0$ \\
				\hline
				$\Psi_+\Psi_- \rightarrow Xh_1$ & $4.7\cdot 10^{-6}$ & $-5.3\cdot 10^{-6}$ & $0$ \\
				$\Psi_+\Psi_- \rightarrow Xh_2$ & $4.4\cdot 10^{-4}$ & $-4.7\cdot 10^{-4}$ & $0$ \\
				$\Psi_+h_1 \rightarrow X\Psi_-$ & $3.6\cdot 10^{-4}$ & $1.8\cdot 10^{-4}$ & $0$ \\
				$\Psi_+h_2 \rightarrow X\Psi_-$ & $4.3\cdot 10^{-2}$ & $-3.9\cdot 10^{-3}$ & $0$ \\
				$X\Psi_+ \rightarrow \Psi_-h_1$ & $4\cdot 10^{-5}$ & $3.3\cdot 10^{-4}$ & $0$ \\
				$X\Psi_+ \rightarrow \Psi_-h_2$ & $2.6\cdot 10^{-3}$ & $3.6\cdot 10^{-2}$ & $0$ \\
				\hline
			\end{tabular}
}
		\end{minipage}
		\begin{minipage}{.45\textwidth}\centering
		\includegraphics[width=\textwidth]{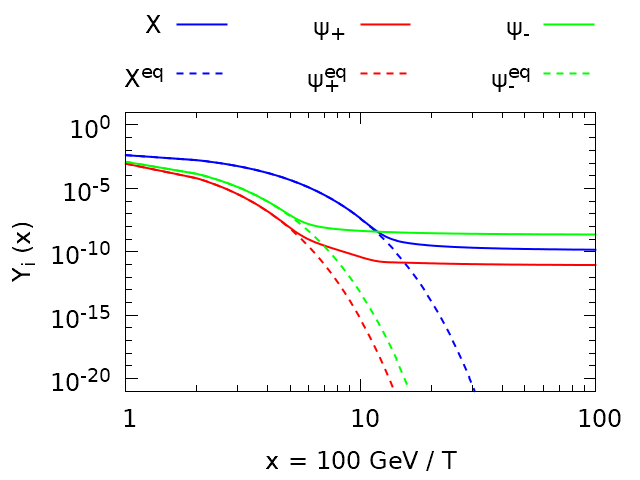}
\scalebox{.85}{
			\begin{tabular}{|c|c|c|c|}
				\hline
				process & $a_N$ & $a_{N+1}$ & $N$ \\
				\hline
				$XX \rightarrow \text{SM}$ & $1.3\cdot 10^{-2}$ & $-8\cdot 10^{-4}$ & $0$ \\
				$\psi_+\psi_+ \rightarrow \text{SM}$ & $2.4\cdot 10^{-7}$ & $-5.2\cdot 10^{-7}$ & $1$ \\
				$\psi_-\psi_- \rightarrow \text{SM}$ & $2.1\cdot 10^{-5}$ & $-4.7\cdot 10^{-5}$ & $1$ \\
				\hline
				$\Psi_+\Psi_+ \rightarrow XX$ & $4.8\cdot 10^{-4}$ & $1.6\cdot 10^{-4}$ & $0$ \\
				$\Psi_-\Psi_- \rightarrow XX$ & $3\cdot 10^{-4}$ & $2.6\cdot 10^{-4}$ & $0$ \\
				$\Psi_+,\Psi_+ \rightarrow \Psi_-\Psi_-$ & $5.5\cdot 10^{-3}$ & $8.4\cdot 10^{-3}$ & $0$ \\
				\hline
				$\Psi_+\Psi_- \rightarrow Xh_1$ & $6.1\cdot 10^{-6}$ & $-7.4\cdot 10^{-6}$ & $0$ \\
				$\Psi_+\Psi_- \rightarrow Xh_2$ & $5.8\cdot 10^{-4}$ & $-6.8\cdot 10^{-4}$ & $0$ \\
				$\Psi_+h_1 \rightarrow X\Psi_-$ & $6.5\cdot 10^{-5}$ & $6.4\cdot 10^{-4}$ & $0$ \\
				$\Psi_+h_2 \rightarrow X\Psi_-$ & $1.7\cdot 10^{-2}$ & $2.7\cdot 10^{-2}$ & $0$ \\
				$X\Psi_+ \rightarrow \Psi_-h_1$ & $2.5\cdot 10^{-5}$ & $1.8\cdot 10^{-4}$ & $0$ \\
				$X\Psi_+ \rightarrow \Psi_-h_2$ & $2\cdot 10^{-3}$ & $1.9\cdot 10^{-2}$ & $0$ \\
				\hline
			\end{tabular}
}
		\end{minipage}\\~\\
		\begin{tabular}{|c|c|c|c|}\hline
		$Y(x=100)$&left&middle&right\\
		\hline
		$Y_X$&$1.8\e{-9}$&$4.3\e{-9}$&$1.4\e{-10}$\\
		$Y_+$&$4.3\e{-14}$&$7.0\e{-15}$&$8.9\e{-12}$\\
		$Y_-$&$2.0\e{-9}$&$2.0\e{-9}$&$2.3\e{-9}$\\
		\hline\end{tabular}
	\end{tabular}}
		\caption{The case of 3 stable components with one of them much lighter. Parameters of the plots are: $m_+=350\gev$, $m_-=300\gev$, $m_2=150\gev$, $\sin\alpha=0.1$, $g_x=0.1$, $m_X$ is $100\gev$ (left), $130\gev$ (middle), $160\gev$ (right).}
		\label{fig:3-1dm1-3}
	\end{figure}

	\begin{figure}[H]
	\makebox[\linewidth][c]{\centering\begin{tabular}{c}
		\begin{minipage}{.45\textwidth}\centering
			\includegraphics[width=\textwidth]{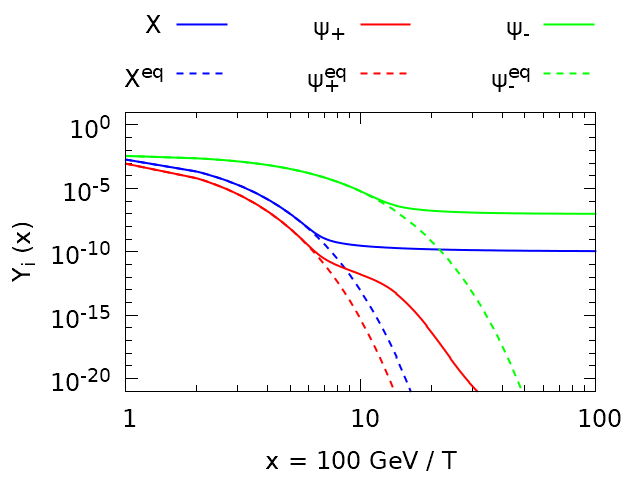}
\scalebox{.85}{
			\begin{tabular}{|c|c|c|c|}
				\hline
				process & $a_N$ & $a_{N+1}$ & $N$ \\
				\hline
				$XX \rightarrow \text{SM}$ & $8.5\cdot 10^{-3}$ & $-8.2\cdot 10^{-3}$ & $0$ \\
				$\psi_+\psi_+ \rightarrow \text{SM}$ & $8.8\cdot 10^{-5}$ & $-1.4\cdot 10^{-4}$ & $1$ \\
				$\psi_-\psi_- \rightarrow \text{SM}$ & $7.4\cdot 10^{-4}$ & $-9.9\cdot 10^{-3}$ & $1$ \\
				\hline
				$\Psi_+\Psi_+ \rightarrow XX$ & $1.8\cdot 10^{-3}$ & $5.2\cdot 10^{-3}$ & $0$ \\
				$XX \rightarrow \Psi_-\Psi_-$ & $3.4\cdot 10^{-4}$ & $-3\cdot 10^{-4}$ & $0$ \\
				$\Psi_+,\Psi_+ \rightarrow \Psi_-\Psi_-$ & $2.6\cdot 10^{-3}$ & $4.7\cdot 10^{-4}$ & $0$ \\
				\hline
				$\Psi_+\Psi_- \rightarrow Xh_1$ & $3.3\cdot 10^{-5}$ & $-9.8\cdot 10^{-5}$ & $0$ \\
				$Xh_2 \rightarrow \Psi_+\Psi_-$ & $5.3\cdot 10^{-1}$ & $-1.34\cdot 10^{3}$ & $1$ \\
				$\Psi_+\Psi_- \rightarrow Xh_2$ & $5.8\cdot 10^{-1}$ & $-1.47\cdot 10^{3}$ & $1$ \\
				$\Psi_+h_1 \rightarrow X\Psi_-$ & $1.7\cdot 10^{-5}$ & $1.8\cdot 10^{-4}$ & $0$ \\
				$\Psi_+h_2 \rightarrow X\Psi_-$ & $2.3\cdot 10^{-3}$ & $1.6\cdot 10^{-2}$ & $0$ \\
				$X\Psi_+ \rightarrow \Psi_-h_1$ & $3.6\cdot 10^{-5}$ & $-7.4\cdot 10^{-6}$ & $0$ \\
				$X\Psi_+ \rightarrow \Psi_-h_2$ & $3.5\cdot 10^{-3}$ & $-7\cdot 10^{-4}$ & $0$ \\
				\hline
			\end{tabular}
}
		\end{minipage}
		\begin{minipage}{.45\textwidth}\centering
		\includegraphics[width=\textwidth]{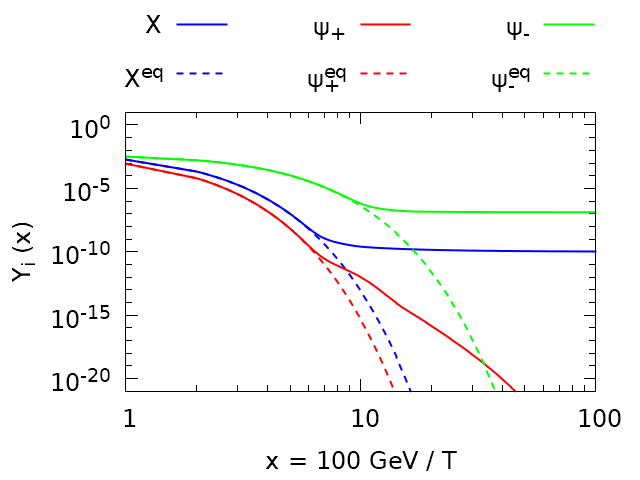}
\scalebox{.85}{
			\begin{tabular}{|c|c|c|c|}
				\hline
				process & $a_N$ & $a_{N+1}$ & $N$ \\
				\hline
				$XX \rightarrow \text{SM}$ & $8.5\cdot 10^{-3}$ & $-8.2\cdot 10^{-3}$ & $0$ \\
				$\psi_+\psi_+ \rightarrow \text{SM}$ & $4.9\cdot 10^{-5}$ & $-7.9\cdot 10^{-5}$ & $1$ \\
				$\psi_-\psi_- \rightarrow \text{SM}$ & $8\cdot 10^{-5}$ & $-4.3\cdot 10^{-4}$ & $1$ \\
				\hline
				$\Psi_+\Psi_+ \rightarrow XX$ & $1.3\cdot 10^{-3}$ & $3.5\cdot 10^{-3}$ & $0$ \\
				$XX \rightarrow \Psi_-\Psi_-$ & $3.4\cdot 10^{-4}$ & $-2.9\cdot 10^{-4}$ & $0$ \\
				$\Psi_+,\Psi_+ \rightarrow \Psi_-\Psi_-$ & $2.3\cdot 10^{-3}$ & $7.2\cdot 10^{-4}$ & $0$ \\
				\hline
				$\Psi_+\Psi_- \rightarrow Xh_1$ & $3.2\cdot 10^{-5}$ & $-8\cdot 10^{-5}$ & $0$ \\
				$\Psi_+\Psi_- \rightarrow Xh_2$ & $2.7\cdot 10^{-3}$ & $-5.6\cdot 10^{-3}$ & $0$ \\
				$\Psi_+h_1 \rightarrow X\Psi_-$ & $1.3\cdot 10^{-5}$ & $1.3\cdot 10^{-4}$ & $0$ \\
				$\Psi_+h_2 \rightarrow X\Psi_-$ & $2\cdot 10^{-3}$ & $1.2\cdot 10^{-2}$ & $0$ \\
				$X\Psi_+ \rightarrow \Psi_-h_1$ & $3.1\cdot 10^{-5}$ & $-1.4\cdot 10^{-6}$ & $0$ \\
				$X\Psi_+ \rightarrow \Psi_-h_2$ & $3\cdot 10^{-3}$ & $-9.7\cdot 10^{-5}$ & $0$ \\
				\hline
			\end{tabular}
}
		\end{minipage}
		\begin{minipage}{.45\textwidth}\centering
		\includegraphics[width=\textwidth]{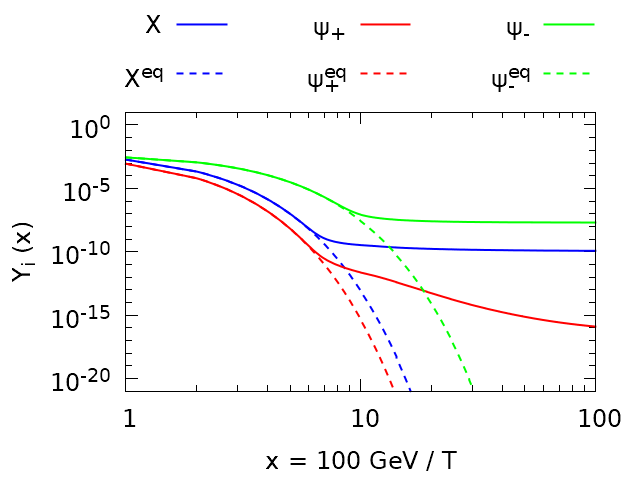}
\scalebox{.85}{
			\begin{tabular}{|c|c|c|c|}
				\hline
				process & $a_N$ & $a_{N+1}$ & $N$ \\
				\hline
				$XX \rightarrow \text{SM}$ & $8.5\cdot 10^{-3}$ & $-8.2\cdot 10^{-3}$ & $0$ \\
				$\psi_+\psi_+ \rightarrow \text{SM}$ & $2.5\cdot 10^{-5}$ & $-4\cdot 10^{-5}$ & $1$ \\
				$\psi_-\psi_- \rightarrow \text{SM}$ & $1.2\cdot 10^{-3}$ & $-1.3\cdot 10^{-3}$ & $1$ \\
				\hline
				$\Psi_+\Psi_+ \rightarrow XX$ & $9.6\cdot 10^{-4}$ & $2.5\cdot 10^{-3}$ & $0$ \\
				$XX \rightarrow \Psi_-\Psi_-$ & $3.4\cdot 10^{-4}$ & $-2.8\cdot 10^{-4}$ & $0$ \\
				$\Psi_+,\Psi_+ \rightarrow \Psi_-\Psi_-$ & $2\cdot 10^{-3}$ & $9.5\cdot 10^{-4}$ & $0$ \\
				\hline
				$\Psi_+\Psi_- \rightarrow Xh_1$ & $2.9\cdot 10^{-5}$ & $-6.3\cdot 10^{-5}$ & $0$ \\
				$\Psi_+\Psi_- \rightarrow Xh_2$ & $2.6\cdot 10^{-3}$ & $-5.2\cdot 10^{-3}$ & $0$ \\
				$\Psi_+h_1 \rightarrow X\Psi_-$ & $4.3\cdot 10^{-6}$ & $9.9\cdot 10^{-5}$ & $0$ \\
				$\Psi_+h_2 \rightarrow X\Psi_-$ & $1.4\cdot 10^{-3}$ & $9.5\cdot 10^{-3}$ & $0$ \\
				$X\Psi_+ \rightarrow \Psi_-h_1$ & $2.7\cdot 10^{-5}$ & $4.5\cdot 10^{-6}$ & $0$ \\
				$X\Psi_+ \rightarrow \Psi_-h_2$ & $2.6\cdot 10^{-3}$ & $4.9\cdot 10^{-4}$ & $0$ \\
				\hline
			\end{tabular}
}
		\end{minipage}\\~\\
		\begin{tabular}{|c|c|c|c|}\hline
		$Y(x=100)$&left&middle&right\\
		\hline
		$Y_X$&$1.1\e{-10}$&$1.0\e{-10}$&$1.2\e{-10}$\\
		$Y_+$&$1.6\e{-26}$&$2.1\e{-27}$&$1.2\e{-16}$\\
		$Y_-$&$9.8\e{-8}$&$1.3\e{-7}$&$2.0\e{-8}$\\
		\hline\end{tabular}
	\end{tabular}}
		\caption{The case of 3 stable components with one of them much lighter. Parameters of the plots are: $m_X=300\gev$, $m_+=350\gev$, $m_2=150\gev$, $\sin\alpha=0.1$, $g_x=0.1$, $m_-$ is $100\gev$ (left), $130\gev$ (middle), $160\gev$ (right).}
		\label{fig:3-1dm4-6}
	\end{figure}

The next three figures, \ref{fig:3dm1-3}, \ref{fig:3dm4-6} and \ref{fig:3dm7-9}, illustrate the case of three stable components with similar masses. The channel of annihilation into $h_2$, which becomes relevant when the mass of the given particle exceeds $150\gev$, leads to steeply decreasing of the corresponding yield.

	\begin{figure}[H]
	\makebox[\linewidth][c]{\centering\begin{tabular}{c}
		\begin{minipage}{.45\textwidth}\centering
			\includegraphics[width=\textwidth]{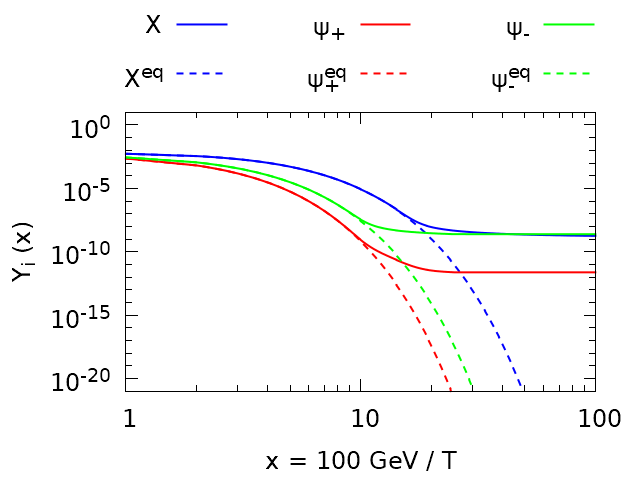}
\scalebox{.85}{
			\begin{tabular}{|c|c|c|c|}
				\hline
				process & $a_N$ & $a_{N+1}$ & $N$ \\
				\hline
				$XX \rightarrow \text{SM}$ & $1.9\cdot 10^{-3}$ & $-1.5\cdot 10^{-2}$ & $0$ \\
				$\psi_+\psi_+ \rightarrow \text{SM}$ & $4.7\cdot 10^{-5}$ & $-1.4\cdot 10^{-4}$ & $1$ \\
				$\psi_-\psi_- \rightarrow \text{SM}$ & $8.3\cdot 10^{-4}$ & $-3.2\cdot 10^{-4}$ & $1$ \\
				\hline
				$\Psi_+\Psi_+ \rightarrow XX$ & $1.6\cdot 10^{-3}$ & $1.1\cdot 10^{-3}$ & $0$ \\
				$\Psi_-\Psi_- \rightarrow XX$ & $7.6\cdot 10^{-4}$ & $1.9\cdot 10^{-3}$ & $0$ \\
				$\Psi_+,\Psi_+ \rightarrow \Psi_-\Psi_-$ & $1.4\cdot 10^{-2}$ & $3.3\cdot 10^{-2}$ & $0$ \\
				\hline
				$\Psi_+\Psi_- \rightarrow Xh_1$ & $1.6\cdot 10^{-5}$ & $-3.3\cdot 10^{-5}$ & $0$ \\
				$\Psi_+\Psi_- \rightarrow Xh_2$ & $1.4\cdot 10^{-3}$ & $-2.6\cdot 10^{-3}$ & $0$ \\
				$\Psi_+h_1 \rightarrow X\Psi_-$ & $7\cdot 10^{-4}$ & $-1.9\cdot 10^{-5}$ & $0$ \\
				$\Psi_+h_2 \rightarrow X\Psi_-$ & $7.4\cdot 10^{-2}$ & $-3.9\cdot 10^{-3}$ & $0$ \\
				$X\Psi_+ \rightarrow \Psi_-h_1$ & $3.4\cdot 10^{-5}$ & $7.9\cdot 10^{-4}$ & $0$ \\
				$\Psi_-h_2 \rightarrow X\Psi_+$ & $6.6\cdot 10^{-3}$ & $2.1\cdot 10^{-1}$ & $0$ \\
				\hline
			\end{tabular}
}
		\end{minipage}
		\begin{minipage}{.45\textwidth}\centering
		\includegraphics[width=\textwidth]{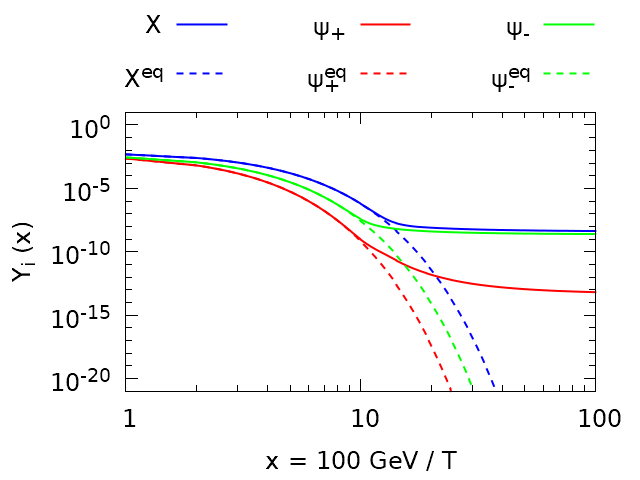}
\scalebox{.85}{
			\begin{tabular}{|c|c|c|c|}
				\hline
				process & $a_N$ & $a_{N+1}$ & $N$ \\
				\hline
				$XX \rightarrow \text{SM}$ & $3.9\cdot 10^{-4}$ & $-9.8\cdot 10^{-4}$ & $0$ \\
				$\psi_+\psi_+ \rightarrow \text{SM}$ & $1.7\cdot 10^{-5}$ & $-5\cdot 10^{-5}$ & $1$ \\
				$\psi_-\psi_- \rightarrow \text{SM}$ & $2.9\cdot 10^{-4}$ & $-1.2\cdot 10^{-4}$ & $1$ \\
				\hline
				$\Psi_+\Psi_+ \rightarrow XX$ & $1.4\cdot 10^{-3}$ & $2.1\cdot 10^{-3}$ & $0$ \\
				$\Psi_-\Psi_- \rightarrow XX$ & $4.1\cdot 10^{-4}$ & $3.4\cdot 10^{-3}$ & $0$ \\
				$\Psi_+,\Psi_+ \rightarrow \Psi_-\Psi_-$ & $7.8\cdot 10^{-3}$ & $1.5\cdot 10^{-2}$ & $0$ \\
				\hline
				$\Psi_+\Psi_- \rightarrow Xh_1$ & $2.6\cdot 10^{-5}$ & $-5.7\cdot 10^{-5}$ & $0$ \\
				$\Psi_+\Psi_- \rightarrow Xh_2$ & $2.2\cdot 10^{-3}$ & $-4.4\cdot 10^{-3}$ & $0$ \\
				$\Psi_+h_1 \rightarrow X\Psi_-$ & $2.3\cdot 10^{-4}$ & $5\cdot 10^{-4}$ & $0$ \\
				$\Psi_+h_2 \rightarrow X\Psi_-$ & $3.2\cdot 10^{-2}$ & $1.8\cdot 10^{-2}$ & $0$ \\
				$X\Psi_+ \rightarrow \Psi_-h_1$ & $3.5\cdot 10^{-5}$ & $3.4\cdot 10^{-4}$ & $0$ \\
				$X\Psi_+ \rightarrow \Psi_-h_2$ & $1.6\cdot 10^{-3}$ & $3.5\cdot 10^{-2}$ & $0$ \\
				\hline
			\end{tabular}
}
		\end{minipage}
		\begin{minipage}{.45\textwidth}\centering
		\includegraphics[width=\textwidth]{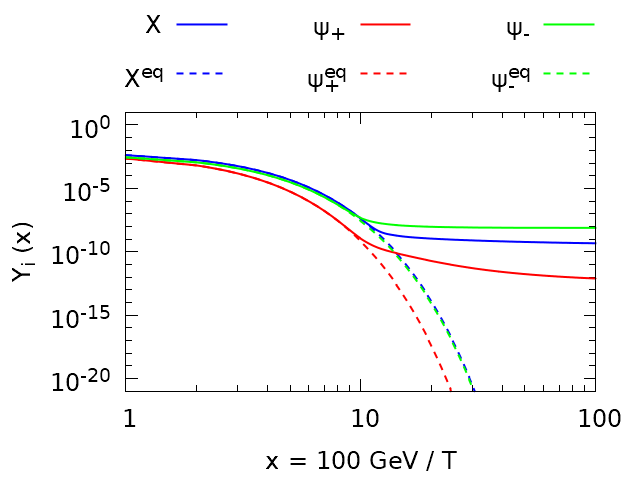}
\scalebox{.85}{
			\begin{tabular}{|c|c|c|c|}
				\hline
				process & $a_N$ & $a_{N+1}$ & $N$ \\
				\hline
				$XX \rightarrow \text{SM}$ & $1.3\cdot 10^{-2}$ & $-8\cdot 10^{-4}$ & $0$ \\
				$\psi_+\psi_+ \rightarrow \text{SM}$ & $7.7\cdot 10^{-6}$ & $-2.3\cdot 10^{-5}$ & $1$ \\
				$\psi_-\psi_- \rightarrow \text{SM}$ & $1.3\cdot 10^{-4}$ & $-6.1\cdot 10^{-5}$ & $1$ \\
				\hline
				$\Psi_+\Psi_+ \rightarrow XX$ & $1\cdot 10^{-3}$ & $4.1\cdot 10^{-3}$ & $0$ \\
				$XX \rightarrow \Psi_-\Psi_-$ & $2.9\cdot 10^{-1}$ & $-4.82\cdot 10^{2}$ & $2$ \\
				$\Psi_-\Psi_- \rightarrow XX$ & $6.4\cdot 10^{-1}$ & $-1.08\cdot 10^{3}$ & $2$ \\
				$\Psi_+,\Psi_+ \rightarrow \Psi_-\Psi_-$ & $4.6\cdot 10^{-3}$ & $7.8\cdot 10^{-3}$ & $0$ \\
				\hline
				$\Psi_+\Psi_- \rightarrow Xh_1$ & $3.7\cdot 10^{-5}$ & $-8.2\cdot 10^{-5}$ & $0$ \\
				$\Psi_+\Psi_- \rightarrow Xh_2$ & $2.9\cdot 10^{-3}$ & $-5.3\cdot 10^{-3}$ & $0$ \\
				$\Psi_+h_1 \rightarrow X\Psi_-$ & $1\cdot 10^{-5}$ & $4.6\cdot 10^{-4}$ & $0$ \\
				$\Psi_+h_2 \rightarrow X\Psi_-$ & $9.4\cdot 10^{-3}$ & $4\cdot 10^{-2}$ & $0$ \\
				$X\Psi_+ \rightarrow \Psi_-h_1$ & $2.8\cdot 10^{-5}$ & $1.8\cdot 10^{-4}$ & $0$ \\
				$X\Psi_+ \rightarrow \Psi_-h_2$ & $2\cdot 10^{-3}$ & $1.9\cdot 10^{-2}$ & $0$ \\
				\hline
			\end{tabular}
}
		\end{minipage}\\~\\
		\begin{tabular}{|c|c|c|c|}\hline
		$Y(x=100)$&left&middle&right\\
		\hline
		$Y_X$&$1.8\e{-9}$&$4.3\e{-9}$&$4.6\e{-10}$\\
		$Y_+$&$2.4\e{-12}$&$6.4\e{-14}$&$7.7\e{-13}$\\
		$Y_-$&$2.4\e{-9}$&$2.5\e{-9}$&$7.6\e{-9}$\\
		\hline\end{tabular}
	\end{tabular}}
		\caption{The case of 3 stable components with similar masses. Parameters of the plots are: $m_+=200\gev$, $m_-=160\gev$, $m_2=150\gev$, $\sin\alpha=0.1$, $g_x=0.1$, $m_X$ is $100\gev$ (left), $130\gev$ (middle), $160\gev$ (right).}
		\label{fig:3dm1-3}
	\end{figure}

	\begin{figure}[H]
	\makebox[\linewidth][c]{\centering\begin{tabular}{c}
		\begin{minipage}{.45\textwidth}\centering
			\includegraphics[width=\textwidth]{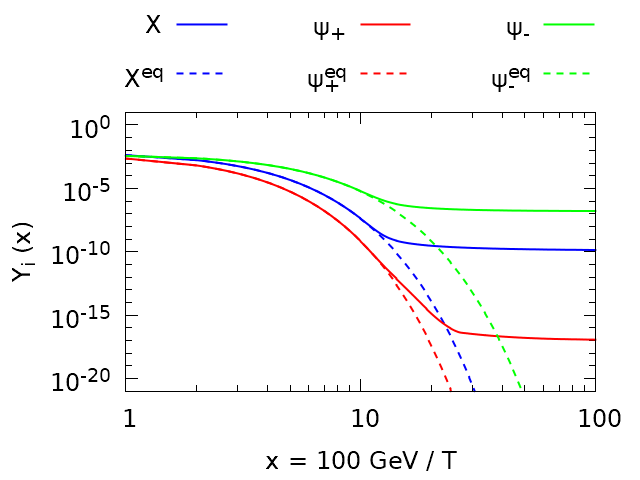}
\scalebox{.85}{
			\begin{tabular}{|c|c|c|c|}
				\hline
				process & $a_N$ & $a_{N+1}$ & $N$ \\
				\hline
				$XX \rightarrow \text{SM}$ & $1.3\cdot 10^{-2}$ & $-8\cdot 10^{-4}$ & $0$ \\
				$\psi_+\psi_+ \rightarrow \text{SM}$ & $2.6\cdot 10^{-5}$ & $-8.7\cdot 10^{-5}$ & $1$ \\
				$\psi_-\psi_- \rightarrow \text{SM}$ & $4.2\cdot 10^{-4}$ & $-5.6\cdot 10^{-3}$ & $1$ \\
				\hline
				$\Psi_+\Psi_+ \rightarrow XX$ & $2.8\cdot 10^{-3}$ & $8.6\cdot 10^{-3}$ & $0$ \\
				$XX \rightarrow \Psi_-\Psi_-$ & $8.7\cdot 10^{-4}$ & $-9.7\cdot 10^{-4}$ & $0$ \\
				$\Psi_+,\Psi_+ \rightarrow \Psi_-\Psi_-$ & $7\cdot 10^{-3}$ & $6.4\cdot 10^{-3}$ & $0$ \\
				\hline
				$\Psi_+\Psi_- \rightarrow Xh_1$ & $3.8\cdot 10^{-5}$ & $-5.8\cdot 10^{-5}$ & $0$ \\
				$Xh_2 \rightarrow \Psi_+\Psi_-$ & $2.8\cdot 10^{-3}$ & $1.5\cdot 10^{-3}$ & $0$ \\
				$\Psi_+h_1 \rightarrow X\Psi_-$ & $1.5\cdot 10^{-4}$ & $3.4\cdot 10^{-4}$ & $0$ \\
				$\Psi_+h_2 \rightarrow X\Psi_-$ & $1.8\cdot 10^{-2}$ & $2.6\cdot 10^{-2}$ & $0$ \\
				$X\Psi_+ \rightarrow \Psi_-h_1$ & $8.1\cdot 10^{-5}$ & $5.3\cdot 10^{-5}$ & $0$ \\
				$X\Psi_+ \rightarrow \Psi_-h_2$ & $7.4\cdot 10^{-3}$ & $6.4\cdot 10^{-3}$ & $0$ \\
				\hline
			\end{tabular}
}
		\end{minipage}
		\begin{minipage}{.45\textwidth}\centering
		\includegraphics[width=\textwidth]{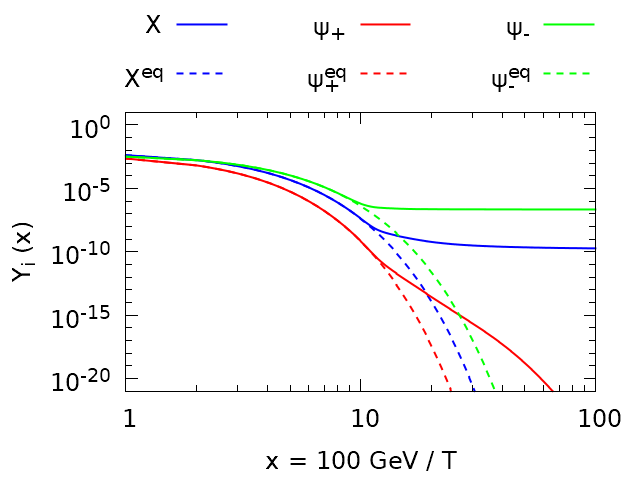}
\scalebox{.85}{
			\begin{tabular}{|c|c|c|c|}
				\hline
				process & $a_N$ & $a_{N+1}$ & $N$ \\
				\hline
				$XX \rightarrow \text{SM}$ & $1.3\cdot 10^{-2}$ & $-8\cdot 10^{-4}$ & $0$ \\
				$\psi_+\psi_+ \rightarrow \text{SM}$ & $5.3\cdot 10^{-6}$ & $-1.5\cdot 10^{-5}$ & $1$ \\
				$\psi_-\psi_- \rightarrow \text{SM}$ & $2.9\cdot 10^{-5}$ & $-1.5\cdot 10^{-4}$ & $1$ \\
				\hline
				$\Psi_+\Psi_+ \rightarrow XX$ & $1.7\cdot 10^{-3}$ & $5.5\cdot 10^{-3}$ & $0$ \\
				$XX \rightarrow \Psi_-\Psi_-$ & $5.9\cdot 10^{-4}$ & $2.7\cdot 10^{-4}$ & $0$ \\
				$\Psi_+,\Psi_+ \rightarrow \Psi_-\Psi_-$ & $5.6\cdot 10^{-3}$ & $8\cdot 10^{-3}$ & $0$ \\
				\hline
				$\Psi_+\Psi_- \rightarrow Xh_1$ & $4.1\cdot 10^{-5}$ & $-9.5\cdot 10^{-5}$ & $0$ \\
				$\Psi_+\Psi_- \rightarrow Xh_2$ & $2.8\cdot 10^{-3}$ & $-2.1\cdot 10^{-3}$ & $0$ \\
				$\Psi_+h_1 \rightarrow X\Psi_-$ & $1.1\cdot 10^{-4}$ & $3.8\cdot 10^{-4}$ & $0$ \\
				$\Psi_+h_2 \rightarrow X\Psi_-$ & $1.6\cdot 10^{-2}$ & $2.2\cdot 10^{-2}$ & $0$ \\
				$X\Psi_+ \rightarrow \Psi_-h_1$ & $5.2\cdot 10^{-5}$ & $1.2\cdot 10^{-4}$ & $0$ \\
				$X\Psi_+ \rightarrow \Psi_-h_2$ & $4.4\cdot 10^{-3}$ & $1.4\cdot 10^{-2}$ & $0$ \\
				\hline
			\end{tabular}
}
		\end{minipage}
		\begin{minipage}{.45\textwidth}\centering
		\includegraphics[width=\textwidth]{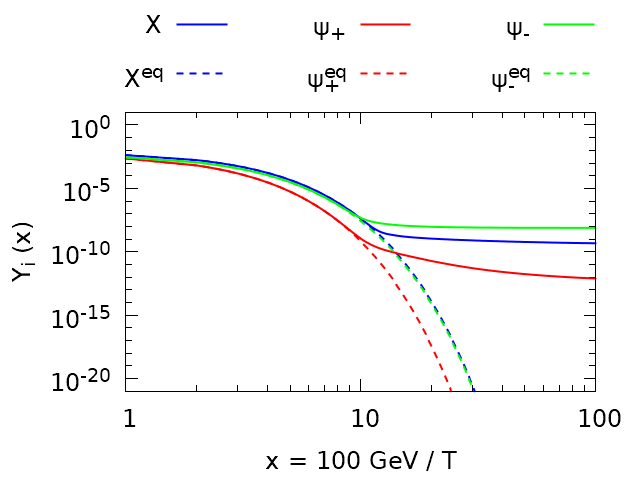}
\scalebox{.85}{
			\begin{tabular}{|c|c|c|c|}
				\hline
				process & $a_N$ & $a_{N+1}$ & $N$ \\
				\hline
				$XX \rightarrow \text{SM}$ & $1.3\cdot 10^{-2}$ & $-8\cdot 10^{-4}$ & $0$ \\
				$\psi_+\psi_+ \rightarrow \text{SM}$ & $7.7\cdot 10^{-6}$ & $-2.3\cdot 10^{-5}$ & $1$ \\
				$\psi_-\psi_- \rightarrow \text{SM}$ & $1.3\cdot 10^{-4}$ & $-6.1\cdot 10^{-5}$ & $1$ \\
				\hline
				$\Psi_+\Psi_+ \rightarrow XX$ & $1\cdot 10^{-3}$ & $4.1\cdot 10^{-3}$ & $0$ \\
				$XX \rightarrow \Psi_-\Psi_-$ & $2.9\cdot 10^{-1}$ & $-4.82\cdot 10^{2}$ & $2$ \\
				$\Psi_-\Psi_- \rightarrow XX$ & $6.4\cdot 10^{-1}$ & $-1.08\cdot 10^{3}$ & $2$ \\
				$\Psi_+,\Psi_+ \rightarrow \Psi_-\Psi_-$ & $4.6\cdot 10^{-3}$ & $7.8\cdot 10^{-3}$ & $0$ \\
				\hline
				$\Psi_+\Psi_- \rightarrow Xh_1$ & $3.7\cdot 10^{-5}$ & $-8.2\cdot 10^{-5}$ & $0$ \\
				$\Psi_+\Psi_- \rightarrow Xh_2$ & $2.9\cdot 10^{-3}$ & $-5.3\cdot 10^{-3}$ & $0$ \\
				$\Psi_+h_1 \rightarrow X\Psi_-$ & $1\cdot 10^{-5}$ & $4.6\cdot 10^{-4}$ & $0$ \\
				$\Psi_+h_2 \rightarrow X\Psi_-$ & $9.4\cdot 10^{-3}$ & $4\cdot 10^{-2}$ & $0$ \\
				$X\Psi_+ \rightarrow \Psi_-h_1$ & $2.8\cdot 10^{-5}$ & $1.8\cdot 10^{-4}$ & $0$ \\
				$X\Psi_+ \rightarrow \Psi_-h_2$ & $2\cdot 10^{-3}$ & $1.9\cdot 10^{-2}$ & $0$ \\
				\hline
			\end{tabular}
}
		\end{minipage}\\~\\
		\begin{tabular}{|c|c|c|c|}\hline
		$Y(x=100)$&left&middle&right\\
		\hline
		$Y_X$&$1.4\e{-10}$&$1.8\e{-10}$&$4.6\e{-10}$\\
		$Y_+$&$1.1\e{-17}$&$1.2\e{-25}$&$7.7\e{-13}$\\
		$Y_-$&$1.6\e{-7}$&$2.1\e{-7}$&$7.6\e{-9}$\\
		\hline\end{tabular}
	\end{tabular}}
		\caption{The case of 3 stable components with similar masses. Parameters of the plots are: $m_X=160\gev$, $m_+=200\gev$, $m_2=150\gev$, $\sin\alpha=0.1$, $g_x=0.1$, $m_-$ is $100\gev$ (left), $130\gev$ (middle), $160\gev$ (right).}
		\label{fig:3dm4-6}
	\end{figure}

	\begin{figure}[H]
	\makebox[\linewidth][c]{\centering\begin{tabular}{c}
		\begin{minipage}{.45\textwidth}\centering
			\includegraphics[width=\textwidth]{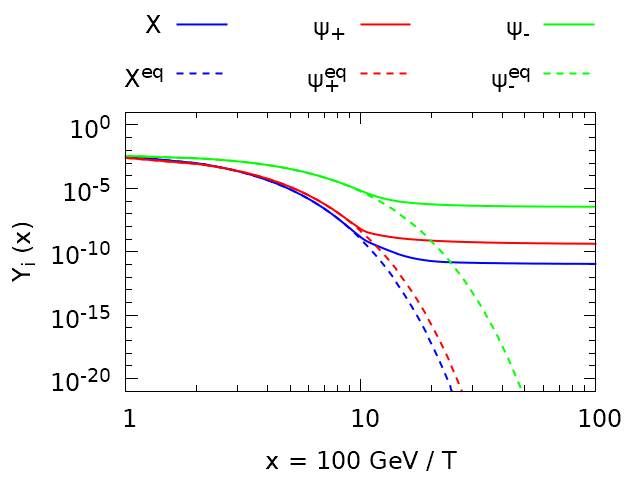}
\scalebox{.85}{
			\begin{tabular}{|c|c|c|c|}
				\hline
				process & $a_N$ & $a_{N+1}$ & $N$ \\
				\hline
				$XX \rightarrow \text{SM}$ & $1.4\cdot 10^{-2}$ & $-1.8\cdot 10^{-2}$ & $0$ \\
				$\psi_+\psi_+ \rightarrow \text{SM}$ & $4.8\cdot 10^{-6}$ & $-1.6\cdot 10^{-5}$ & $1$ \\
				$\psi_-\psi_- \rightarrow \text{SM}$ & $1.7\cdot 10^{-4}$ & $-2.3\cdot 10^{-3}$ & $1$ \\
				\hline
				$XX \rightarrow \Psi_+\Psi_+$ & $3\cdot 10^{-3}$ & $-7.5\cdot 10^{-3}$ & $0$ \\
				$XX \rightarrow \Psi_-\Psi_-$ & $1.5\cdot 10^{-3}$ & $-2.6\cdot 10^{-3}$ & $0$ \\
				$\Psi_+,\Psi_+ \rightarrow \Psi_-\Psi_-$ & $3.3\cdot 10^{-3}$ & $4.6\cdot 10^{-3}$ & $0$ \\
				\hline
				$Xh_1 \rightarrow \Psi_+\Psi_-$ & $1.1\cdot 10^{-4}$ & $-3.2\cdot 10^{-4}$ & $0$ \\
				$Xh_2 \rightarrow \Psi_+\Psi_-$ & $7.2\cdot 10^{-3}$ & $-1.9\cdot 10^{-2}$ & $0$ \\
				$\Psi_+h_1 \rightarrow X\Psi_-$ & $2.2\cdot 10^{-6}$ & $1.5\cdot 10^{-4}$ & $0$ \\
				$\Psi_+h_2 \rightarrow X\Psi_-$ & $2.5\cdot 10^{-3}$ & $2.4\cdot 10^{-2}$ & $0$ \\
				$X\Psi_+ \rightarrow \Psi_-h_1$ & $3.9\cdot 10^{-5}$ & $5.5\cdot 10^{-5}$ & $0$ \\
				$X\Psi_+ \rightarrow \Psi_-h_2$ & $3.5\cdot 10^{-3}$ & $6\cdot 10^{-3}$ & $0$ \\
				\hline
			\end{tabular}
}
		\end{minipage}
		\begin{minipage}{.45\textwidth}\centering
		\includegraphics[width=\textwidth]{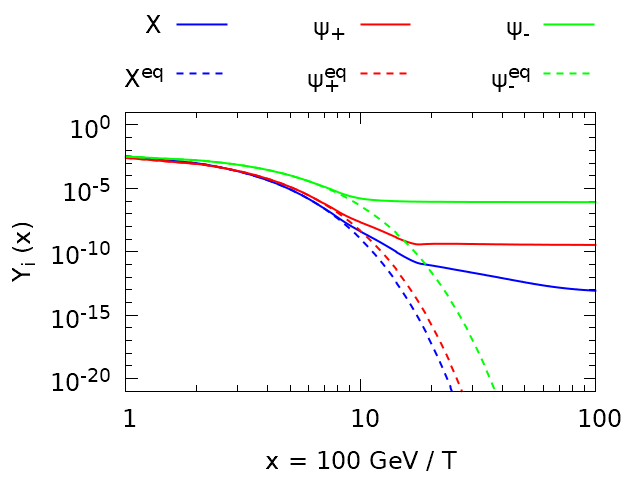}
\scalebox{.85}{
			\begin{tabular}{|c|c|c|c|}
				\hline
				process & $a_N$ & $a_{N+1}$ & $N$ \\
				\hline
				$XX \rightarrow \text{SM}$ & $1.4\cdot 10^{-2}$ & $-1.8\cdot 10^{-2}$ & $0$ \\
				$\psi_+\psi_+ \rightarrow \text{SM}$ & $5.5\cdot 10^{-6}$ & $-1.3\cdot 10^{-5}$ & $1$ \\
				$\psi_-\psi_- \rightarrow \text{SM}$ & $9.5\cdot 10^{-6}$ & $-4.9\cdot 10^{-5}$ & $1$ \\
				\hline
				$XX \rightarrow \Psi_+\Psi_+$ & $1.6\cdot 10^{-3}$ & $-1.1\cdot 10^{-3}$ & $0$ \\
				$XX \rightarrow \Psi_-\Psi_-$ & $1.4\cdot 10^{-3}$ & $-2.4\cdot 10^{-3}$ & $0$ \\
				$\Psi_+,\Psi_+ \rightarrow \Psi_-\Psi_-$ & $2.5\cdot 10^{-3}$ & $4.6\cdot 10^{-3}$ & $0$ \\
				\hline
				$Xh_1 \rightarrow \Psi_+\Psi_-$ & $8.4\cdot 10^{-5}$ & $2.7\cdot 10^{-5}$ & $0$ \\
				$Xh_2 \rightarrow \Psi_+\Psi_-$ & $7.3\cdot 10^{-3}$ & $-1.5\cdot 10^{-2}$ & $0$ \\
				$X\Psi_- \rightarrow \Psi_+h_1$ & $3.6\cdot 10^{-6}$ & $9.3\cdot 10^{-5}$ & $0$ \\
				$\Psi_+h_2 \rightarrow X\Psi_-$ & $3.53\cdot 10^{0}$ & $-5.95\cdot 10^{3}$ & $2$ \\
				$X\Psi_- \rightarrow \Psi_+h_2$ & $1.25\cdot 10^{0}$ & $-2.1\cdot 10^{3}$ & $2$ \\
				$X\Psi_+ \rightarrow \Psi_-h_1$ & $2.7\cdot 10^{-5}$ & $8\cdot 10^{-5}$ & $0$ \\
				$X\Psi_+ \rightarrow \Psi_-h_2$ & $2.3\cdot 10^{-3}$ & $8.4\cdot 10^{-3}$ & $0$ \\
				\hline
			\end{tabular}
}
		\end{minipage}
		\begin{minipage}{.45\textwidth}\centering
		\includegraphics[width=\textwidth]{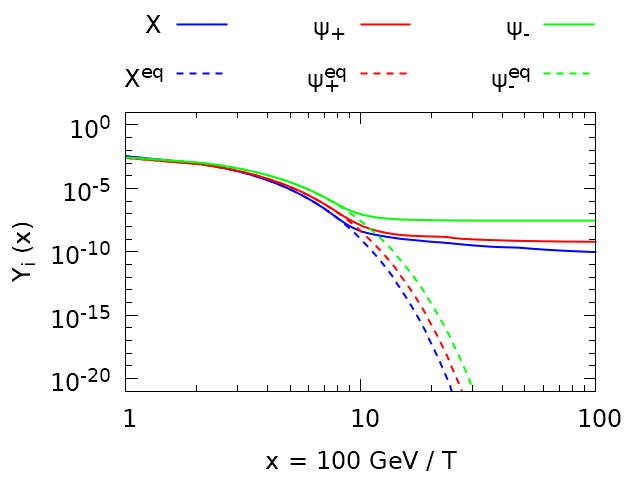}
\scalebox{.85}{
			\begin{tabular}{|c|c|c|c|}
				\hline
				process & $a_N$ & $a_{N+1}$ & $N$ \\
				\hline
				$XX \rightarrow \text{SM}$ & $1.4\cdot 10^{-2}$ & $-1.8\cdot 10^{-2}$ & $0$ \\
				$\psi_+\psi_+ \rightarrow \text{SM}$ & $2.7\cdot 10^{-6}$ & $-7.6\cdot 10^{-6}$ & $1$ \\
				$\psi_-\psi_- \rightarrow \text{SM}$ & $9.3\cdot 10^{-6}$ & $-2.9\cdot 10^{-6}$ & $1$ \\
				\hline
				$XX \rightarrow \Psi_+\Psi_+$ & $9.1\cdot 10^{-4}$ & $6.6\cdot 10^{-4}$ & $0$ \\
				$XX \rightarrow \Psi_-\Psi_-$ & $1.1\cdot 10^{-3}$ & $-1.2\cdot 10^{-3}$ & $0$ \\
				$\Psi_+,\Psi_+ \rightarrow \Psi_-\Psi_-$ & $1.6\cdot 10^{-3}$ & $3.7\cdot 10^{-3}$ & $0$ \\
				\hline
				$\Psi_+\Psi_- \rightarrow Xh_1$ & $4.7\cdot 10^{-5}$ & $4.5\cdot 10^{-5}$ & $0$ \\
				$Xh_2 \rightarrow \Psi_+\Psi_-$ & $4.3\cdot 10^{-3}$ & $1.5\cdot 10^{-2}$ & $0$ \\
				$X\Psi_- \rightarrow \Psi_+h_1$ & $7.8\cdot 10^{-6}$ & $1.1\cdot 10^{-4}$ & $0$ \\
				$X\Psi_- \rightarrow \Psi_+h_2$ & $4\cdot 10^{-4}$ & $9.8\cdot 10^{-3}$ & $0$ \\
				$X\Psi_+ \rightarrow \Psi_-h_1$ & $1.6\cdot 10^{-5}$ & $9.7\cdot 10^{-5}$ & $0$ \\
				$X\Psi_+ \rightarrow \Psi_-h_2$ & $1.3\cdot 10^{-3}$ & $1\cdot 10^{-2}$ & $0$ \\
				\hline
			\end{tabular}
}
		\end{minipage}\\~\\
		\begin{tabular}{|c|c|c|c|}\hline
		$Y(x=100)$&left&middle&right\\
		\hline
		$Y_X$&$1.1\e{-11}$&$8.3\e{-14}$&$9.4\e{-11}$\\
		$Y_+$&$4.2\e{-10}$&$3.4\e{-10}$&$6.1\e{-10}$\\
		$Y_-$&$3.5\e{-7}$&$8.2\e{-7}$&$2.8\e{-8}$\\
		\hline\end{tabular}
	\end{tabular}}
		\caption{The case of 3 stable components with similar masses. Parameters of the plots are: $m_X=200\gev$, $m_+=180\gev$, $m_2=150\gev$, $\sin\alpha=0.1$, $g_x=0.1$, $m_-$ is $100\gev$ (left), $130\gev$ (middle), $160\gev$ (right).}
		\label{fig:3dm7-9}
	\end{figure}
\end{subsubsection}
\end{subsection}
\begin{subsection}{Influence of $\sin\alpha$ and the Higgs portal cancellation}
Since the range of $\sin\alpha$ is limited to $|\sin\alpha|\lsim 0.3$ \cite{bib:sina}, therefore the results are less sensitive to $\sin\alpha$ than to other parameters. Dependence of results on the mixing angle increases with the difference between two Higgs particles' masses, $m_1=125\gev$ and $m_2$. The reason is, that for the most\footnote{Except the annihilation into two Higgs particles -- value of the three-Higgses vertex depends non-trivially on the mixing angle, in particular, matrix elements involving $h_1$ and $h_2$ propagators have different absolute values.} channels of DM$\to$SM annihilation the corresponding Feynman diagrams contain virtual Higgs particle mediating between DM and SM. Product of $h_i$'s couplings to dark particles and to the Standard Model is proportional to $\R_{1i}\R_{2i}$ ($\R$ is the scalar sector mixing matrix defined in eq. (\ref{eq:rot})). To calculate the cross-section, one has to sum over the virtual Higgs being $h_1$ or $h_2$, which contribute with different signs (because of the minus sign in $\R$). The sum is proportional to the difference of the $h_1$ and $h_2$ propagators, multiplied by $\sin\alpha\cos\alpha$, which is a function increasing with $\sin\alpha$ in the range of small $\alpha$. Figs. \ref{fig:sina7-9}, \ref{fig:sina1-3} and \ref{fig:sina4-6} show this tendency -- annihilation cross-section rises, hence the final yield decreases. Note that in the first case shown, masses of the Higgs particles are equal, therefore diagrams containing Higgs particle coupling to both SM and DM vanish -- the only annihilation channel that survives is annihilation into $h_ih_j$, $i=1,2$, $j=1,2$, which is not very sensitive to changes of $\sin\alpha$. Hence there is no visible difference between different $\sin\alpha$ cases.

	\begin{figure}[H]
	\makebox[\linewidth][c]{\centering\begin{tabular}{c}
		\begin{minipage}{.45\textwidth}\centering
			\includegraphics[width=\textwidth]{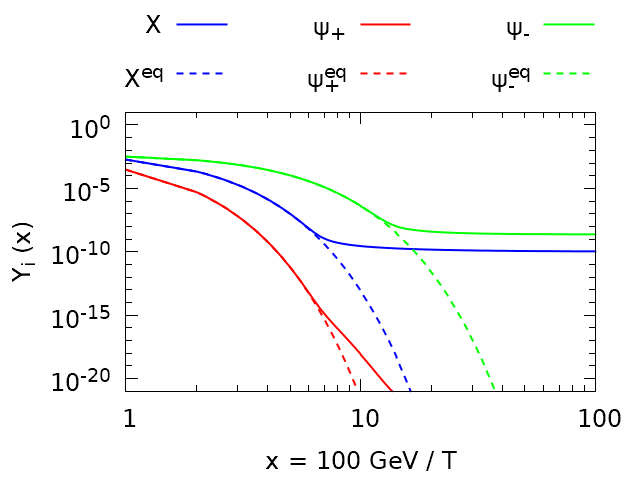}
\scalebox{.85}{
			\begin{tabular}{|c|c|c|c|}
				\hline
				process & $a_N$ & $a_{N+1}$ & $N$ \\
				\hline
				$XX \rightarrow \text{SM}$ & $9.1\cdot 10^{-3}$ & $-8.9\cdot 10^{-3}$ & $0$ \\
				$\psi_+\psi_+ \rightarrow \text{SM}$ & $1.8\cdot 10^{-4}$ & $-2\cdot 10^{-4}$ & $1$ \\
				$\psi_-\psi_- \rightarrow \text{SM}$ & $1.9\cdot 10^{-2}$ & $5.2\cdot 10^{-2}$ & $1$ \\
				\hline
				$\Psi_+\Psi_+ \rightarrow XX$ & $7.9\cdot 10^{-4}$ & $8.8\cdot 10^{-4}$ & $0$ \\
				$XX \rightarrow \Psi_-\Psi_-$ & $1.7\cdot 10^{-4}$ & $-1.3\cdot 10^{-4}$ & $0$ \\
				$\Psi_+,\Psi_+ \rightarrow \Psi_-\Psi_-$ & $3.3\cdot 10^{-3}$ & $-5.8\cdot 10^{-5}$ & $0$ \\
				\hline
				$\Psi_+\Psi_- \rightarrow Xh_1$ & $3\cdot 10^{-6}$ & $9.5\cdot 10^{-6}$ & $0$ \\
				$\Psi_+\Psi_- \rightarrow Xh_2$ & $3\cdot 10^{-4}$ & $9.4\cdot 10^{-4}$ & $0$ \\
				$\Psi_+h_1 \rightarrow X\Psi_-$ & $3.3\cdot 10^{-5}$ & $5.6\cdot 10^{-4}$ & $0$ \\
				$\Psi_+h_2 \rightarrow X\Psi_-$ & $3.2\cdot 10^{-3}$ & $5.5\cdot 10^{-2}$ & $0$ \\
				$X\Psi_+ \rightarrow \Psi_-h_1$ & $5.7\cdot 10^{-5}$ & $-2.9\cdot 10^{-5}$ & $0$ \\
				$X\Psi_+ \rightarrow \Psi_-h_2$ & $5.6\cdot 10^{-3}$ & $-2.9\cdot 10^{-3}$ & $0$ \\
				\hline
				$\Psi_+ \rightarrow X\Psi_-$ & \multicolumn{3}{|c|}{$1.4\cdot 10^{-2}$} \\
				\hline
			\end{tabular}
}
		\end{minipage}
		\begin{minipage}{.45\textwidth}\centering
		\includegraphics[width=\textwidth]{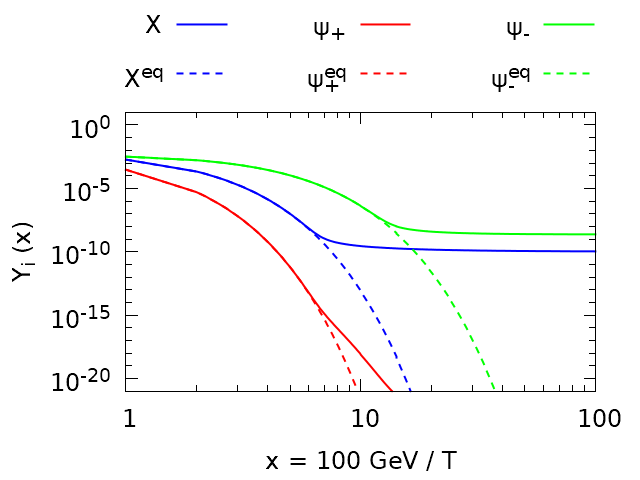}
\scalebox{.85}{
			\begin{tabular}{|c|c|c|c|}
				\hline
				process & $a_N$ & $a_{N+1}$ & $N$ \\
				\hline
				$XX \rightarrow \text{SM}$ & $9.1\cdot 10^{-3}$ & $-8.9\cdot 10^{-3}$ & $0$ \\
				$\psi_+\psi_+ \rightarrow \text{SM}$ & $1.8\cdot 10^{-4}$ & $-2\cdot 10^{-4}$ & $1$ \\
				$\psi_-\psi_- \rightarrow \text{SM}$ & $1.9\cdot 10^{-2}$ & $5.2\cdot 10^{-2}$ & $1$ \\
				\hline
				$\Psi_+\Psi_+ \rightarrow XX$ & $7.9\cdot 10^{-4}$ & $8.8\cdot 10^{-4}$ & $0$ \\
				$XX \rightarrow \Psi_-\Psi_-$ & $1.7\cdot 10^{-4}$ & $-1.3\cdot 10^{-4}$ & $0$ \\
				$\Psi_+,\Psi_+ \rightarrow \Psi_-\Psi_-$ & $3.3\cdot 10^{-3}$ & $-5.8\cdot 10^{-5}$ & $0$ \\
				\hline
				$\Psi_+\Psi_- \rightarrow Xh_1$ & $1.2\cdot 10^{-5}$ & $3.8\cdot 10^{-5}$ & $0$ \\
				$\Psi_+\Psi_- \rightarrow Xh_2$ & $2.9\cdot 10^{-4}$ & $9.1\cdot 10^{-4}$ & $0$ \\
				$\Psi_+h_1 \rightarrow X\Psi_-$ & $1.3\cdot 10^{-4}$ & $2.2\cdot 10^{-3}$ & $0$ \\
				$\Psi_+h_2 \rightarrow X\Psi_-$ & $3.1\cdot 10^{-3}$ & $5.3\cdot 10^{-2}$ & $0$ \\
				$X\Psi_+ \rightarrow \Psi_-h_1$ & $2.3\cdot 10^{-4}$ & $-1.2\cdot 10^{-4}$ & $0$ \\
				$X\Psi_+ \rightarrow \Psi_-h_2$ & $5.5\cdot 10^{-3}$ & $-2.9\cdot 10^{-3}$ & $0$ \\
				\hline
				$\Psi_+ \rightarrow X\Psi_-$ & \multicolumn{3}{|c|}{$1.4\cdot 10^{-2}$} \\
				\hline
			\end{tabular}
}
		\end{minipage}
		\begin{minipage}{.45\textwidth}\centering
		\includegraphics[width=\textwidth]{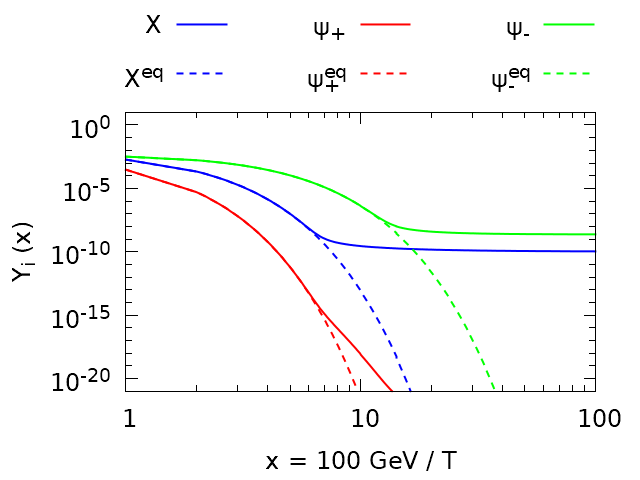}
\scalebox{.85}{
			\begin{tabular}{|c|c|c|c|}
				\hline
				process & $a_N$ & $a_{N+1}$ & $N$ \\
				\hline
				$XX \rightarrow \text{SM}$ & $9.1\cdot 10^{-3}$ & $-8.7\cdot 10^{-3}$ & $0$ \\
				$\psi_+\psi_+ \rightarrow \text{SM}$ & $1.8\cdot 10^{-4}$ & $-2.1\cdot 10^{-4}$ & $1$ \\
				$\psi_-\psi_- \rightarrow \text{SM}$ & $1.9\cdot 10^{-2}$ & $5.3\cdot 10^{-2}$ & $1$ \\
				\hline
				$\Psi_+\Psi_+ \rightarrow XX$ & $7.9\cdot 10^{-4}$ & $8.8\cdot 10^{-4}$ & $0$ \\
				$XX \rightarrow \Psi_-\Psi_-$ & $1.7\cdot 10^{-4}$ & $-1.3\cdot 10^{-4}$ & $0$ \\
				$\Psi_+,\Psi_+ \rightarrow \Psi_-\Psi_-$ & $3.3\cdot 10^{-3}$ & $-5.8\cdot 10^{-5}$ & $0$ \\
				\hline
				$\Psi_+\Psi_- \rightarrow Xh_1$ & $2.7\cdot 10^{-5}$ & $8.5\cdot 10^{-5}$ & $0$ \\
				$\Psi_+\Psi_- \rightarrow Xh_2$ & $2.7\cdot 10^{-4}$ & $8.6\cdot 10^{-4}$ & $0$ \\
				$\Psi_+h_1 \rightarrow X\Psi_-$ & $2.9\cdot 10^{-4}$ & $5\cdot 10^{-3}$ & $0$ \\
				$\Psi_+h_2 \rightarrow X\Psi_-$ & $3\cdot 10^{-3}$ & $5.1\cdot 10^{-2}$ & $0$ \\
				$X\Psi_+ \rightarrow \Psi_-h_1$ & $5.1\cdot 10^{-4}$ & $-2.7\cdot 10^{-4}$ & $0$ \\
				$X\Psi_+ \rightarrow \Psi_-h_2$ & $5.2\cdot 10^{-3}$ & $-2.7\cdot 10^{-3}$ & $0$ \\
				\hline
				$\Psi_+ \rightarrow X\Psi_-$ & \multicolumn{3}{|c|}{$1.4\cdot 10^{-2}$} \\
				\hline
			\end{tabular}
}
		\end{minipage}\\~\\
		\begin{tabular}{|c|c|c|c|}\hline
		$Y(x=100)$&left&middle&right\\
		\hline
		$Y_X$&$1.1\e{-10}$&$1.1\e{-10}$&$1.1\e{-10}$\\
		$Y_+$&$2.4\e{-50}$&$2.4\e{-50}$&$2.4\e{-50}$\\
		$Y_-$&$2.3\e{-9}$&$2.3\e{-9}$&$2.3\e{-9}$\\
		\hline\end{tabular}
	\end{tabular}}
		\caption{Influence of $\sin\alpha$ on the shape of solutions of the Boltzmann equation. Parameters of the plots are: $m_X=300\gev$, $m_+=500\gev$, $m_-=130\gev$, $m_2=125\gev$, $g_x=0.1$. Sine of the mixing angle $\sin\alpha$ is $0.1$ (left), $0.2$ (middle), $0.3$ (right).}
		\label{fig:sina7-9}
	\end{figure}

	\begin{figure}[H]
	\makebox[\linewidth][c]{\centering\begin{tabular}{c}
		\begin{minipage}{.45\textwidth}\centering
			\includegraphics[width=\textwidth]{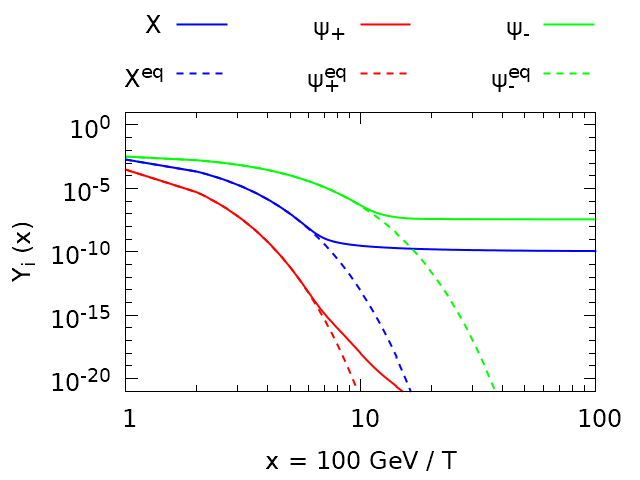}
\scalebox{.85}{
			\begin{tabular}{|c|c|c|c|}
				\hline
				process & $a_N$ & $a_{N+1}$ & $N$ \\
				\hline
				$XX \rightarrow \text{SM}$ & $8.5\cdot 10^{-3}$ & $-8.2\cdot 10^{-3}$ & $0$ \\
				$\psi_+\psi_+ \rightarrow \text{SM}$ & $1.7\cdot 10^{-4}$ & $-1.9\cdot 10^{-4}$ & $1$ \\
				$\psi_-\psi_- \rightarrow \text{SM}$ & $2.2\cdot 10^{-4}$ & $-1.2\cdot 10^{-3}$ & $1$ \\
				\hline
				$\Psi_+\Psi_+ \rightarrow XX$ & $7.9\cdot 10^{-4}$ & $8.8\cdot 10^{-4}$ & $0$ \\
				$XX \rightarrow \Psi_-\Psi_-$ & $1.8\cdot 10^{-4}$ & $-1.4\cdot 10^{-4}$ & $0$ \\
				$\Psi_+,\Psi_+ \rightarrow \Psi_-\Psi_-$ & $3.3\cdot 10^{-3}$ & $-5.8\cdot 10^{-5}$ & $0$ \\
				\hline
				$\Psi_+\Psi_- \rightarrow Xh_1$ & $3\cdot 10^{-6}$ & $9.5\cdot 10^{-6}$ & $0$ \\
				$\Psi_+\Psi_- \rightarrow Xh_2$ & $3.5\cdot 10^{-4}$ & $1\cdot 10^{-3}$ & $0$ \\
				$\Psi_+h_1 \rightarrow X\Psi_-$ & $3.3\cdot 10^{-5}$ & $5.6\cdot 10^{-4}$ & $0$ \\
				$\Psi_+h_2 \rightarrow X\Psi_-$ & $3.6\cdot 10^{-3}$ & $3.8\cdot 10^{-2}$ & $0$ \\
				$X\Psi_+ \rightarrow \Psi_-h_1$ & $5.7\cdot 10^{-5}$ & $-2.9\cdot 10^{-5}$ & $0$ \\
				$X\Psi_+ \rightarrow \Psi_-h_2$ & $5.6\cdot 10^{-3}$ & $-3.1\cdot 10^{-3}$ & $0$ \\
				\hline
				$\Psi_+ \rightarrow X\Psi_-$ & \multicolumn{3}{|c|}{$1.4\cdot 10^{-2}$} \\
				\hline
			\end{tabular}
}
		\end{minipage}
		\begin{minipage}{.45\textwidth}\centering
		\includegraphics[width=\textwidth]{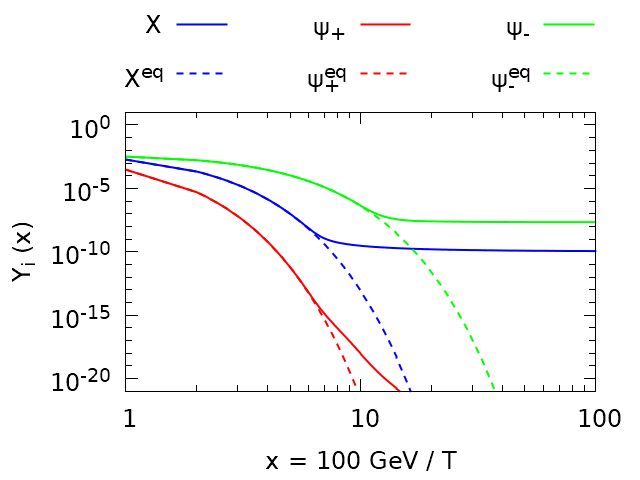}
\scalebox{.85}{
			\begin{tabular}{|c|c|c|c|}
				\hline
				process & $a_N$ & $a_{N+1}$ & $N$ \\
				\hline
				$XX \rightarrow \text{SM}$ & $8.5\cdot 10^{-3}$ & $-8.1\cdot 10^{-3}$ & $0$ \\
				$\psi_+\psi_+ \rightarrow \text{SM}$ & $1.7\cdot 10^{-4}$ & $-1.9\cdot 10^{-4}$ & $1$ \\
				$\psi_-\psi_- \rightarrow \text{SM}$ & $7.8\cdot 10^{-4}$ & $-4.9\cdot 10^{-3}$ & $1$ \\
				\hline
				$\Psi_+\Psi_+ \rightarrow XX$ & $7.9\cdot 10^{-4}$ & $8.8\cdot 10^{-4}$ & $0$ \\
				$XX \rightarrow \Psi_-\Psi_-$ & $1.8\cdot 10^{-4}$ & $-1.4\cdot 10^{-4}$ & $0$ \\
				$\Psi_+,\Psi_+ \rightarrow \Psi_-\Psi_-$ & $3.3\cdot 10^{-3}$ & $-7.8\cdot 10^{-5}$ & $0$ \\
				\hline
				$\Psi_+\Psi_- \rightarrow Xh_1$ & $1.2\cdot 10^{-5}$ & $3.8\cdot 10^{-5}$ & $0$ \\
				$\Psi_+\Psi_- \rightarrow Xh_2$ & $3.4\cdot 10^{-4}$ & $1\cdot 10^{-3}$ & $0$ \\
				$\Psi_+h_1 \rightarrow X\Psi_-$ & $1.3\cdot 10^{-4}$ & $2.2\cdot 10^{-3}$ & $0$ \\
				$\Psi_+h_2 \rightarrow X\Psi_-$ & $3.5\cdot 10^{-3}$ & $3.7\cdot 10^{-2}$ & $0$ \\
				$X\Psi_+ \rightarrow \Psi_-h_1$ & $2.3\cdot 10^{-4}$ & $-1.2\cdot 10^{-4}$ & $0$ \\
				$X\Psi_+ \rightarrow \Psi_-h_2$ & $5.4\cdot 10^{-3}$ & $-2.9\cdot 10^{-3}$ & $0$ \\
				\hline
				$\Psi_+ \rightarrow X\Psi_-$ & \multicolumn{3}{|c|}{$1.4\cdot 10^{-2}$} \\
				\hline
			\end{tabular}
}
		\end{minipage}
		\begin{minipage}{.45\textwidth}\centering
		\includegraphics[width=\textwidth]{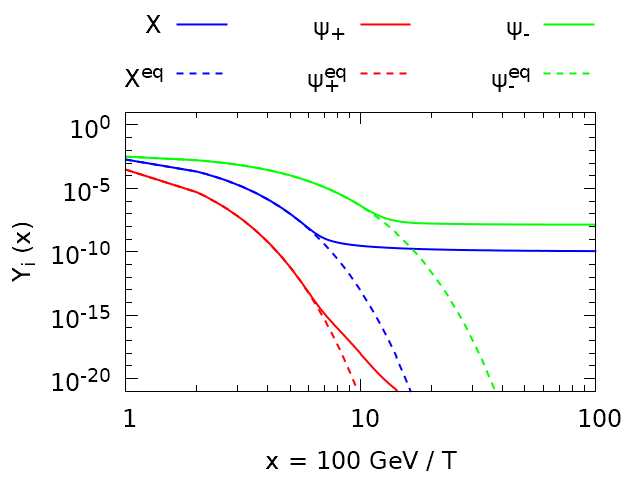}
\scalebox{.85}{
			\begin{tabular}{|c|c|c|c|}
				\hline
				process & $a_N$ & $a_{N+1}$ & $N$ \\
				\hline
				$XX \rightarrow \text{SM}$ & $8.6\cdot 10^{-3}$ & $-8.3\cdot 10^{-3}$ & $0$ \\
				$\psi_+\psi_+ \rightarrow \text{SM}$ & $1.7\cdot 10^{-4}$ & $-2\cdot 10^{-4}$ & $1$ \\
				$\psi_-\psi_- \rightarrow \text{SM}$ & $1.5\cdot 10^{-3}$ & $-1.1\cdot 10^{-2}$ & $1$ \\
				\hline
				$\Psi_+\Psi_+ \rightarrow XX$ & $7.9\cdot 10^{-4}$ & $8.8\cdot 10^{-4}$ & $0$ \\
				$XX \rightarrow \Psi_-\Psi_-$ & $1.8\cdot 10^{-4}$ & $-1.3\cdot 10^{-4}$ & $0$ \\
				$\Psi_+,\Psi_+ \rightarrow \Psi_-\Psi_-$ & $3.3\cdot 10^{-3}$ & $-7.8\cdot 10^{-5}$ & $0$ \\
				\hline
				$\Psi_+\Psi_- \rightarrow Xh_1$ & $2.7\cdot 10^{-5}$ & $8.5\cdot 10^{-5}$ & $0$ \\
				$\Psi_+\Psi_- \rightarrow Xh_2$ & $3.2\cdot 10^{-4}$ & $9.5\cdot 10^{-4}$ & $0$ \\
				$\Psi_+h_1 \rightarrow X\Psi_-$ & $2.9\cdot 10^{-4}$ & $5\cdot 10^{-3}$ & $0$ \\
				$\Psi_+h_2 \rightarrow X\Psi_-$ & $3.3\cdot 10^{-3}$ & $3.5\cdot 10^{-2}$ & $0$ \\
				$X\Psi_+ \rightarrow \Psi_-h_1$ & $5.1\cdot 10^{-4}$ & $-2.7\cdot 10^{-4}$ & $0$ \\
				$X\Psi_+ \rightarrow \Psi_-h_2$ & $5.2\cdot 10^{-3}$ & $-2.7\cdot 10^{-3}$ & $0$ \\
				\hline
				$\Psi_+ \rightarrow X\Psi_-$ & \multicolumn{3}{|c|}{$1.4\cdot 10^{-2}$} \\
				\hline
			\end{tabular}
}
		\end{minipage}\\~\\
		\begin{tabular}{|c|c|c|c|}\hline
		$Y(x=100)$&left&middle&right\\
		\hline
		$Y_X$&$1.1\e{-10}$&$1,1\e{-10}$&$1.1\e{-10}$\\
		$Y_+$&$4.0\e{-49}$&$2.3\e{-49}$&$1.5\e{-49}$\\
		$Y_-$&$3.6\e{-8}$&$2.1\e{-8}$&$1.4\e{-8}$\\
		\hline\end{tabular}
	\end{tabular}}
		\caption{Influence of $\sin\alpha$ on the shape of solutions of the Boltzmann equation. Parameters of the plots are: $m_X=300\gev$, $m_+=500\gev$, $m_-=130\gev$,$m_2=150\gev$, $g_x=0.1$. Sine of the mixing angle $\sin\alpha$ is $0.1$ (left), $0.2$ (middle), $0.3$ (right).}
		\label{fig:sina1-3}
	\end{figure}

	\begin{figure}[H]
	\makebox[\linewidth][c]{\centering\begin{tabular}{c}
		\begin{minipage}{.45\textwidth}\centering
			\includegraphics[width=\textwidth]{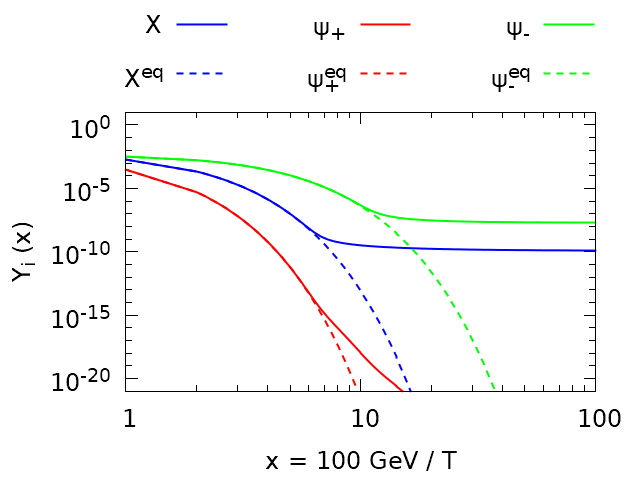}
\scalebox{.85}{
			\begin{tabular}{|c|c|c|c|}
				\hline
				process & $a_N$ & $a_{N+1}$ & $N$ \\
				\hline
				$XX \rightarrow \text{SM}$ & $7.7\cdot 10^{-3}$ & $-7.2\cdot 10^{-3}$ & $0$ \\
				$\psi_+\psi_+ \rightarrow \text{SM}$ & $1.6\cdot 10^{-4}$ & $-1.8\cdot 10^{-4}$ & $1$ \\
				$\psi_-\psi_- \rightarrow \text{SM}$ & $2.3\cdot 10^{-3}$ & $-1.5\cdot 10^{-2}$ & $1$ \\
				\hline
				$\Psi_+\Psi_+ \rightarrow XX$ & $7.9\cdot 10^{-4}$ & $8.8\cdot 10^{-4}$ & $0$ \\
				$XX \rightarrow \Psi_-\Psi_-$ & $1.9\cdot 10^{-4}$ & $-1.5\cdot 10^{-4}$ & $0$ \\
				$\Psi_+,\Psi_+ \rightarrow \Psi_-\Psi_-$ & $3.3\cdot 10^{-3}$ & $-5.8\cdot 10^{-5}$ & $0$ \\
				\hline
				$\Psi_+\Psi_- \rightarrow Xh_1$ & $3\cdot 10^{-6}$ & $9.5\cdot 10^{-6}$ & $0$ \\
				$\Psi_+\Psi_- \rightarrow Xh_2$ & $4.7\cdot 10^{-4}$ & $1.1\cdot 10^{-3}$ & $0$ \\
				$\Psi_+h_1 \rightarrow X\Psi_-$ & $3.3\cdot 10^{-5}$ & $5.6\cdot 10^{-4}$ & $0$ \\
				$\Psi_+h_2 \rightarrow X\Psi_-$ & $4.2\cdot 10^{-3}$ & $2.6\cdot 10^{-2}$ & $0$ \\
				$X\Psi_+ \rightarrow \Psi_-h_1$ & $5.7\cdot 10^{-5}$ & $-2.9\cdot 10^{-5}$ & $0$ \\
				$X\Psi_+ \rightarrow \Psi_-h_2$ & $5.6\cdot 10^{-3}$ & $-2.9\cdot 10^{-3}$ & $0$ \\
				\hline
				$\Psi_+ \rightarrow X\Psi_-$ & \multicolumn{3}{|c|}{$1.4\cdot 10^{-2}$} \\
				\hline
			\end{tabular}
}
		\end{minipage}
		\begin{minipage}{.45\textwidth}\centering
		\includegraphics[width=\textwidth]{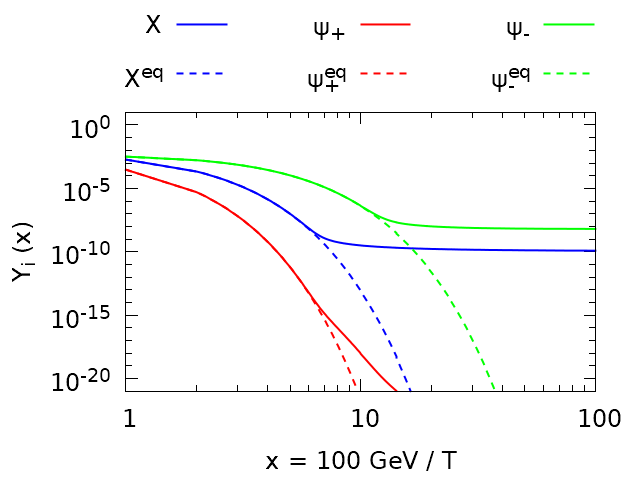}
\scalebox{.85}{
			\begin{tabular}{|c|c|c|c|}
				\hline
				process & $a_N$ & $a_{N+1}$ & $N$ \\
				\hline
				$XX \rightarrow \text{SM}$ & $7.8\cdot 10^{-3}$ & $-7.5\cdot 10^{-3}$ & $0$ \\
				$\psi_+\psi_+ \rightarrow \text{SM}$ & $1.7\cdot 10^{-4}$ & $-2\cdot 10^{-4}$ & $1$ \\
				$\psi_-\psi_- \rightarrow \text{SM}$ & $8.4\cdot 10^{-3}$ & $-6\cdot 10^{-2}$ & $1$ \\
				\hline
				$\Psi_+\Psi_+ \rightarrow XX$ & $7.9\cdot 10^{-4}$ & $8.8\cdot 10^{-4}$ & $0$ \\
				$XX \rightarrow \Psi_-\Psi_-$ & $1.9\cdot 10^{-4}$ & $-1.5\cdot 10^{-4}$ & $0$ \\
				$\Psi_+,\Psi_+ \rightarrow \Psi_-\Psi_-$ & $3.3\cdot 10^{-3}$ & $-5.8\cdot 10^{-5}$ & $0$ \\
				\hline
				$\Psi_+\Psi_- \rightarrow Xh_1$ & $1.2\cdot 10^{-5}$ & $3.8\cdot 10^{-5}$ & $0$ \\
				$\Psi_+\Psi_- \rightarrow Xh_2$ & $4.6\cdot 10^{-4}$ & $1.1\cdot 10^{-3}$ & $0$ \\
				$\Psi_+h_1 \rightarrow X\Psi_-$ & $1.3\cdot 10^{-4}$ & $2.2\cdot 10^{-3}$ & $0$ \\
				$\Psi_+h_2 \rightarrow X\Psi_-$ & $4\cdot 10^{-3}$ & $2.6\cdot 10^{-2}$ & $0$ \\
				$X\Psi_+ \rightarrow \Psi_-h_1$ & $2.3\cdot 10^{-4}$ & $-1.2\cdot 10^{-4}$ & $0$ \\
				$X\Psi_+ \rightarrow \Psi_-h_2$ & $5.4\cdot 10^{-3}$ & $-2.9\cdot 10^{-3}$ & $0$ \\
				\hline
				$\Psi_+ \rightarrow X\Psi_-$ & \multicolumn{3}{|c|}{$1.4\cdot 10^{-2}$} \\
				\hline
			\end{tabular}
}
		\end{minipage}
		\begin{minipage}{.45\textwidth}\centering
		\includegraphics[width=\textwidth]{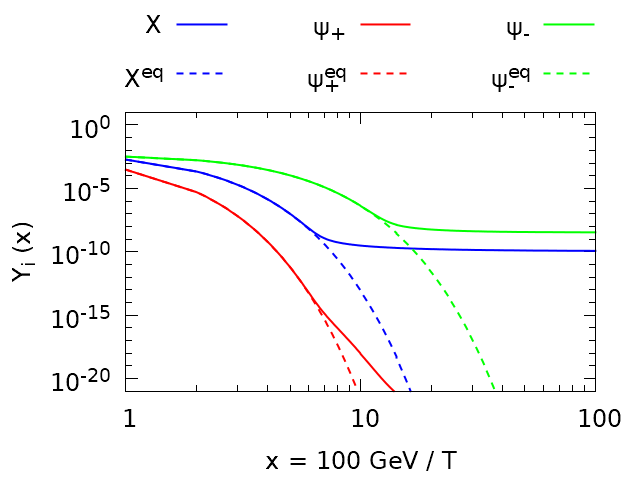}
\scalebox{.85}{
			\begin{tabular}{|c|c|c|c|}
				\hline
				process & $a_N$ & $a_{N+1}$ & $N$ \\
				\hline
				$XX \rightarrow \text{SM}$ & $8.1\cdot 10^{-3}$ & $-7.8\cdot 10^{-3}$ & $0$ \\
				$\psi_+\psi_+ \rightarrow \text{SM}$ & $1.7\cdot 10^{-4}$ & $-1.9\cdot 10^{-4}$ & $1$ \\
				$\psi_-\psi_- \rightarrow \text{SM}$ & $1.7\cdot 10^{-2}$ & $-1.3\cdot 10^{-1}$ & $1$ \\
				\hline
				$\Psi_+\Psi_+ \rightarrow XX$ & $7.9\cdot 10^{-4}$ & $8.8\cdot 10^{-4}$ & $0$ \\
				$XX \rightarrow \Psi_-\Psi_-$ & $1.9\cdot 10^{-4}$ & $-1.5\cdot 10^{-4}$ & $0$ \\
				$\Psi_+,\Psi_+ \rightarrow \Psi_-\Psi_-$ & $3.3\cdot 10^{-3}$ & $-5.8\cdot 10^{-5}$ & $0$ \\
				\hline
				$\Psi_+\Psi_- \rightarrow Xh_1$ & $2.7\cdot 10^{-5}$ & $8.5\cdot 10^{-5}$ & $0$ \\
				$\Psi_+\Psi_- \rightarrow Xh_2$ & $4.3\cdot 10^{-4}$ & $1\cdot 10^{-3}$ & $0$ \\
				$\Psi_+h_1 \rightarrow X\Psi_-$ & $2.9\cdot 10^{-4}$ & $5\cdot 10^{-3}$ & $0$ \\
				$\Psi_+h_2 \rightarrow X\Psi_-$ & $3.8\cdot 10^{-3}$ & $2.4\cdot 10^{-2}$ & $0$ \\
				$X\Psi_+ \rightarrow \Psi_-h_1$ & $5.1\cdot 10^{-4}$ & $-2.7\cdot 10^{-4}$ & $0$ \\
				$X\Psi_+ \rightarrow \Psi_-h_2$ & $5.1\cdot 10^{-3}$ & $-2.7\cdot 10^{-3}$ & $0$ \\
				\hline
				$\Psi_+ \rightarrow X\Psi_-$ & \multicolumn{3}{|c|}{$1.4\cdot 10^{-2}$} \\
				\hline
			\end{tabular}
}
		\end{minipage}\\~\\
		\begin{tabular}{|c|c|c|c|}\hline
		$Y(x=100)$&left&middle&right\\
		\hline
		$Y_X$&$1.2\e{-10}$&$1.2\e{-10}$&$1.2\e{-10}$\\
		$Y_+$&$2.4\e{-49}$&$7.4\e{-50}$&$3.9\e{-50}$\\
		$Y_-$&$2.0\e{-8}$&$6.3\e{-9}$&$3.4\e{-9}$\\
		\hline\end{tabular}
	\end{tabular}}
		\caption{Influence of $\sin\alpha$ on the shape of solutions of the Boltzmann equation. Parameters of the plots are: $m_X=300\gev$, $m_+=500\gev$, $m_-=130\gev$,$m_2=180\gev$, $g_x=0.1$. Sine of the mixing angle $\sin\alpha$ is $0.1$ (left), $0.2$ (middle), $0.3$ (right).}
		\label{fig:sina4-6}
	\end{figure}
\end{subsection}
\begin{subsection}{The role of $g_x$}
In fig. \ref{fig:gx1-3} we can see a strong sensitivity of the shape of plots to the value of $g_x$. As $g_x$ rises, $y_x\sim g_x$ increases and $v_x\sim g_x^{-1}$ decreases, leading to larger cross-section for annihilation into the Standard Model.
	\begin{figure}[H]
	\makebox[\linewidth][c]{\centering\begin{tabular}{c}
		\begin{minipage}{.45\textwidth}\centering
			\includegraphics[width=\textwidth]{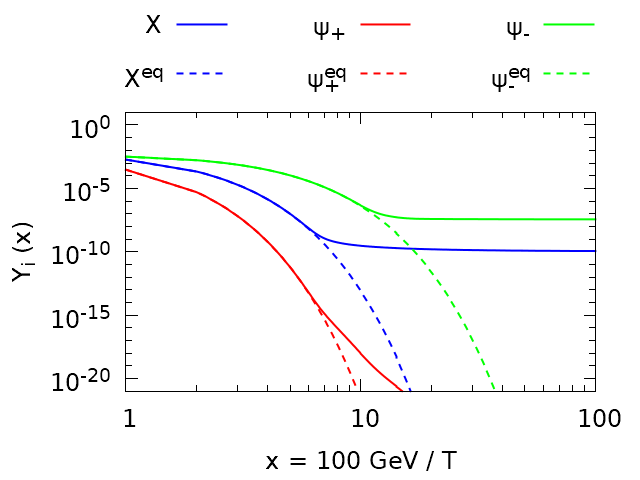}
\scalebox{.85}{
			\begin{tabular}{|c|c|c|c|}
				\hline
				process & $a_N$ & $a_{N+1}$ & $N$ \\
				\hline
				$XX \rightarrow \text{SM}$ & $8.5\cdot 10^{-3}$ & $-8.2\cdot 10^{-3}$ & $0$ \\
				$\psi_+\psi_+ \rightarrow \text{SM}$ & $1.7\cdot 10^{-4}$ & $-1.9\cdot 10^{-4}$ & $1$ \\
				$\psi_-\psi_- \rightarrow \text{SM}$ & $2.2\cdot 10^{-4}$ & $-1.2\cdot 10^{-3}$ & $1$ \\
				\hline
				$\Psi_+\Psi_+ \rightarrow XX$ & $7.9\cdot 10^{-4}$ & $8.8\cdot 10^{-4}$ & $0$ \\
				$XX \rightarrow \Psi_-\Psi_-$ & $1.8\cdot 10^{-4}$ & $-1.4\cdot 10^{-4}$ & $0$ \\
				$\Psi_+,\Psi_+ \rightarrow \Psi_-\Psi_-$ & $3.3\cdot 10^{-3}$ & $-5.8\cdot 10^{-5}$ & $0$ \\
				\hline
				$\Psi_+\Psi_- \rightarrow Xh_1$ & $3\cdot 10^{-6}$ & $9.5\cdot 10^{-6}$ & $0$ \\
				$\Psi_+\Psi_- \rightarrow Xh_2$ & $3.5\cdot 10^{-4}$ & $1\cdot 10^{-3}$ & $0$ \\
				$\Psi_+h_1 \rightarrow X\Psi_-$ & $3.3\cdot 10^{-5}$ & $5.6\cdot 10^{-4}$ & $0$ \\
				$\Psi_+h_2 \rightarrow X\Psi_-$ & $3.6\cdot 10^{-3}$ & $3.8\cdot 10^{-2}$ & $0$ \\
				$X\Psi_+ \rightarrow \Psi_-h_1$ & $5.7\cdot 10^{-5}$ & $-2.9\cdot 10^{-5}$ & $0$ \\
				$X\Psi_+ \rightarrow \Psi_-h_2$ & $5.6\cdot 10^{-3}$ & $-3.1\cdot 10^{-3}$ & $0$ \\
				\hline
				$\Psi_+ \rightarrow X\Psi_-$ & \multicolumn{3}{|c|}{$1.4\cdot 10^{-2}$} \\
				\hline
			\end{tabular}
}
		\end{minipage}
		\begin{minipage}{.45\textwidth}\centering
		\includegraphics[width=\textwidth]{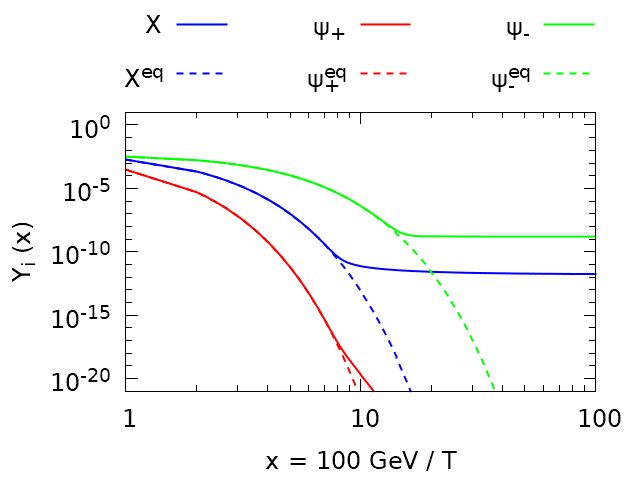}
\scalebox{.85}{
			\begin{tabular}{|c|c|c|c|}
				\hline
				process & $a_N$ & $a_{N+1}$ & $N$ \\
				\hline
				$XX \rightarrow \text{SM}$ & $6.8\cdot 10^{-1}$ & $-6.6\cdot 10^{-1}$ & $0$ \\
				$\psi_+\psi_+ \rightarrow \text{SM}$ & $1.4\cdot 10^{-2}$ & $-1.6\cdot 10^{-2}$ & $1$ \\
				$\psi_-\psi_- \rightarrow \text{SM}$ & $1.7\cdot 10^{-3}$ & $-1.2\cdot 10^{-2}$ & $1$ \\
				\hline
				$\Psi_+\Psi_+ \rightarrow XX$ & $6.4\cdot 10^{-2}$ & $7\cdot 10^{-2}$ & $0$ \\
				$XX \rightarrow \Psi_-\Psi_-$ & $1.5\cdot 10^{-2}$ & $-1.1\cdot 10^{-2}$ & $0$ \\
				$\Psi_+,\Psi_+ \rightarrow \Psi_-\Psi_-$ & $2.6\cdot 10^{-1}$ & $-5.8\cdot 10^{-3}$ & $0$ \\
				\hline
				$\Psi_+\Psi_- \rightarrow Xh_1$ & $2.4\cdot 10^{-4}$ & $7.7\cdot 10^{-4}$ & $0$ \\
				$\Psi_+\Psi_- \rightarrow Xh_2$ & $2.8\cdot 10^{-2}$ & $8.4\cdot 10^{-2}$ & $0$ \\
				$\Psi_+h_1 \rightarrow X\Psi_-$ & $2.6\cdot 10^{-3}$ & $4.5\cdot 10^{-2}$ & $0$ \\
				$\Psi_+h_2 \rightarrow X\Psi_-$ & $2.9\cdot 10^{-1}$ & $3.06\cdot 10^{0}$ & $0$ \\
				$X\Psi_+ \rightarrow \Psi_-h_1$ & $4.6\cdot 10^{-3}$ & $-2.3\cdot 10^{-3}$ & $0$ \\
				$X\Psi_+ \rightarrow \Psi_-h_2$ & $4.5\cdot 10^{-1}$ & $-2.3\cdot 10^{-1}$ & $0$ \\
				\hline
				$\Psi_+ \rightarrow X\Psi_-$ & \multicolumn{3}{|c|}{$1.3\cdot 10^{-1}$} \\
				\hline
			\end{tabular}
}
		\end{minipage}
		\begin{minipage}{.45\textwidth}\centering
		\includegraphics[width=\textwidth]{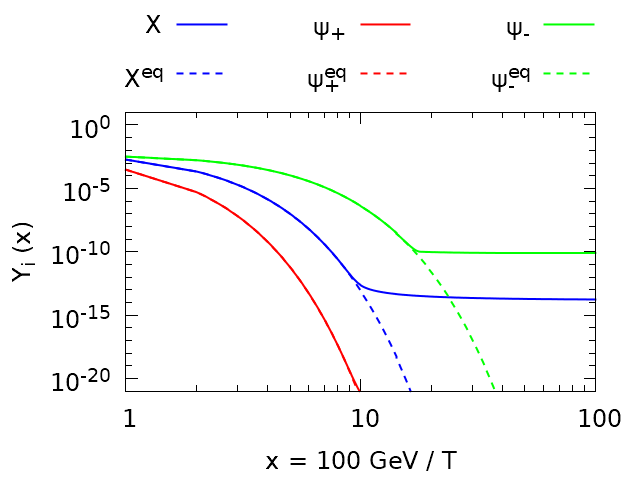}
\scalebox{.85}{
			\begin{tabular}{|c|c|c|c|}
				\hline
				process & $a_N$ & $a_{N+1}$ & $N$ \\
				\hline
				$XX \rightarrow \text{SM}$ & $8.45\cdot 10^{1}$ & $-8.15\cdot 10^{1}$ & $0$ \\
				$\psi_+\psi_+ \rightarrow \text{SM}$ & $1.73\cdot 10^{0}$ & $-1.98\cdot 10^{0}$ & $1$ \\
				$\psi_-\psi_- \rightarrow \text{SM}$ & $2.2\cdot 10^{-2}$ & $-1.2\cdot 10^{-1}$ & $1$ \\
				\hline
				$\Psi_+\Psi_+ \rightarrow XX$ & $7.92\cdot 10^{0}$ & $8.76\cdot 10^{0}$ & $0$ \\
				$XX \rightarrow \Psi_-\Psi_-$ & $1.8\cdot 10^{0}$ & $-1.36\cdot 10^{0}$ & $0$ \\
				$\Psi_+,\Psi_+ \rightarrow \Psi_-\Psi_-$ & $3.26\cdot 10^{1}$ & $-5.8\cdot 10^{-1}$ & $0$ \\
				\hline
				$\Psi_+\Psi_- \rightarrow Xh_1$ & $3\cdot 10^{-2}$ & $9.5\cdot 10^{-2}$ & $0$ \\
				$\Psi_+\Psi_- \rightarrow Xh_2$ & $3.52\cdot 10^{0}$ & $1.03\cdot 10^{1}$ & $0$ \\
				$\Psi_+h_1 \rightarrow X\Psi_-$ & $3.3\cdot 10^{-1}$ & $5.56\cdot 10^{0}$ & $0$ \\
				$\Psi_+h_2 \rightarrow X\Psi_-$ & $3.64\cdot 10^{1}$ & $3.78\cdot 10^{2}$ & $0$ \\
				$X\Psi_+ \rightarrow \Psi_-h_1$ & $5.7\cdot 10^{-1}$ & $-2.9\cdot 10^{-1}$ & $0$ \\
				$X\Psi_+ \rightarrow \Psi_-h_2$ & $5.61\cdot 10^{1}$ & $-3.12\cdot 10^{1}$ & $0$ \\
				\hline
				$\Psi_+ \rightarrow X\Psi_-$ & \multicolumn{3}{|c|}{$1.44\cdot 10^{0}$} \\
				\hline
			\end{tabular}
}
		\end{minipage}\\~\\
		\begin{tabular}{|c|c|c|c|}\hline
		$Y(x=100)$&left&middle&right\\
		\hline
		$Y_X$&$1.1\e{-10}$&$1.7\e{-12}$&$1.7\e{-14}$\\
		$Y_+$&$4.0\e{-49}$&$2.6\e{-52}$&$1.3\e{-55}$\\
		$Y_-$&$3.6\e{-8}$&$1.6\e{-9}$&$7.9\e{-11}$\\
		\hline\end{tabular}
	\end{tabular}}
		\caption{Influence of $g_x$ on the shape of solutions of the Boltzmann equation. Parameters of the plots are: $m_X=300\gev$, $m_+=500\gev$, $m_-=130\gev$, $m_2=150\gev$, $\sin\alpha=0.1$. Coupling constant $g_x$ is $0.1$ (left), $0.3$ (middle), $1.0$ (right).}
		\label{fig:gx1-3}
	\end{figure}
\end{subsection}
\end{section}
\begin{section}{Summary}
In this thesis we have introduced a simple but QFT-consistent and renormalizable extension of the Standard Model with an additional $U(1)_X$ group. The model provides four new particles (a gauge boson, two Majorana fermions and an additional Higgs-like particle). Among the first three of them, 2 or 3 are stable, depending on the values of the parameters, and can be cold dark matter candidates.\\

We have also provided a detailed derivation of the Boltzmann equation and carefully discussed all the assumptions that are made.\\

A \protect\url{C++} code that solves set of the Boltzmann equations for multi-component dark matter has been developed and verified against \micromegas \cite{bib:micromegas} for the 2-component case. We have investigated the influence of the values of the model parameters (masses of the extra particles, Higgs sector mixing angle and $U(1)_X$ coupling constant) on the shape of solutions of the Boltzmann equations.\\

Analysis of our model provides a laboratory to study non-standard effects for interactions of dark matter, such as Higgs portal cancelling and influence of the semi-annihilation processes. It is also possible to analyze the influence of interactions between various components within the dark sector on the final yield of dark matter -- for further reading, see \cite{bib:vfdm}.
\end{section}

\renewcommand\refname{

\end{document}
\section{References}}
\begin{thebibliography}{50}
\bibitem{bib:kt}  E. W. Kolb, M. S. Turner \emph{The Early Universe},
\href{http://inspirehep.net/record/299778}{\emph{Front.Phys.} 69 (1990) 1-547}
\bibitem{bib:structure} J. R. Primack \emph{Dark Matter and Structure Formation in the Universe}, \arxiv{astro-ph/9707285}
\bibitem{bib:zwicky} F. Zwicky \emph{Die Rotverschiebung von extragalaktischen Nebeln},
\href{http://doi.org/10.1007/s10714-008-0707-4}{\emph{Helvetica Physica Acta} 6 (1933) 110–127}
\bibitem{bib:rotcurve} E. Corbelli, P. Salucci \emph{The Extended Rotation Curve and the Dark Matter Halo of M33}, \arxiv{astro-ph/9909252} 
\bibitem{bib:macho} A. Dar et al. \emph{Baryonic Dark Matter and Big Bang Nucleosynthesis},
\href{https://doi.org/ 10.1086/176078}{\emph{Astrophys. J.} 449 (1995) 550}, \arxiv{astro-ph/9405010}
\bibitem{bib:dm} G. Bertone et al. \emph{Particle Dark Matter: Evidence, Candidates and Constraints},
\href{https://doi.org/10.1016/j.physrep.2004.08.031}{\emph{Phys. Rept.} 405 (2005) 279-390}, \arxiv{hep-ph/0404175}
\bibitem{bib:planck} {\bf Planck} Collaboration, P. A. R. Ade et al.
\emph{Planck 2015 results. XIII. Cosmological parameters},
\href{https://doi.org/10.1051/0004-6361/201525830}{\emph{A\&A} 594 (2016) A13}, \arxiv{1502.01589}
\bibitem{bib:wmap} J.-P. Uzan et. al. \emph{WMAP data and the curvature of space},
\href{https://doi.org/10.1046/j.1365-8711.2003.07043.x}{\emph{Mon. Not. Roy. Astron. Soc.} 344 (2003) L65}, \arxiv{astro-ph/0302597}
\bibitem{bib:multidm1} J. S. Bullock, M. Boylan-Kolchin \emph{Small-Scale Challenges to the $\Lambda$CDM Paradigm},
\href{https://doi.org/ 10.1146/annurev-astro-091916-055313}{\emph{ARA\&A} 55 (2017) 343-387}, \arxiv{1707.04256}
\bibitem{bib:multidm2} S. Tulin, H.-B. Yu \emph{Dark Matter Self-interactions and Small Scale Structure}, \arxiv{1705.02358}
\bibitem{bib:cusp} J. F. Navarro et al. \emph{The Structure of Cold Dark Matter Halos},
\href{https://doi.org/10.1086/177173}{\emph{Astrophys. J.} 462 (1996) 563-575}, \arxiv{astro-ph/9508025}
\bibitem{bib:cusp2} W. J. G. de Blok \emph{The Core-Cusp Problem}, \arxiv{0910.3538}
\bibitem{bib:imperfect1} M. Heller et al. \emph{Imperfect fluid Friedmannian cosmology},
\href{https://doi.org/10.1007/BF00645597}{\emph{Astrophys. Space Sci.} 20 (1973) 205}
\bibitem{bib:imperfect2} L. Buoninfante, G. Lambiase \emph{Cosmology with bulk viscosity and the gravitino problem}, \arxiv{1610.01827}
\bibitem{bib:imperfect3} X H Meng, X Dou \emph{Friedmann cosmology with bulk viscosity: a concrete model for dark energy},
\href{https://doi.org/ 10.1088/0253-6102/52/2/36}{\emph{Commun. Theor. Phys.} 52 (2009) 377-382}, \arxiv{0812.4904}
\bibitem{bib:curv} P. Schneider \emph{Extragalactic Astronomy and Cosmology: An Introduction}, Berlin; Heidelberg: Springer-Verlag (2015)
\bibitem{bib:baumann} D. Baumann's lectures on cosmology, \protect\url{http://www.damtp.cam.ac.uk/user/db275/Cosmology/Lectures.pdf}
\bibitem{bib:neutr} S. Hannestad, J. Madsen \emph{Neutrino decoupling in the early Universe},
\href{https://doi.org/10.1103/PhysRevD.52.1764}{\emph{Phys. Rev. D} 52 (1995) 1764-1769}, \arxiv{astro-ph/9506015}
\bibitem{bib:mdbg} M. Duch, B. Grządkowski \emph{Resonance enhancement of dark matter interactions: the case for early kinetic decoupling and velocity dependent resonance width}, \arxiv{1705.10777}
\bibitem{bib:decoWIMP} T. Bringmann, S. Hofmann \emph{Thermal decoupling of WIMPs from first principles},
\href{https://doi.org/10.1088/1475-7516/2007/04/016}{\emph{JCAP} 0704 (2007) 016}, \arxiv{hep-ph/0612238}
\bibitem{bib:chemPot} R. Baierlein \emph{The elusive chemical potential},
\href{https://doi.org/10.1119/1.1336839}{\emph{Am. J. Phys.} 69 (2001) 423}
\bibitem{bib:cpt1} O. W. Greenberg \emph{CPT violation implies violation of Lorentz invariance},
\href{https://doi.org/10.1103/PhysRevLett.89.231602}{\emph{Phys. Rev. Lett.} 89 (2002) 231602}, \arxiv{hep-ph/0201258}
\bibitem{bib:cpt2}  A. Tureanu \emph{CPT and Lorentz invariance: their relation and
violation}
\href{https://doi.org/10.1088/1742-6596/474/1/012031}{\emph{J. Phys.: Conf. Ser.} 474 (2013) 012031}
\bibitem{bib:cp-k} {\bf NA48} collaboration, V. Fanti et al. \emph{A new measurement of direct CP violation in two pion decays of the neutral kaon}
\arxiv{hep-ex/9909022}
\bibitem{bib:cp-b} {\bf LHCb collaboration}, R. Aaij et al. \emph{First observation of CP violation in the decays of Bs mesons}
\arxiv{1304.6173}
\bibitem{bib:vfdm} A. Ahmed et al. \emph{Multi-component dark matter: the vector and fermion case}
\arxiv{1710.01853}
\bibitem{bib:wein} S. Weinberg \emph{Goldstone bosons as fractional cosmic neutrinos},
\href{https://doi.org/10.1103/PhysRevLett.110.241301}{\emph{Phys. Rev. Lett.} 110 (2013) 241301}, \arxiv{1305.1971}
\bibitem{bib:micromegas} G. Belanger et al. \emph{micrOMEGAs : a code for the calculation of Dark Matter properties in generic models of particle interaction}, \arxiv{1402.0787}
\bibitem{bib:calchep} A. Belyaev et al. \emph{CalcHEP 3.4 for collider physics within and beyond the Standard Model}, \arxiv{1207.6082}
\bibitem{bib:sina} G. Chalons et al. \emph{The Higgs singlet extension at LHC Run 2}, \arxiv{1611.03007}
\bibitem{bib:gnuplot} \protect\url{http://www.gnuplot.info/}
\end{thebibliography}
